\documentclass[11pt]{article}
\usepackage{amssymb}
\usepackage{amsmath}
\usepackage{amscd}
\usepackage{latexsym}
\usepackage{graphics}
\usepackage{color}
\usepackage{mathtools}
\usepackage{cite}
\usepackage{tikz}
\usepackage{hyperref}
\usetikzlibrary{arrows}
\input xy
\xyoption{all}

\catcode`@=11
\renewcommand{\section}
{\@startsection{section}{1}{0pt}{\medskipamount}{\medskipamount}{\large\bf}}
\makeatletter\renewcommand{\subsection}
{\@startsection{subsection}{2}{\z@}{-3.25ex plus -1ex minus -.2ex}
{1.5ex plus .2ex}{\it }}

\numberwithin{equation}{section}
\catcode`@=12

\def\a{\alpha}
\def\b{\beta}
\def\g{\gamma}

\def\e{\epsilon}
\def\l{\lambda}

\def\s{\sigma}

\def\m{\mu}
\def\n{\nu}

\def\beq{\begin{equation}}
\def\eeq{\end{equation}}
\def\bea{\begin{eqnarray}}
\def\eea{\end{eqnarray}}

\renewcommand{\e}{\,\mathrm{e}\,}
\newcommand{\id}{{1\!\!1}}
\newcommand{\im}{\,\mathrm{i}\,}
\newcommand{\diff}{\mathrm{d}}

\newcommand{\R}{{\mathbb{R}}}

\newcommand{\C}{{\mathbb{C}}}
\newcommand{\Z}{{\mathbb{Z}}}

\newcommand{\ca}{{\cal{A}}}
\newcommand{\cf}{{\cal{F}}}

\newcommand{\mbf}[1]{{\boldsymbol {#1} }}

\def\>{\rangle}
\def\<{\langle}
\def\+{\dagger}
\def\={\ =\ }

\newcommand{\lb}{\left(}
\newcommand{\rb}{\right)}

\newcommand{\com}[2]{\big[#1,#2\big]}

\def\ang1{$15$} 
\def\ang2{$15$}
\def\haken{\mathbin{\hbox to 6pt{%
\vrule height0.4pt width5pt depth0pt
\kern-.4pt
\vrule height6pt width0.4pt depth0pt\hss}}}
\begin{document}

\begin{titlepage}
\setcounter{page}{0}
\begin{flushright}
EMPG--17--08\\
\end{flushright}

\vskip 1.5cm

\begin{center}

{\LARGE\bf Sasakian quiver gauge theories and instantons \\[4pt]
on cones over round and squashed seven-spheres}

\vspace{15mm}

{\large Jakob C. Geipel${}^1$}, \ \ 
{\large Olaf Lechtenfeld${}^{1,2}$}, \ \ 
{\large Alexander D. Popov${}^{1}$} \ \ and \ \ 
{\large Richard J. Szabo${}^3$}
\\[5mm]
\noindent ${}^1$
{\em Institut f\"ur Theoretische Physik}\\ 
{\em Leibniz Universit\"at Hannover}\\
{\em Appelstra\ss e 2, 30167 Hannover, Germany}\\
Email: {\tt jakob.geipel@itp.uni-hannover.de, alexander.popov@itp.uni-hannover.de}
\\[5mm]
\noindent ${}^2$
{\em Riemann Center for Geometry and Physics}\\
{\em Leibniz Universit\"at Hannover}\\
{\em Appelstra\ss e 2, 30167 Hannover, Germany}\\
Email: {\tt olaf.lechtenfeld@itp.uni-hannover.de}
\\[5mm]
\noindent ${}^3$
{\em Department of Mathematics\\
Heriot-Watt University\\
Colin Maclaurin Building, Riccarton, Edinburgh EH14 4AS, U.K.}\\
and
{\em Maxwell Institute for Mathematical Sciences, Edinburgh, U.K.}\\
and
{\em The Higgs Centre for Theoretical Physics, Edinburgh, U.K.}\\
{Email: {\tt R.J.Szabo@hw.ac.uk}}

\vspace{15mm}

\begin{abstract}
\noindent
We study quiver gauge theories on the round and squashed seven-spheres, and orbifolds thereof. They arise by imposing $G$-equivariance on the 
homogeneous space $G/H=\mathrm{SU}(4)/\mathrm{SU}(3)$ endowed with
its Sasaki-Einstein structure, and $G/H=\mathrm{Sp}(2)/\mathrm{Sp}(1)$
as a 3-Sasakian
manifold. In both cases we describe the equivariance conditions and
the resulting quivers. We further study the moduli 
spaces of instantons on the metric cones over these spaces by using
the known description for Hermitian Yang-Mills instantons on Calabi-Yau cones.
It is shown that the moduli space of instantons on the hyper-K\"ahler
cone can be described as the intersection of three Hermitian Yang-Mills moduli spaces. We
also study moduli spaces of translationally invariant instantons on the metric cone
$\mathbb{R}^8/\mathbb{Z}_k$ over $S^7/\mathbb{Z}_k$.
\end{abstract}

\end{center}
\end{titlepage}

{\baselineskip=12pt 
\tableofcontents}

\newpage

\section{Introduction\label{sec:intro}}
\label{sect:intro}

Equivariant dimensional reduction and quiver gauge theories in the sense of \cite{Popov:2005ik,Lechtenfeld:2007st,Dolan:2009ie,Dolan:2009nz,Dolan:2010ur}
have been extended recently from applications on K\"ahler coset spaces \cite{Lechtenfeld:2008nh,Lechtenfeld:2006wu,Rahn:2009yt} to homogeneous Sasaki-Einstein manifolds. This \emph{Sasakian quiver gauge theory} \cite{Lechtenfeld:2014fza} is motivated by the close relation between 
K\"ahler and Sasakian geometry, as well as by the prominent role of Sasaki-Einstein manifolds
in the AdS/CFT correspondence. The examples covered so far
include the sphere $S^5$ (and orbifolds thereof)
\cite{Lechtenfeld:2015ona} and the conifold $T^{1,1}$ \cite{Geipel:2016uij} in five dimensions, and 
the Aloff-Wallach space $X_{1,1}$ \cite{Geipel:2016hpk} in seven
dimensions which
also involves aspects of its 3-Sasakian structure. The aim of this paper is to compare Sasakian and 3-Sasakian quiver gauge theories on the 
round and squashed
sphere $S^7$ in seven dimensions, realized respectively as the homogeneous spaces $\mathrm{SU}(4)/\mathrm{SU}(3)$
and $\mathrm{Sp}(2)/\mathrm{Sp}(1)$. Developing these two
examples then essentially exhausts the list of physically interesting
Sasakian quiver gauge theories.

As Sasaki-Einstein manifolds, the round and squashed seven-spheres
naturally appear in the context of AdS$_4$/CFT$_3$ duality in 
M-theory. The effective 
$\mathcal{N}=8$ supergravity theory on $\mathrm{AdS}_4 \times  S^7$ is
broken to an $\mathcal{N}=1$ theory when the usual round metric of
$S^7$ is deformed to that of the squashed seven-sphere
\cite{Awada:1982pk,Duff:1983ajq}. After orbifolding the 
seven-sphere by the cyclic group $\mathbb{Z}_{k}$, these backgrounds describe the near
horizon geometry of coincident M2-branes
situated at the conical singularity of an eight-dimensional cone $X^8$ over
$S^7/\mathbb{Z}_{k}$. Their low energy effective worldvolume theories
are three-dimensional superconformal Chern-Simons theories at level
$k$, with $\mathcal{N}=6$ supersymmetry and global R-symmetry group
$\mathrm{SU}(4)$ for the round metric on $S^7$~\cite{Aharony:2008ug}, and with
$\mathcal{N}=1$ supersymmetry and global R-symmetry group
$\mathrm{Sp}(2)\subset\mathrm{SU}(4)$ for the squashed metric on $S^7$~\cite{Ooguri:2008dk}. In
these theories the Chern-Simons gauge fields on $\mathbb{R}^3$ couple
to scalar and spinor fields, so that our equivariant dimensional
reductions over the eleven-dimensional spacetimes $\mathbb{R}^3\times
X^8$ lead to Sasakian quiver gauge theories whose Higgs branch moduli spaces could
shed light on the generic vacuum structure of the possible low energy descriptions.

In the general context of dimensional reduction of gauge theories, a natural
condition to impose on connections over internal homogeneous spaces $G/H$ is equivariance with 
respect to the group $G$; this is known as \emph{equivariant dimensional reduction}. 
It has a natural close relation with the representation theory of quivers.
Quiver gauge theories allow one to organise the physical degrees of freedom that are present 
in a chosen representation of the group $G$ inside the structure group
in terms of directed graphs which represent
the quivers. In this way, they take into account more general solutions
to the equivariance condition
than the {scalar solution}, used for instance in~\cite{Harland:2011zs,Haupt:2011mg,Ivanova:2012vz}. An equivariant connection
is then characterized by a quiver for a chosen representation of $G$, and imposing instanton equations on this connection yields relations for
the quiver and gradient flow equations. In this paper we shall
follow the same route for the examples we consider: We will first discuss the equivariance conditions in detail and then describe the moduli spaces
of instantons, which determine vacuum moduli spaces for the supersymmetric field theories discussed above. 

This paper is organized as follows. In Section~\ref{sec:Review} we
briefly review pertinent aspects of quiver gauge theories in the context of
equivariant dimensional reduction and their relation to moduli spaces
of instantons on homogeneous manifolds. The core of the present work is the description of 
Sasakian quiver gauge theory on the round seven-sphere in Section \ref{sect:sasaki_quiver} and of 3-Sasakian quiver gauge theory on
the squashed seven-sphere in Section \ref{sect:3Sasaki_QGT}. In both
cases we describe the geometry of the coset spaces $G/H$, derive the equivariance 
conditions and illustrate them with explicit examples for some
low-dimensional representations of $G$. Then we consider instantons on the metric cones,
building on the general theory for Calabi-Yau cones, and show that the moduli spaces of instantons on hyper-K\"ahler cones can 
be reduced to the intersection of a $\C P^1$-family of Hermitian
Yang-Mills moduli spaces. In Section~\ref{sec:translationinvariant} we
study moduli spaces of
translationally invariant instantons on the metric cone
$\R^8/\mathbb{Z}_k$ over $S^7/\mathbb{Z}_k$. Finally, in Section~\ref{sec:Summary} we conclude
with an overview of the main achievements of this paper. Two
appendices at the end of the paper contain technical details about
geometric structures, connections, and explicit representations of the Lie algebras of
$G=\mathrm{SU}(4)$ and $G=\mathrm{Sp}(2)$ which are used throughout the main text.

\section{Quiver gauge theory and equivariant dimensional reduction\label{sec:Review}}
In this section we briefly review some technical preliminaries that
are needed in this paper. We begin by reviewing the theory of equivariant vector bundles and their relation to quiver 
representations, and then relate the equivariance condition to the
conventional approach used in studies of connections on homogeneous
spaces. We also discuss the generalized instanton equation.
\subsection{Equivariant vector bundles}
Let us begin with the basics of quiver gauge theory that we will apply in the remainder of this article.
For details on the  physical motivation and an outline of the construction, we refer to the reviews 
\cite{Lechtenfeld:2007st,Dolan:2010ur}, whereas a rigorous mathematical account can be 
found for example in \cite{AlvarezConsul:2001uk}.

The general setup is that of a gauge theory on a product $M^d \times
G/H$ of a $d$-dimensional Riemannian manifold $M^d$ and a homogeneous
manifold $G/H$. The natural objects in geometric considerations of
gauge theories are principal fibre bundles, but in this paper we will
work with (associated) complex
vector bundles; the formulation of equivariant dimensional reduction
and the corresponding quiver gauge theories in the setting of principal
bundles can be found
in~\cite{Szabo:2014zua,Alvarez-Consul:2016ibn}. Thus let $\pi: \mathcal{E}\rightarrow M^d \times G/H$ be a Hermitian vector bundle of rank $r$, and assume that the group $G$ 
acts trivially on the Riemannian manifold $M^d$. The bundle
$\mathcal{E}$ is $G$-equivariant if the $G$-action on the base
manifold and on $\mathcal{E}$ 
commute with the projection map $\pi$, and if it induces isomorphisms among the fibres. 
$G$-equivariant bundles $\mathcal{E}\rightarrow M^d \times G/H$ are in one-to-one correspondence with $H$-equivariant 
bundles $E\rightarrow M^d$, with the correspondence given by
induction of vector bundles $\mathcal{E}=G\times_HE$ \cite{AlvarezConsul:2001uk}.

Since the subgroup $H$ acts trivially on the base space, the fibres ${E}_x\cong \C^r$ carry $H$-representations due to equivariance. 
We assume that this $H$-representation stems from a $G$-representation $\mathcal{D}$ which decomposes under restriction to $H$ as 
\bea
\label{eq:restriction_D}
\mathcal{D}\big|_{H}\= \bigoplus_{j=0}^m\, \rho_j
\eea
into representations $\rho_j$ of $H$.\footnote{We will denote the
  representations of the corresponding Lie algebras with the same
  symbols.} Then the structure group of the bundle is broken as
\bea
\mathrm{U}(r) \ \longrightarrow \ \prod_{j=0}^m \, \mathrm{U}(r_j) \qquad
\text{with} \quad \sum_{j=0}^m\, r_j \=r \ .
\eea
The bundle $E\to M^d$ decomposes in the same way under the action of
$H$ as a Whitney sum and admits an isotopical decomposition
\bea
\label{eq:iso_dec}
E \= \bigoplus_{j=0}^m \, E_j \otimes V_j  \ , 
\eea
where the vector space $V_j$ carries the representation $\rho_j$ and
$H$ acts trivially on the vector bundles $E_j\to M^d$. The
induction map $\mathcal{E}= G \times_H E$ yields an isotopical
decomposition of the bundle $\mathcal{E}$ as well, where
the action of the group $G$ then connects 
different summands of the decomposition \eqref{eq:iso_dec}
 by bundle maps. 

This induces a representation (in the category of vector bundles) of a quiver $\mathcal{Q} = (\mathcal{Q}_0, \mathcal{Q}_1)$ \cite{derksen2005quiver}: 
For each representation $\rho_j$ one associates a vertex $v_j\in\mathcal{Q}_0$, representing a 
vector bundle $E_j$, 
 and, if the $G$-action 
connects $\rho_j$ and $\rho_i$, an arrow $\phi_{ij}\in\mathcal{Q}_1$
between vertices $v_j$ and $v_i$ is associated, representing a homomorphism from $E_j$ to $E_i$; such an entity is sometimes also called a quiver bundle~\cite{AlvarezConsul:2001uk}. 
In this way, equivariant bundles over the homogeneous space $G/H$
correspond to linear quiver 
representations (in the category of vector spaces), and their construction reduces to studying representations of $G$ and their weight diagrams after suitably collapsing along the generators 
of the subgroup $H$.\footnote{This collapsing is nothing other than
  obtaining the $H$-representations $\rho_i$ from the weight diagram.}

\subsection{Instantons on homogeneous spaces\label{sec:insthomsp}}

The condition giving rise to the quiver, which is referred to as the \emph{equivariance condition}, ensures the invariance of gauge connections on 
coset spaces, and therefore naturally occurs when studying instantons  over reductive homogeneous  spaces $G/H$ where the Lie algebras $\mathfrak{g}$ and $\mathfrak{h}$ of the Lie groups $G$ and $H$ decompose according to
\bea
\mathrm{span}\langle I_{\m}\rangle \ = \ \mathfrak{g} \= \mathfrak{m} \oplus \mathfrak{h} \ = \ \mathrm{span}\langle I_{a}\rangle 
\oplus \mathrm{span}\langle I_{j}\rangle \ .
\eea
The typical  approach we will apply here is to start from a generalized instanton and express the connection locally in terms of matrices. On Riemannian manifolds with Killing spinors, a connection
$\Gamma$ is called a \emph{generalized instanton} \cite{CORRIGAN1983452,WARD1984381} if its curvature $\cf_{\Gamma} = \diff \Gamma + \Gamma \wedge \Gamma$ satisfies 
\cite{Harland:2011zs}
\bea
\label{eq:eq_inst}
\star\, \cf_{\Gamma} \= - \cf_{\Gamma} \wedge \star\, Q \ ,
\eea
where $Q$ is an invariant 4-form constructed as a bilinear in the Killing spinors. Solutions to this {first}-order equation also satisfy 
the {second}-order Yang-Mills equation. A potentially occurring torsion term vanishes for our cases of interest -- Sasaki-Einstein and 3-Sasakian geometries -- 
due to the properties of the Killing spinors in these instances. A special instanton solution is constructed in~\cite{Harland:2011zs} which is based solely on the geometry
induced by the Killing spinors. This instanton is referred to as the \emph{canonical connection}, and its explicit expression is used below.

Given any instanton $\Gamma$ (not necessarily the canonical one) the
general form of a gauge connection $\ca$ on $M^d\times G/H$
(see e.g. \cite{Bauer:2010fia,Ivanova:2012vz,Haupt:2011mg}) 
then reads
\bea
\label{eq:equ_conn}
\ca \=  A +\Gamma + \sum_{a=1}^{\dim(\mathfrak{m})}\, X_a \otimes e^a \ ,
\eea
where $A$ is a connection on the vector bundle $E$ over the Riemannian
manifold $M^d$ and
$\{e^a\}$ is a local frame on $T_e(G/H)\cong\mathfrak{m}$. For our highly symmetric cases, 
where the quotient of the isometry group $G$ by the structure group
$H$ of the principal bundle $G\to G/H$
coincides with the given realization as a homogeneous space
$G/H$,\footnote{This is the guiding principle of the construction in
  \cite{Ivanova:2012vz}.} the canonical connection is simply given by 
\bea
\Gamma \= \sum_{j=1}^{\dim(\mathfrak{h})}\, I_j \otimes e^j 
\eea
in \eqref{eq:equ_conn}.\footnote{Notice that the generators of $H$ 
as well as the connection $A$ on $M^d$ have a block diagonal form with respect to the isotopical decomposition. 
Hence they commute when evaluating the curvature $\cf=\diff\ca+\ca\wedge\ca$.}
To yield an invariant connection over the homogeneous space, the
matrices $X_a$ must act on $I_i$ in the same way that the generators of $\mathfrak{g}$ do, 
\bea
\label{eq:equivariance_condition}
\com{I_i}{X_a} \= f_{ia}^b\, X_b \ ,
\eea
which means that no terms containing the mixed 2-forms $e^i \wedge e^a$ may appear in the curvature $\cf = \diff \ca + \ca \wedge \ca$. This leads to the 
quivers discussed before with the bundle endomorphisms $X_a$
represented by the arrows. The Cartan generators contained in the subalgebra $\mathfrak{h}$ determine the shape of the quiver, once one has chosen a representation for them.
Collapsing the weight diagram along the ladder operators of
$\mathfrak{h}$, one obtains the decomposition \eqref{eq:restriction_D}, and the $G$-action then yields the arrows of the
quiver. This procedure will be illustrated for the two examples 
$G=\mathrm{SU}(4)$ with $H=\mathrm{SU}(3)$ and $G=\mathrm{Sp}(2)$ with $H=\mathrm{Sp}(1)$ in the remainder of this paper.

Because \eqref{eq:equ_conn} with \eqref{eq:equivariance_condition} only ensure the equivariance of the connection, one still has to impose the instanton
equation on the curvature. On cones over Sasaki-Einstein manifolds,
one can use the Hermitian Yang-Mills equation to impose the instanton condition. This will be exploited later on.

The equivariance condition determines the general block form of the matrices $X_a$, expressing the gauge connection, which induces the 
quivers. The homomorphisms from $\mathrm{Hom}(E_j,E_i)$ as entries in these matrices, or equivalently as arrows of the quivers,
are restricted by relations induced by the instanton equation. If one considers only rank one vector bundles $E_i$ and takes all 
functions occurring in each $X_a$ to be the same, then one obtains the scalar solution $X_a = \l_a (x)\, I_a$ which has been previously used for 
explicit constructions~\cite{Harland:2011zs, Haupt:2015wdq}. 


\section{Sasakian quiver gauge theory on the round seven-sphere}

\label{sect:sasaki_quiver}
In this section we consider quiver gauge theory on $S^7 \cong
\mathrm{SU}(4)/\mathrm{SU}(3)$, regarded as a Sasaki-Einstein manifold. 
Since the canonical connection, the structure equations and the
instanton equations of any particular odd-dimensional sphere can be easily generalized to 
all odd-dimensional spheres, the exposition will closely follow that of the five-sphere in \cite{Lechtenfeld:2015ona}. We start by describing 
the geometry of the homogeneous space and orbifolds thereof, including the canonical connection with respect to the Sasaki-Einstein structure.
We then use this to derive the general form for the equivariant connection and provide some explicit examples of the quivers
induced by the equivariance condition. We conclude by formulating the Hermitian Yang-Mills equation on the metric cone and describing the 
moduli space of solutions.   
\subsection{Geometry of $S^7 \cong \mathrm{SU}(4)/\mathrm{SU}(3)$}
\label{sect:geometry_s7}
\paragraph{Local section.} The geometric description of $S^7 \cong \mathrm{SU}(4)/\mathrm{SU}(3)$ used in this paper is based on 
its realization as a circle bundle over the K\"ahler 3-fold $\C P^3$
via a commuting diagram of fibrations
\bea
\label{eq:s7_fibration}
\small
\label{eq:s7_bundle}
\begin{tikzpicture}[->,scale=1.1]
    \node (a) at (-2,0) {$\mathrm{SU}(4)$};
    \node (b) at (1,0) {$S^7$};
    \node (c) at (1,-1) {$\C P^3$};

     \path[thin,->](a) edge node [above]{$\scriptstyle{\mathrm{SU}(3)}$} (b);
     \path[thin,->](b) edge node [right]{$\scriptstyle{\mathrm{U}(1)}$} (c);
     \path[thin,->](a) edge node [sloped, anchor=center, below]{$\scriptstyle{\mathrm{S}( \mathrm{U}(3) \times \mathrm{U}(1))}$} (c);
 
\end{tikzpicture}
\normalsize
\eea
This allows for the introduction of local coordinates on $S^7$ by considering a section of the bundle $\mathrm{SU}(4)\rightarrow \C P^3$ 
in the following way, which is analogous to the procedure employed in \cite{Lechtenfeld:2006wu,Lechtenfeld:2015ona}.
In a patch $\mathcal{U}_0 \coloneqq \left\{ [z^0:z^1:z^2:z^3] \in \C P^3\ \big|\ z^0 \neq 0\right\}\subset\C P^3$, one defines
\bea
\label{eq:local_coord}
Y \ = \ (y^1,y^2,y^3)^\top
\coloneqq
 \lb \frac{z^1}{z^0}, \frac{z^2}{z^0}, \frac{z^3}{z^0} \rb^\top
\eea
and the complex $4\times4$ matrix
\bea
V \ \coloneqq \ \frac1\gamma\, \begin{pmatrix}
             1     & {Y}^{\+}\\
             - {Y} & \Lambda
            \end{pmatrix} \qquad \text{with} \quad \Lambda \ \coloneqq \
 \gamma \, \id_3 - \frac{1}{1+\gamma} \, Y \, Y^{\+} \qquad \mbox{and} \qquad
\gamma \ \coloneqq \ \sqrt{1+ Y^{\+}\,Y} \ .
\eea
By definition, they have the properties 
\bea
\Lambda \, Y \= Y \ , \quad Y^{\+} \, \Lambda \= Y^{\+} \qquad
\mbox{and} \qquad \Lambda^2 \= \gamma^2 \, \id_3 - Y\, Y^{\+} \ ,
\eea
so that $V$ indeed is an element of $\mathrm{SU}(4)$, i.e. $V^{\+}\,
V= V \, V^{\+}=\id_4$. Hence the matrix $V$ is a local section of the bundle
$\mathrm{SU}(4) \to \C P^3$. The Maurer-Cartan form $A_0 \coloneqq
V^{-1} \, \diff V$ provides left $\mathrm{SU}(4)$-invariant
\mbox{1-forms} on $\C P^3$:
\bea
A_0 \= V^{\+} \, \diff V \ \eqqcolon \ 
\begin{pmatrix}
                                 - 3a       & \beta^{\+}\\
                                  - {\beta} & B
\end{pmatrix}
 \qquad \text{with} \quad
a \= -\frac{1}{2\gamma^2} \, \lb  Y^{\+} \, \diff Y - \diff Y^{\+} \,
Y\rb \ , \\[4pt]
\nonumber \beta \=  \frac{1}{\gamma^2}\, \Lambda \, \diff Y \qquad \text{and} \qquad
B \= \frac{1}{\gamma^2}\,  \Big(  Y \, \diff Y^{\+} + \Lambda \, \diff
\Lambda -\frac12\, \diff\big(Y^\dag\,Y\big)\, \id_3\Big) \ .
\eea
The flatness of the connection $A_0$ yields the equations
\bea
&& \nonumber 3 \, \diff a \= - \beta^{\+} \wedge \beta \= \sum_{\a=1}^3\, \beta^\a
\wedge \bar{\beta}^{\bar\a} \ , \quad
\diff \beta \= - 3a\wedge \beta - B \wedge \beta \quad \text{and}
\quad \diff B \= \beta \wedge \beta^{\+} - B \wedge B \ . \\ &&
\eea
A section of the $\mathrm{SU}(3)$-bundle $\mathrm{SU}(4) \to S^7$ can now be obtained by including
 the additional $\mathrm{U}(1)$ factor in the fibration \eqref{eq:s7_fibration} as
\bea
\label{eq:bundle_s7}
S^7 \ni \lb y^1,y^2,y^3, \phi \rb \ \longmapsto \ \tilde{V} \coloneqq
V \, \mathrm{diag} \big( \e^{3\im \phi}, \e^{-\im \phi}, \e^{-\im
  \phi}, \e^{-\im \phi}\big) \ .
\eea
The corresponding canonical flat connection 
$\tilde{A}_0 \coloneqq \tilde{V}^{\+} \, \diff \tilde{V}$ reads
\small
\bea
\label{eq:can_flat_s7}
\nonumber \tilde{A}_0 &=& \begin{pmatrix}
                -3 a + 3 \im \diff \phi  & \e^{-4 \im \phi}\, \beta^{\+}\\
                 - \beta \e^{4 \im \phi} & B - \im \diff \phi\, \id_3 
                           \end{pmatrix}\\[4pt]
&\eqqcolon&
\begin{pmatrix}
  3 \im \m_7 \, e^7 & \zeta_1 \, \Theta^1 & \zeta_2\, \Theta^2 &
  \zeta_3 \, \Theta^3\\
  - \zeta_1 \, \bar\Theta^{\bar{1}} & - \im \m_7 \, e^7 + 2 \im \m_8\,  e^8 & \lambda_4\, \Theta^4 & \lambda_5\, \Theta^5\\
  -\zeta_2\, \bar\Theta^{\bar{2}} & -\lambda_4\, \bar\Theta^{\bar{4}} & - \im
  \m_7 \, e^7 - \im \m_8 \, e^8 - \im \m_9\, e^9& \lambda_6 \Theta^6\\
  - \zeta_3 \bar\Theta^{\bar{3}} & -\lambda_5 \, \bar\Theta^{\bar{5}} &
  \lambda_6 \, \bar\Theta^{\bar{6}}& -\im \m_7\, e^7 -\im \m_8\, e^8 + \im
   \m_9 \, e^9
\end{pmatrix}
\eea
\normalsize
which defines left $\mathrm{SU}(4)$-invariant \mbox{1-forms} and the corresponding generators 
of $\mathrm{SU}(4)$ in the fundamental representation. We also introduce real 1-forms 
\bea
e^{2\a -1} - \im e^{2\a} \ \coloneqq \ \Theta^{\a} \qquad \text{for} \quad
\a\=1,2,3 \ ,
\eea
and an orthonormal frame metric 
\bea
\label{eq:metric_s7}
\diff s^2 \= \sum_{\m=1}^7 \, e^{\m} \otimes e^{\m}
\eea
on $T_e S^7 \cong \mathrm{span} \langle e^1, \ldots, e^7\rangle \cong \mathfrak{m}$,
 where the generators of the reductive homogeneous space split according to
\bea
\mathfrak{su}(4) \ = \ \mathfrak{g} \= \mathfrak{m} \oplus
\mathfrak{h} \ \cong \ T_e S^7 \oplus \mathfrak{su}(3)
\eea
with respect to a left-invariant metric. The real parameters $\m_i$ and $\zeta_i$ in the definition of the 1-forms \eqref{eq:can_flat_s7} will be fixed by
the condition that \eqref{eq:metric_s7} defines a Sasaki-Einstein metric. 
\paragraph{Sasaki-Einstein geometry.} 
Among the available equivalent definitions of a Sasaki-Einstein manifold, we use
the one that declares the Riemannian manifold $S^7$ to be
Sasaki-Einstein if its eight-dimensional metric cone 
$\lb \mathbb{R}^+ \times S^7, g_{\rm cone}\rb$ is Calabi-Yau, i.e.~a Ricci-flat K\"ahler manifold,
with the cone metric\footnote{For explicit calculations we will use
  the conformally equivalent cylinder metric $\diff s^2_{\rm cyl}$.}
\bea
\diff {s}^2_{\rm cone} \= r^2\, \diff s^2 + \diff r \otimes \diff r \=
r^2 \, \Big(\, \diff s^2 + \frac{\diff r}{r} \otimes \frac{\diff r}{r}\,
\Big) \= r^2 \, \diff {s}^2_{\rm cyl} \ .
\eea
One introduces a complex structure $J$ by declaring the 1-forms $\Theta^{\a}$ for $\a=1,2,3$ and the 1-form
\bea
\Theta^0 \ \coloneqq \ \frac{\diff r}{r} - \im e^7 \ \eqqcolon \ e^{\tau} - \im e^7
\eea 
to be holomorphic, ${J}\Theta^{\a}=\im \Theta^{\a}$. In terms of these 1-forms, the cone metric reads
\bea
\diff s^2_{\rm cone} \= r^2 \, \sum_{\a=0}^3\, \Theta^{\a} \otimes \bar\Theta^{\bar{\a}}
\eea
and it yields the K\"ahler form 
\bea
\Omega^{1,1} \ \coloneqq \ - \tfrac{\im}{2} \, r^2 \,
\big(\Theta^{0\bar{0}}+ \Theta^{1\bar{1}}+ \Theta^{2\bar{2}}+
\Theta^{3\bar{3}} \big) \ =: \ - \tfrac{\im}{2} \, r^2 \,
\Theta^{0\bar{0}} + \omega \ ,
\eea
where we generally denote
$\Theta^{\alpha\bar\beta\gamma\cdots}=\Theta^\alpha\wedge
\bar\Theta^{\bar\beta}\wedge \Theta^{\gamma}\wedge\cdots$, etc.
Using the structure equations \eqref{eq:str_s7_parameter} induced by flatness of the connection \eqref{eq:can_flat_s7}, one shows that the closure of this form
requires 
\bea
\label{eq:se_cond1}
\zeta \ \coloneqq \ \zeta_1 \= \zeta_2 \= \zeta_3 \qquad \text{and}
\qquad \m_7 \= \tfrac{1}{3}\, \zeta^2 \ .
\eea
Furthermore, in order for the cone to be Calabi-Yau, the holomorphic 4-form 
\bea
\Omega^{4,0} \ \coloneqq \ r^4 \, \Theta^{1} \wedge \Theta^{2} \wedge \Theta^{3} \wedge \Theta^{0}
\eea
must be closed. This condition yields
\bea
\label{eq:se_cond2}
\zeta^2 \= 1 \qquad \text{and} \qquad \m_7 \=\tfrac{1}{3} \ ,
\eea
and for definiteness we fix the undetermined parameters $\l_i=1$ for $i=4,5,6$, $\m_8=\frac{1}{6}$ and 
$\m_9=\frac{1}{2}$. This choice leads to the structure equations 
\bea
\label{eq:structure_s7_1}
\nonumber  \diff \Theta^1 &=& - \tfrac{4}{3} \im e^7 \wedge \Theta^1 +
\tfrac{1}{3} \im e^8 \wedge \Theta^1 + \Theta^{2\bar{4}}+
\Theta^{3\bar{5}} \ ,\\[4pt]
\nonumber \diff \Theta^2  &=& - \tfrac{4}{3} \im e^7 \wedge \Theta^2 -\tfrac{1}{6} \im e^8 \wedge \Theta^2 - \tfrac{1}{2} \im e^9 \wedge \Theta^2 - 
                                  \Theta^{14}+ \Theta^{3\bar{6}} \ ,\\[4pt]
\nonumber \diff \Theta^3 &=& - \tfrac{4}{3} \im e^7 \wedge \Theta^3- \tfrac{1}{6} \im e^8 \wedge \Theta^3 + \tfrac{1}{2} \im e^9 \wedge \Theta^3- \Theta^{15}
                                - \Theta^{26} \ ,\\[4pt]
\diff e^7 &=& - \im
\big(\Theta^{1\bar{1}}+\Theta^{2\bar{2}}+\Theta^{3\bar{3}} \big) \= 2\, \omega \ .
\eea
\paragraph{Canonical connection.}
According to the general construction in \cite{Harland:2011zs}, the
Sasaki-Einstein metric provides an instanton solution. From the
last structure equation 
in \eqref{eq:structure_s7_1} we see that, as expected, the 1-form
$\eta \coloneqq e^7$ is the contact form dual to the Reeb vector field
of the circle fibration, and $\omega$ is the 
K\"ahler form of the leaf space underlying the Sasaki-Einstein structure.  The \emph{canonical connection} of such a structure, 
in the sense of \cite{Harland:2011zs},
is determined by the 3-form
\bea
P \ \coloneqq \ \eta \wedge \omega \= \tfrac{1}{2}\, \eta \wedge \diff
\eta \= e^7 \wedge \lb e^{12}+e^{34}+e^{56}\rb \ ,
\eea
where $e^{\mu\nu\cdots}=e^\mu\wedge e^\nu\wedge\cdots$,
and the torsion
\bea
T^7 \= P_{7\m\n}\, e^{\m\n} \qquad \text{and} \qquad T^a\=
\tfrac{2}{3} \, P_{a\m\n}\, e^{\m\n} \qquad \mbox{for} \quad a\=1,\dots,6 \ .
\eea
With these torsion components, the structure equations take the form
\bea
\nonumber \diff \Theta^1 &=& -2\, \Theta^1 \wedge \im e^8 +
\Theta^{2\bar{4}} +\Theta^{3\bar{5}} +\lb T^1 - \im T^2 \rb \ ,\\[4pt]
\nonumber \diff \Theta^2 &=&  \Theta^2 \wedge \im e^8 + \Theta^2
\wedge \im e^9 -\Theta^{14} +\Theta^{3\bar{6}} +\lb T^3-\im T^4\rb \ ,\\[4pt]
\nonumber \diff \Theta^3 &=& \Theta^3 \wedge \im e^8 - \Theta^3 \wedge
\im e^9 - \Theta^{15} -\Theta^{26} +\lb T^5-\im T^6\rb \ ,\\[4pt]
\diff e^7 &=& T^7 \ .
\eea
By writing $\diff e^{\m}=-\Gamma_{\n}^{\m} \wedge e^{\n} +T^{\m}$ one obtains the connection matrix
\bea
\Gamma\= 
\begin{pmatrix}
 -2 \im e^8 	& \bar\Theta^{\bar{4}} 	& \bar\Theta^{\bar{5}} 	&0\\
-\Theta^4    	&      \im e^8 +\im e^9 & \bar\Theta^{\bar{6}} 	&0\\
-\Theta^5    	& -\Theta^6   		&  \im e^8 - \im e^9 	&0\\
0 & 0& 0 &0
\end{pmatrix} \ .
\eea
Consequently, the canonical connection of the Sasaki-Einstein structure on $S^7$ 
is given by\footnote{This connection coincides with the one obtained by declaring the torsion to be
given by $T( X,Y) = - \com{X}{Y}_{\mathfrak{m}}$ for vector fields $X,Y \in TS^7$.} 
\bea
\label{eq:can_con_s7}
\Gamma \= I_8 \, e^8  + I_9 \, e^9  +I_4^+ \, \Theta^4 + I_5^+ \,
\Theta^5  +I_6^+ \, \Theta^6  + I_{\bar{4}}^- \, \bar\Theta^{\bar{4}}
+I_{\bar{5}}^- \, \bar\Theta^{\bar{5}}  +I_{\bar{6}}^- \, \bar\Theta^{\bar{6}}
\ \eqqcolon \ \sum_{\m=8}^{15} \, \hat{I}_{\m}\, {e}^{\m} \ .
\eea
The canonical connection involves all the generators of $\mathfrak{h}=\mathfrak{su}(3)$ in this way because the realization
of $S^7$ as the homogeneous space $\mathrm{SU}(4)/\mathrm{SU}(3)$ is
isomorphic to the quotient of its isometry group $\mathrm{SU}(4)$ by the structure group 
$\mathrm{SU}(3)$ of the principal bundle $\mathrm{SU}(4)\to S^7$, as
used in the general construction of \cite{Ivanova:2012vz}. It may be
compared to the more involved situation
on the Aloff-Wallach space $X_{1,1}$~\cite{Geipel:2016hpk}, where the
representation as a coset space is related to the 3-Sasakian structure
instead.

The curvature $\cf_{\Gamma} = \diff \Gamma +\Gamma \wedge \Gamma$ of the canonical connection is given by
\bea
\label{eq:curv_can_con_s7}
\cf_{\Gamma} &=& I_4^+ \, \Theta^{\bar{1}2} + I_5^+\, \Theta^{\bar{1}3} + I_6^+\, \Theta^{\bar{2}3} + I_{\bar{4}}^-\, \Theta^{1\bar{2}}  + I_{\bar{5}}^-\, \Theta^{1\bar{3}} 
+I_{\bar{6}}^-\, \Theta^{2\bar{3}}\\
\nonumber && + \, I_8 \, \big( 2 \im \Theta^{1\bar{1}}- \im \Theta^{2\bar{2}}- \im \Theta^{3\bar{3}}\big) 
+ I_9 \, \big( -\im \Theta^{2\bar{2}}+ \im \Theta^{3\bar{3}}\big)
\eea
and it indeed solves the instanton equation \eqref{eq:eq_inst} with the 4-form $Q$ defined as \cite{Harland:2011zs}
\bea
Q \ \coloneqq \ \tfrac{1}{2} \, \omega \wedge \omega \= -\tfrac{1}{4}
\, \big(\Theta^{1\bar{1}2\bar{2}} +\Theta^{1\bar{1}3\bar{3}}
+\Theta^{2\bar{2}3\bar{3}} \big)
     \= e^{1234}+e^{1256}+e^{3456} \ .
\eea
By the geometric construction using the Killing spinor equations, such an instanton solution implies the usual torsion-free Yang-Mills equation.
In the following, we will study connections and instanton solutions based on this canonical connection.

\paragraph{Orbifolds.}
We conclude by briefly describing the corresponding geometry of the orbifold $S^7/\mathbb{Z}_{k}$, closely following
the treatment of \cite{Lechtenfeld:2015ona}. For compatibility with the bundle structure of the homogeneous space
$S^7 = \mathrm{SU}(4)/\mathrm{SU}(3)$, the cyclic group
$\mathbb{Z}_{k}$ is embedded in $G= \mathrm{SU}(4)$ in a way that it commutes 
with $H=\mathrm{SU}(3)$, i.e. 
the action of $\mathbb{Z}_k$ is embedded in the $\mathrm{U}(1)$-factor associated to the contact direction generated by $I_7$. Hence we
 modify the section (\ref{eq:bundle_s7}) to
\bea
\label{eq:bundle_s7_z}
S^7/\mathbb{Z}_{k} \ni \big(y^1,y^2,y^3, {\phi}/{k} \big) \
\longmapsto \ V' \coloneqq V \,
\mathrm{diag} \big( \e^{3\im {\phi}/k}, \e^{-\im {\phi}/k} \,
\id_3\big) \ .
\eea
As $\mathbb{Z}_{k}$-action on the coordinates of $\C^4$, we use 
\bea
\label{eq:def_z-action}
h_k \cdot \mbf{z} \ \coloneqq \ \mathrm{diag} \big( \zeta_{k}^3,
\zeta_{k}^{-1} \, \id_3\big)\, \mbf{z} \qquad \text{with} \quad \zeta_{k}
\ \coloneqq \ 
\e^{{2 \pi \im}/k} \ ,
\eea
where $h_k$ is a generator of $\mathbb{Z}_{k}$ and $\mbf{z}\in \C^4$. Recalling the definition (\ref{eq:local_coord}) of the local coordinates $y^{\a}$ based on a quotient of $\C ^4$
in a local patch, the action of $\mathbb{Z}_{k}$ on them is given by
\bea
\label{eq:z_transformation}
y^{\a} \ \longmapsto \ \frac{\zeta_{k}^{-1}\, z^{\a}}{\zeta_{k}^{3}
  \,z^0} = \zeta_{k}^{-4}\, y^{\a} \ , \qquad 
\bar{y}^{\bar{\a}} \ \longmapsto \ \frac{\zeta_{k} \,
 \bar z^{\bar{\a}}}{\zeta_{k}^{-3}\, \bar z^{\bar{0}}} =
\zeta_{k}^{4}\, \bar y^{\bar{\a}} \qquad \text{and} 
\qquad e^7 \ \longmapsto \  e^7
\eea
with $\a=1,2,3$; for details see \cite{Lechtenfeld:2015ona}. The local section (\ref{eq:bundle_s7_z}) provides the very same structure equations as 
those of $S^7$, with the replacement $\phi \to \phi_k \coloneqq {\phi}/{k}$ 
(and correspondingly for the dual 1-form $\eta$). In particular,
they still define a Sasaki-Einstein manifold, and -- with a slight abuse of notation -- we will use the same symbols as before also for the orbifold case.

\subsection{Quivers}
\label{section:quiver_diagrams}
The equivariance condition (\ref{eq:equivariance_condition}) enables
us to depict the allowed 
endomorphisms as arrows in a quiver, starting from the
weight diagram of chosen $\mathrm{SU}(4)$-representations.  
In the following we will consider five explicit examples of this
construction and elaborate on some generic features of 
Sasakian quiver gauge theories on odd-dimensional spheres.
\paragraph{Fundamental representation $\mbf{\underline{4}}$\,.} The weight diagram of the fundamental 
representation $\mbf{\underline{4}}$ of $\mathrm{SU}(4)$ is the tetrahedron \eqref{fig:weight_4} consisting of the four vertices
$(3,0,0)$, $(-1,2,0)$, $(-1,-1,-1)$ and $(-1,-1,1)$. Under restriction to $\mathrm{SU}(3)$ it decomposes into the trivial representation and the 
fundamental representation of $\mathrm{SU}(3)$,
\bea
\label{eq:dec_4_su4}
\mbf{\underline{4}}\, \big|_{\mathrm{SU}(3)} \=
\mbf{\underline{(3,0,0)_{1}}} \ \oplus \ \mbf{\underline{(-1,-1,1)_3}}
\ ,
\eea
where the subscripts indicate the dimension of the representation and the triples label the quantum numbers of the highest weight states.
Collapsing the weight diagram along the ladder operators of $\mathrm{SU}(3)$ in this way and implementing the equivariance conditions yields
the quiver
\bea
\label{fig:quiver4_fund}
    \begin{tikzpicture}[->,scale=1.4]
    \node (a) at (0,0) {{\small${(-1,-1,1)}$}};
    \node (b) at (2.5,0) {{\small${(3,0,0)}$}};
    \node (labela) at (0,.8) {{$\scriptstyle\psi_{-1}$}};
    \node (labelb) at (2.5,.8) {{$\scriptstyle\psi_{3}$}};
    
    \path[->](b) edge node [above]{$\scriptstyle\phi$} (a);
    \draw [->](b) to [out=65,in=115,looseness=7] (b) node[right] {$ $};
    \draw [->](a) to [out=65,in=115,looseness=7] (a) node[right] {$ $};
\end{tikzpicture}
\eea
Defining $\phi_{(\a)}:=X_{2\a-1}+\im X_{2\a}$ for $\a=1,2,3$, the endomorphism part
of the gauge connection $\ca$ from \eqref{eq:equ_conn} reads
\bea
\label{eq:Higgs_fund}
\phi_{(\a)} \= \begin{pmatrix}
                0 & 0\\
                \phi \otimes I_{\a} & \mathbf{0}_3
               \end{pmatrix}
\qquad \text{with} \quad I_{1} \ \coloneqq \ \begin{pmatrix}
                                           1,0,0
                                           \end{pmatrix}^{\top} \ , \ 
I_{2}\coloneqq \begin{pmatrix}
                                           0,1,0
                                           \end{pmatrix}^{\top} \ , \ 
I_{3}\coloneqq \begin{pmatrix}
                                           0,0,1
                                           \end{pmatrix}^{\top}
\eea
and
\bea
\label{eq:Higgs_fund2}
X_7 \= \begin{pmatrix}
         \psi_{3} & 0\\
         0        & \psi_{-1} \otimes \id_3
       \end{pmatrix} \ .
\eea
Here the constant matrices $I_{\a}$ stem from the collapsing of the weight diagram along the subalgebra and realize the part living on the representation spaces
$V_j$ of the isotopical decomposition \eqref{eq:iso_dec}. The
homomorphisms $\phi$, $\psi_3$ and $\psi_{-1}$, which are represented
by the arrows $\mathcal{Q}_1$ of the quiver, 
are morphisms between the vector bundles $E_j$ attached to the vertices $\mathcal{Q}_0$. 

The quiver \eqref{fig:quiver4_fund} is precisely the
higher-dimensional analogue of the corresponding quiver in the case of
the five-sphere $S^5$~\cite{Lechtenfeld:2015ona}. 
Due to the straightforward generalization of sections of $S^{2n+1}
\cong \mathrm{SU}(n{+}1)/\mathrm{SU}(n)$ for all $n\geq2$ and the 
fact that the fundamental representation $\mbf{\underline{{n{+}1}}}$
of $\mathrm{SU}(n{+}1)$ splits under restriction to $\mathrm{SU}(n)$
into the fundamental representation and
the trivial representation,
\bea
\mbf{\underline{n{+}1}}\,\big|_{\mathrm{SU}(n)} \= \mbf{\underline{
    n}} \ \oplus \ \mbf{\underline{ 1}} \ ,
\eea
the quiver (\ref{fig:quiver4_fund}) will govern the solutions of the
equivariance conditions on all odd-dimensional spheres $S^{2n+1}$ for $n\geq2$.

\paragraph{\bf Representation $\mbf{\underline{6}}$\,.} 
Due to the accidental isomorphism\footnote{The Dynkin diagrams $\mathrm{A}_3$ and $\mathrm{D}_3$ coincide. For the 
representation theory of $\mathrm{SU}(4)$ and $\mathrm{SL}(4,\C)$,
respectively, see for instance \cite{fulton2013representation}.}
$\mathrm{Spin}(6) \cong \mathrm{SU}(4)$, there is also an irreducible six-dimensional 
representation $\mbf{\underline{6}}$ of $\mathrm{SU}(4)$. Its weight diagram is the octahedron (\ref{fig:weight_6}), and the representation 
decomposes under restriction into the fundamental and anti-fundamental representation of $\mathrm{SU}(3)$,
\bea
\label{eq:dec_6_su4}
\underline{\mbf{6}}\,\big|_{\mathrm{SU}(3)} \=
\mbf{\underline{(2,-1,1)_3}} \ \oplus \ \mbf{\underline{(-2,-2,0)_3}}
\ . 
\eea
This symmetric splitting yields the quiver 
\bea
\label{fig:quiver_6}
    \begin{tikzpicture}[->,scale=1.4]
    \node (a) at (0,0) {{\small${(-2,-2,0)}$}};
    \node (b) at (2.5,0) {{\small${(2,-1,1)}$}};
    \node (labela) at (0,.8) {{$\scriptstyle\psi_{-2}$}};
    \node (labelb) at (2.5,.8) {{$\scriptstyle\psi_{2}$}};
    
    \path[->](b) edge node [above]{$\scriptstyle\phi$} (a);
    \draw [->](b) to [out=65,in=115,looseness=7] (b) node[right] {$ $};
    \draw [->](a) to [out=65,in=115,looseness=7] (a) node[right] {$ $};
\end{tikzpicture}
\eea
The endomorphisms are given as
\bea
\phi_{(\a)} = \begin{pmatrix}
                \mathbf{0}_3 & \mathbf{0}_3\\
               \phi \otimes I_{\a} &\mathbf{0}_3
               \end{pmatrix}
\  \text{with} \  I_1 \coloneqq \begin{pmatrix}
                                             0 & 0 & 0\\
                                             0 & 1 & 0\\
                                             0 & 0 & 1
                                            \end{pmatrix} \ , \ 
I_2 \coloneqq \begin{pmatrix}
                                             0 & 1 & 0\\
                                             0 & 0 & 0\\
                                             -1 & 0 & 0
                                            \end{pmatrix} \ , \  
I_3 \coloneqq \begin{pmatrix}
                                             0 & 0 & -1\\
                                             1 & 0 & 0\\
                                             0 & 0 & 0
                                            \end{pmatrix}
\eea
and
\bea
X_7 \= \begin{pmatrix}
        \psi_{2} \otimes\id_3 & \mathbf{0}_3\\
        \mathbf{0}_3  &  \psi_{-2} \otimes \id_3
       \end{pmatrix} \ .
\eea
\paragraph{Representation $\mbf{\underline{10}}$\,.} The
ten-dimensional representation of $\mathrm{SU}(4)$ can be realized by the generators (\ref{eq:rep_10}), 
and its weight diagram is a 
tetrahedron consisting of three layers, according to the $\mathrm{SU}(3)$ decomposition
\bea
\mbf{\underline{10}}\,\big|_{\mathrm{SU}(3)} \=
\mbf{\underline{(-2,-2,2)_6}} \ \oplus \ \mbf{\underline{(2,-1,1)_3}}
\ \oplus \ \mbf{\underline{(6,0,0)_1}} \ .
\eea
The quiver therefore has three vertices
\bea
\label{fig:quiver_10}
    \begin{tikzpicture}[->,scale=1.4]
    \node (a) at (0,0) {{\small${(-2,-2,2)}$}};
    \node (b) at (2.5,0) {{\small${(2,-1,1)}$}};
    \node (c) at (5,0) {{\small${(6,0,0)}$}};
    \node (labela) at (0,.8) {$\scriptstyle\psi_{-2}$};
    \node (labelb) at (2.5,.8) {$\scriptstyle\psi_{2}$};
    \node (labelb) at (5,.8) {$\scriptstyle\psi_{6}$}; 
 
    \path[->](b) edge node [above]{$\scriptstyle\phi_2$} (a);
    \path[->](c) edge node [above]{$\scriptstyle\phi_1$} (b);
    \draw [->](b) to [out=65,in=115,looseness=7] (b) node[right] {$ $};
    \draw [->](a) to [out=65,in=115,looseness=7] (a) node[right] {$ $};
    \draw [->](c) to [out=65,in=115,looseness=7] (c) node[right] {$ $};
\end{tikzpicture}
\eea
and the Higgs fields read
\bea
X_7\=\begin{pmatrix}
      \psi_6 & 0 & 0\\
      0     & \psi_{2} \otimes \id_3 & 0\\
     0      & 0        & \psi_{-2} \otimes\id_{6}     
    \end{pmatrix}\qquad\text{and} \qquad
\phi_{(\a)} \= \begin{pmatrix}
            0  & 0 &0 \\
            \phi_1 \otimes I^1_{\a} & \mathbf{0}_3 & 0\\
            0      & \phi_2 \otimes I^2_{\a} & \mathbf{0}_6
           \end{pmatrix} \ ,
\eea
with $I^1_{\a}$ as in (\ref{eq:Higgs_fund}) and $I^2_{\a}$ given by
\bea
         I^2_1 \= \begin{pmatrix}
                     \sqrt{2} 	& 0	  &0\\
                     0   	& 1 	  &0 \\ 
                     0          & 0       &1\\
                     0          & 0       &0\\  
		     0 & 0 &0\\
		      0 & 0 &0
                    \end{pmatrix} \ , \qquad
I^2_2 \= \begin{pmatrix}
                     0 & 0 &0\\
                     1   & 0 &0 \\
                     0 & 0 &0\\  
		      0 & 0 &1\\
		      0 & \sqrt{2} &0\\ 
                     0    & 0 &0
                    \end{pmatrix}\qquad\text{and}\qquad
I^2_3 \= \begin{pmatrix}
                     0 & 0 &0\\
                     0   & 0 &0 \\ 
                     1    & 0 &0\\
                     0 & 1 &0\\  
		      0 & 0 &0\\
		      0 & 0 &\sqrt{2} 
                    \end{pmatrix} \ .
\eea

\paragraph{\bf Adjoint representation $\mbf{\underline{15}}$\,.} Implementing the equivariance conditions in the 15-dimensional adjoint representation
transforms the weight diagram (\ref{fig:weight_adj}) into the quiver 
\bea
\label{fig:quiver_adj}
    \begin{tikzpicture}[->,scale=1.4]
     \node (a) at (0,0) {{\small${(4,-2,0)}$}};
     \node (b) at (2.5,0) {{\small${(-4,-1,1)}$}};
     \node (c) at (1.25,-1.25) {{\small${(0,0,0)}$}};
     \node (d) at (1.25,1.25) {{\small${(0,-3,1)}$}};
     \node (labela) at (-1.2,0) {$\scriptstyle\psi_{4}$};
     \node (labelb) at (3.8,0) {$\scriptstyle\psi_{-4}$};
     \node (labelc) at (1.25,-2.1) {$\scriptstyle\psi_{0}$};
     \node (labeld) at (1.25,2.1) {$\scriptstyle\tilde{\psi}_{0}$};
    
    \path[->](b) edge node [above]{$\scriptstyle\phi_5$} (a);
    \path[->](c) edge node [below]{$\scriptstyle\ \phi_2$} (b);
    \path[->](a) edge node [below]{$\scriptstyle\phi_1\ $} (c);
    \path[->](a) edge node [above]{$\scriptstyle\phi_3\ $} (d);
    \path[->](d) edge node [above]{$\scriptstyle\ \phi_4$} (b);

    \draw [->](d) to [out=65,in=115,looseness=7] (d) node[right] {$ $};
    \draw [->](c) to [out=-115,in=-65,looseness=7] (c) node[right] {$ $};
    \draw [->](a) to [out=155,in=205,looseness=4] (a) node[right] {$ $};
    \draw [->](b) to [out=-25,in=25,looseness=4] (b) node[right] {$ $};
\end{tikzpicture}
\eea
The Higgs fields are given by
\small
\bea
X_7 \= \begin{pmatrix}
       \psi_0 		&  \mathbf{0}	& 	\mathbf{0}	&  \mathbf{0}\\
       \mathbf{0}	&  \psi_4 \otimes \id_3 &     \mathbf{0}	&   \mathbf{0}	\\
       \mathbf{0}	& \mathbf{0}	& 	\psi_{-4} \otimes \id_3 & \mathbf{0}\\
       \mathbf{0}	& \mathbf{0}	& 	\mathbf{0}	&  \tilde{\psi}_{0} \otimes \id_3
      \end{pmatrix} \ , \  
\phi_{(\a)} \= \begin{pmatrix}
               0 & \phi_1 \otimes I^1_{\a} & \mathbf{0} & \mathbf{0}\\
               \mathbf{0} & \mathbf{0} & \phi_5 \otimes I^5_{\a} & \mathbf{0}\\
                \phi_2 \otimes I^2_{\a} & \mathbf{0} & \mathbf{0} & \phi_4 \otimes I^4_{\a}\\
                \mathbf{0} & \phi_3 \otimes I^3_{\a} & \mathbf{0} & \mathbf{0}
              \end{pmatrix}
\eea
\normalsize
where the explicit forms of the matrices $I^j_{\a}$ follow from collapsing the weight diagram along the generators of $\mathrm{SU}(3)$.

The quiver associated to the adjoint representation also allows for an
immediate generalization to higher-dimensional spheres: The adjoint representation 
$\mbf{\underline{(n{+}1)^2{-}1}}$ of $\mathrm{SU}(n{+}1)$ decomposes under restriction to $\mathrm{SU}(n)$ into the adjoint, fundamental, 
antifundamental,
 and trivial 
representation as
\bea
\mbf{\underline{(n{+}1)^2{-}1}}\, \big|_{\mathrm{SU}(n)}\=
\mbf{\underline{n^2{-}1}} \ \oplus \ \mbf{\underline{n}} \ \oplus \
\mbf{\underline{\bar{n}}} \ 
 \oplus \ \mathbf{\underline{1}} \ . 
\eea 
Thus, after collapsing along the ladder operators of the subalgebra $\mathfrak{su}(n)$, one obtains a quiver consisting of four vertices.


\paragraph{Representation $\mbf{\underline{6}}\oplus \mbf{\underline{4}}$\,.} Our last example stems from a \emph{reducible} representation 
of $G$. Since the arrows of the quiver depend only on the equivariance relations with respect to the Cartan generators 
of the subalgebra $\mathfrak{h}$, the Higgs fields might have more entries than the corresponding ladder operators; see for instance the morphism 
between fundamental and antifundamental representations in the
previous example
of the adjoint representation. For reducible representations, one can expect this effect to be
even more prominent. 

Let us consider as example the direct sum of the $\mathrm{SU}(4)$-respresentations $\mbf{\underline{6}}$ and $\mbf{\underline{{4}}}$\,.
The quiver contains the two individual quivers \eqref{fig:quiver4_fund} and \eqref{fig:quiver_6}, but there can also be morphisms between them
for the case of $S^7$ (without orbifolding):
\bea
\label{fig:quiver_6+4}
    \begin{tikzpicture}[->,scale=1.4]
    \node (a) at (0,0) {{\small${(-2,-2,0)}$}};
    \node (b) at (2.5,0) {{\small${(2,-1,1)}$}};
    \node (c) at (0,-1.5) {{\small${(-1,-1,1)}$}};
    \node (d) at (2.5,-1.5) {{\small${(3,0,0)}$}};

    \node (labela) at (0,.8) {{$\scriptstyle\psi_{-2}$}};
    \node (labelb) at (2.5,.8) {{$\scriptstyle\psi_{2}$}};
    \node (labelc) at (0,-2.3) {{$\scriptstyle\psi_{-1}$}};
    \node (labeld) at (2.5,-2.3) {{$\scriptstyle\psi_{3}$}};
    
    \path[->](b) edge node [above]{$\scriptstyle\phi_1$} (a);
    \path[->](d) edge node [below]{$\scriptstyle\phi_2$} (c);
    \path[->](c) edge node [left]{$\scriptstyle\phi_3$} (a);
    \path[->](d) edge node [right]{$\scriptstyle\phi_4$} (b);
     \path[<->](b) edge node [above]{$\scriptstyle\chi$} (c);
    \path[->](a) edge node [below]{$\scriptstyle\phi_5$} (d);

    \draw [->](b) to [out=65,in=115,looseness=7] (b) node[right] {$ $};
    \draw [->](a) to [out=65,in=115,looseness=7] (a) node[right] {$ $};
    \draw [->](c) to [out=245,in=295,looseness=7] (c) node[right] {$ $};
    \draw [->](d) to [out=245,in=295,looseness=7] (d) node[right] {$ $};
\end{tikzpicture}
\eea
Anticipating Section \ref{sect:I7_equivariance}, note that the
homomorphisms $\phi_3$, $\phi_4$, $\phi_5$ and $\chi$, which are
additional degrees of freedom to those anticipated from
the shape of the ladder operators, have to vanish if one imposes also equivariance with respect to the finite subgroup in the orbifold case. Then 
the quiver decomposes into the direct sum of the relevant quivers \eqref{fig:quiver4_fund} and \eqref{fig:quiver_6},
as is to be expected from the existence of invariant subspaces in a reducible $G$-representation.
\subsection{Yang-Mills-Higgs theories}
Having seen how to obtain an $\mathrm{SU}(4)$-equivariant gauge connection from graphical techniques on the weight diagrams of the Lie algebra $\mathfrak{su}(4)$, 
we now derive the dimensional reduction of the Yang-Mills action 
\bea
S_{\mathrm{YM}} \ \coloneqq \ - \frac{1}{4}\, \int_{M^d \times S^7} \,
\mathrm{tr} ( \cf \wedge \star\, \cf)
\eea
for the connection (\ref{eq:equ_conn}) over $M^d \times S^7$ to a Yang-Mills-Higgs theory on the manifold $M^d$. Using the 
Sasaki-Einstein metric (\ref{eq:metric_s7}), one obtains for the Lagrangian\footnote{Here we denote by $\hat{\m},\hat\n,\dots$ generic directions along $M \times G/H$, 
with $\m, \n,\dots$ directions along $M$ and $\a, \beta,\dots$ directions along the coset.}
\bea \nonumber
 \mathcal{L}_{\mathrm{YM}} &=& - \tfrac{1}{4} \, \sqrt{\hat{g}}\
 \mathrm{tr}\big( \cf_{\hat{\m}\hat{\n}} \, \cf^{\hat{\m}\hat{\n}}
 \big) \\[4pt]
\nonumber          &=& - \tfrac{1}{4} \, \sqrt{\hat{g}}\ \mathrm{tr}
\big( \cf_{\m\n}\, \cf^{\m\n} + 8 \, g^{\m\n}\, \cf_{\a\m}\, \cf_{\n \bar{\a}} 
+ 2 \, g^{\m\n}\, \cf_{\m7}\, \cf_{\n7} \nonumber \\ && \qquad \qquad
\qquad +\, 8 \, \cf_{\a\b}\,
\cf_{\bar{\a}\bar{\b}} + 8 \, \cf_{\a\bar{\b}}\, \cf_{\bar{\a}\b} + 8\,
\cf_{\a7}\, \cf_{\bar{\a}7}\big) \ .
\eea
Inserting the non-vanishing components of the curvature and writing $\left| X\right|^2 \coloneqq X\, X^{\+}$, one ends up with  the result
\small
\bea
\label{eq:action_s7}
 S_{\mathrm{YM}} &=& \mathrm{vol}\big( S^7\big) \, \int_{M^d}\,
 \diff^d y \ \sqrt{g}\ \mathrm{tr} \Big( \, \frac{1}{4}\, F_{\m\n} \, \lb F^{\m\n} \rb^{\+}
+ 2\, \sum_{\m=1}^d\, \ \sum_{\a=1}^3 \, \big|D_{\m}\phi_{(\a)} \big|^2  + \frac{1}{2}\,
\sum_{\m=1}^d \, \big|D_{\m} X_7 \big|^2  \\
\nonumber &&+ \, 2\, \sum_{\a=1}^3 \, \Big|\big[X_7,\phi_{(\a)}\big] +
\frac{3}{4} \im \phi_{(\a)} \Big|^2
+ 4 \, \big|\big[\phi_{(1)},\phi_{(2)}\big] \big|^2 + 4 \, \big|\big[\phi_{(1)},\phi_{(3)}\big] \big|^2    
+4 \, \big|\com{\phi_{(2)}}{\phi_{(3)}} \big|^2 \\
  \nonumber &&+ \, 2 \, \big|\big[\phi_{(1)},\phi_{(1)}^\+\big] - \im X_7 + 2 \im I_8 \big|^2    
 + 2\, \big|\big[\phi_{(2)},\phi_{(2)}^{\+}\big] - \im X_7 - \im I_8 - \im I_9 \big|^2 
+4\, \big|\big[\phi_{(2)},\phi_{(1)}^{\+}\big]+ I_{\bar{4}}^- \big|^2 \\
\nonumber && +\, 4\, \big|\big[\phi_{(3)},\phi_{(1)}^\+\big]+ I_{\bar{5}}^- \big|^2 
+4\, \big|\com{\phi_{(3)}}{\phi_{(2)}^\+}+ I_{\bar{6}}^- \big|^2 + 2\,
\big|\com{\phi_{(3)}}{\phi_{(3)}^{\+}} - \im X_7 - \im I_8 + \im I_9
\big|^2  \, \Big) \ ,
\eea
\normalsize
where $F:=\diff A+A\wedge A$ is the curvature of the gauge connection
$A$ on $M^d$, while $D_{\m} \phi_{(\a)}$ and $D_{\m}X_7$ denote the covariant derivatives of the Higgs fields,
\bea
D \phi_{(\a)} \ \coloneqq \ \diff \phi_{(\a)} + \com{A}{\phi_{(\a)}}  \qquad \text{for} \quad \a\=1,2,3 \qquad \text{and} \qquad 
D X_7 \ \coloneqq \ \diff X_7 + \com{A}{X_7} \ .
\eea
Since we assumed, due to equivariance, that the 
endomorphisms $X_{a}$ are independent of the coordinates of $S^7$,
the integration over the coset space here simply produced its
volume in the metric (\ref{eq:metric_s7}).
\subsection{Orbifold quivers and reduction to $\C P^3$\label{sec:CP3}}
\label{sect:I7_equivariance}
Due to the fibration structure \eqref{eq:s7_bundle} of $S^7$ as a $\mathrm{U}(1)$-bundle over 
the complex projective space $\C P^3$, it is natural to consider the reduction of the Sasakian quiver gauge theory on $S^7$ to that 
of the underlying  K\"ahler coset structure on $\C P^3$. In this reduction,
M-theory on $\mathrm{AdS}_4\times S^7/\Z_k$ becomes
IIA string theory on
$\mathrm{AdS}_4\times\C P^3$~\cite{Nilsson:1984bj}, which in the
't~Hooft limit is dual to
$\mathcal{N}=6$ superconformal Chern-Simons theories with matter
fields~\cite{Aharony:2008ug} to which our constructions apply. Similar
reductions to the underlying K\"ahler leaf spaces have been carried out
for the Sasaki-Einstein manifolds considered in
\cite{Lechtenfeld:2014fza,Lechtenfeld:2015ona,Geipel:2016uij}.

For this, one has to further factor by the generator $I_7$, which corresponds to the Reeb vector field of the Sasakian structure, 
by setting $X_7=I_7$ so that one 
obtains the further equivariance conditions
\bea
\com{\hat{I}_7}{\phi_{(\a)}} \= - 4\, \phi_{(\a)} \qquad \text{for}
\quad \a\=1,2,3 \ .
\eea 
This forces the Higgs fields to have the same form  as the ladder operators of $G$, i.e.~the remaining three complex Higgs fields must act in 
the weight diagrams of $\mathrm{SU}(4)$ as
\bea
\label{eq:eq_rules_cp3}
\nonumber \phi_{(1)}: \lb \n_7,\n_8,\n_9 \rb & \longmapsto & \lb
\n_7-{4}, \n_8+2, \n_9\rb \ ,\\[4pt]
	  \phi_{(2)}: \lb \n_7,\n_8,\n_9 \rb & \longmapsto & \lb
          \n_7-{4}, \n_8-1, \n_9-1\rb \ ,\\[4pt]
\nonumber \phi_{(3)}: \lb \n_7,\n_8,\n_9 \rb & \longmapsto & \lb
\n_7-{4}, \n_8-1, \n_9+1\rb \ .
\eea
Since we removed the contact direction as degree of freedom by setting $X_7=I_7$, the loops in the quivers disappear, as expected. Apart from this, in 
our examples from Section \ref{section:quiver_diagrams} only 
the quiver \eqref{fig:quiver_adj} associated to the adjoint representation is altered: The morphism $\phi_5$ is ruled out by the additional conditions (\ref{eq:eq_rules_cp3})
 with respect to $I_7$. 

The quiver gauge theory on the orbifold $S^7/\mathbb{Z}_{k}$ shares
some features with that on  $\C P^3$, because the action of the
finite subgroup $\Z_k$ on
the fibres is embedded in the $\mathrm{U}(1)$ subgroup generated by $I_7$. Combining this $\mathbb{Z}_{k}$-action on the fibres with the action (\ref{eq:z_transformation}), one has to impose the \mbox{conditions \cite{Lechtenfeld:2014fza,Lechtenfeld:2015ona}}
\bea
\label{eq:eq_cond_z}
\g ( h_k)\, \phi_{(\a)} \,\g (h_k)^{-1} \= \zeta_{k}^4 \, \phi_{(\a)} \qquad \text{and} \qquad
\g (h_k) \, X_7 \, \g (h_k)^{-1} \=  X_7 \ ,
\eea
where $\g $ is the embedding of $\mathbb{Z}_{k}$ in the group $\mathrm{U}(1)$ generated by the chosen representation of $I_7$. 
In order to satisfy this condition for \emph{all} integers $k$, the Higgs fields have to act on the quantum number $\n_7$ in the same way that 
the ladder operators of $\mathrm{SU}(4)$ do, so that --  combining this
condition with $\mathrm{SU}(4)$-equivariance --  their form must be the
same as that of the ladder operators of $G$.  

In contrast to the additional equivariance condition on $\C P^3$, however, the endomorphism $X_7$ is still a degree of freedom. Furthermore, when
considering only a particular given fixed integer $k$,  the condition \eqref{eq:eq_cond_z} might still be satisfied for fields with more entries than the 
ladder operators have, because the equation on the powers of roots of unity holds \emph{modulo} $k$. Therefore other contributions may also
match the condition for a special value of $k$. For details of $\mathbb{Z}_{k}$-equivariance, and 
comparisons with orbifolds of $S^3$ and $S^5$,
see~\cite{Lechtenfeld:2014fza} and~\cite{Lechtenfeld:2015ona} respectively.

\subsection{Hermitian Yang-Mills instantons on the K\"ahler cone}
\label{sect:cone_instantons}
We will now extend the form of the gauge connection to the metric cone $C(S^7)$ and make use of its K\"ahler structure
to formulate instanton equations. Starting from the canonical
connection $\Gamma$,\footnote{The instanton $\Gamma$ also lifts to an
  instanton on the metric cone and the
cylinder.} we now include also the radial direction
\bea
\label{eq:equ_conn_cone}
\ca \=   \Gamma + \sum_{a=1}^7\, X_a\, e^a  + X_{\tau}\, e^{\tau} 
 \ \eqqcolon \ \Gamma + \sum_{\a=0}^3\, \big(Y_{\a}\, \Theta^{\a} +
\bar Y_{\bar{\a}} \, \bar\Theta^{\bar{\a}} \big)
\eea
with $Y_{0} \coloneqq \tfrac{1}{2} \, ( X_{\tau} + \im X_7 )$,
and we assume that the Higgs fields depend only on the radial coordinate, $Y_{\a}=Y_{\a}(r)$ for $\a=0,1,2,3$. Writing the 
\emph{Hermitian Yang-Mills equations}\footnote{They are also
  known as the \emph{Donaldson-Uhlenbeck-Yau equations} and are related to the
stability of holomorphic vector bundles \cite{donaldson1985anti,uhlenbeck1986existence}.}~\cite{Popov:2009nx}
\bea
\label{eq:HYM}
\cf^{2,0} \= 0 \= \cf^{0,2} \qquad \text{and} \qquad \Omega^{1,1} \haken \cf \=0
\eea
for the components of the curvature yields the algebraic conditions
\bea
\label{eq:constraint}
\com{\bar Y_{\bar{1}}}{\bar Y_{\bar{2}}} \= \com{\bar Y_{\bar{1}}}{\bar Y_{\bar{3}}} \= \com{\bar Y_{\bar{2}}}{\bar Y_{\bar{3}}} \= 0
\eea
and the flow equations
\begin{subequations}
\beq
\label{eq:s7_flow1}
r \, \dot{\bar Y}_{\bar{\a}} \= -\tfrac{4}{3}\, \bar Y_{\bar{\a}} + 2\,
\com{\bar Y_{\bar{\a}}}{\bar Y_{\bar{0}}} \qquad \text{for} \quad \a\=1,2,3 \ ,
\eeq
\beq
\label{eq:s7_flow2}
 r \, \big(\dot{Y}_{0} - \dot{\bar Y}_{\bar{0}} \big) \= - 6 \,( Y_0 -
 \bar Y_{\bar{0}}) + 2\, \com{Y_{0}}{\bar Y_{\bar{0}}} +2\,
 \com{Y_{1}}{\bar Y_{\bar{1}}} +2\, \com{Y_{2}}{\bar Y_{\bar{2}}}+
 2\, \com{Y_{3}}{\bar Y_{\bar{3}}} \ ,          
\eeq
\label{eq:s7_flow}\end{subequations}
where a dot indicates the $r$-derivative.
From (\ref{eq:constraint}) one sees that the Hermitian Yang-Mills equations force the complex 
Higgs fields $\phi_{(\a)}$ of the Sasakian quiver gauge theory 
to commute with each other, i.e. they impose relations on the quivers.

\paragraph{Examples.} The utility of the quiver gauge theory is in its
depiction of the matter fields 
as arrows in a quiver. Having chosen a representation of $G$, the quiver is fixed, and the decomposition encoded in the quiver yields the 
form of the system of equations one has to solve. In the following, we
collect the Hermitian Yang-Mills equations for some of the examples
from Section~\ref{section:quiver_diagrams}.

For the fundamental representation \eqref{fig:quiver4_fund}, we only have the arrow $\phi$, and the two loops $\psi_1$ and $\psi_3$. Plugging the Higgs fields \eqref{eq:Higgs_fund} and \eqref{eq:Higgs_fund2}
into the algebraic conditions \eqref{eq:constraint} shows that they
are automatically satisfied without any restriction on the fields. From \eqref{eq:s7_flow1}
we get 
\bea
r \, \dot{\phi} \= -\tfrac{3}{4} \, \phi - \im \phi \, \psi_3 + \im
\psi_1 \, \phi \ .
\eea
From \eqref{eq:s7_flow2}, one obtains two equations for the endomorphisms
\begin{subequations}
 \beq
r \, \dot{\psi}_3 \= - 6 \, \psi_3 + 6 \im \phi^{\dagger} \, \phi \ ,
\eeq
\beq
r \, \dot{\psi}_{-1} \= - 6 \, \psi_{-1} - 2 \im \phi \,
\phi^{\dagger} \ .
\eeq
\end{subequations}
These equations are the analogues of
those for the fundamental representation in the quiver gauge theory
on $S^5$ \cite{Lechtenfeld:2015ona}, as is to be expected from the identical quivers.

Although the quiver \eqref{fig:quiver_6} of the representation $\mbf{\underline{6}}$ looks
formally the same as \eqref{fig:quiver4_fund}, there is a crucial subtlety arising from the different
dimensions of the $\mathrm{SU}(3)$ representations in the
decomposition \eqref{eq:dec_6_su4} compared to the decomposition \eqref{eq:dec_4_su4} of the fundamental representation. While the algebraic
conditions do not provide further constraints, and the flow equations \eqref{eq:s7_flow1} again yield
\bea
r \, \dot{\phi} \= -\tfrac{3}{4}\, \phi -\im  \phi \, \psi_2 + \im
\psi_{-2} \, \phi \ ,
\eea 
the flow equations for the loop contributions are slightly changed. We obtain
\begin{subequations}
 \beq
r \, \dot{\psi}_2 \= - 6 \, \psi_2 + 4 \im  \phi^{\dagger} \, \phi \ ,
\eeq
\beq
r \, \dot{\psi}_{-2} \= - 6 \, \psi_{-2} - 4 \im  \phi \, \phi^{\dagger}
\ ,
\eeq
\end{subequations}
so that both equations have coefficients with the same modulus, in
contrast to those in the example of the fundamental representation.

\paragraph{Constant endomorphisms.} Before we proceed with the general description of the moduli space of the Hermitian Yang-Mills equations under the constraints
imposed by equivariance, we consider the special case of constant endomorphisms $Y_{\bar{\a}}$. In this situation, the radial coordinate $r$ enters the setup simply as a label
of the foliations comprising the underlying Sasaki-Einstein structure along the cone direction. Gauging $X_{\tau}=0$, one obtains from the flow equations the algebraic
conditions
\bea
\com{X_7}{\bar Y_{\bar{\a}}} \= - \tfrac{4}{3}\im \bar Y_{\bar{\a}} \ ,
\eea
which implies the vanishing of many contributions to the action (\ref{eq:action_s7}), as is to be expected of an instanton solution.
As we have seen in Section~\ref{sec:CP3}, this condition can be satisfied, for instance, by the quiver gauge theory on the projective space $\C P^3$.

\paragraph{Moduli space of Hermitian Yang-Mills equations.} 
For the analysis of the flow equations under the given constraints,
one can apply the general results of 
\cite{Sperling:2015sra} concerning Hermitian Yang-Mills instantons on
metric cones over generic $2n{+}1$-dimensional Sasaki-Einstein
manifolds $M^{2n+1}$; we briefly review the main aspects here, refering
to \cite{Sperling:2015sra} and references therein for details. Rescaling the matrices
\bea
\bar Y_{\bar{\a}} \ \eqqcolon \ r^{-{4}/{3}} \, W_{\a} \qquad
\text{for} \quad \a\=1,2,3 \qquad \text{and} \qquad \bar Y_{\bar{0}} \
\eqqcolon \ r^{-6} \, Z
\eea
and changing the argument to $s\coloneqq - \frac{1}{6}\, r^{- 6}$ in
\eqref{eq:s7_flow} leads to the equations
\begin{subequations}
\label{eq:nahm_type}
 \begin{equation}
\label{eq:nahm_type1}
\frac{\diff W_{\a}}{\diff s} \= 2 \, \com{W_{\a}}{Z} \ ,
\end{equation}
\begin{equation}
 \label{eq:nahm_type2}
\frac{\diff }{\diff s} \big(Z+Z^{\+} \big)+
2\, \com{Z}{Z^{\+}} + 
2\,( -6s)^{-{14}/{9}}\, \sum_{\a=1}^3 \, \com{W_{\a}}{W_{\a}^{\+}} \=
0 \ ,
\end{equation}
\end{subequations}
in correspondence with the general results of \cite{Sperling:2015sra} for $n=3$. The first equations \eqref{eq:nahm_type1} are referred to as the \emph{complex equations} 
and the second equation \eqref{eq:nahm_type2} as the \emph{real
  equation}. The description of the moduli space $\mathcal{M}$ of these Nahm-type equations is based on 
the invariance of the complex equations under the gauge transformations
\bea  
W_{\a} \longmapsto W_{\a}^g \ \coloneqq \ g \, W_{\a} \, g^{-1} \qquad
\text{and} \qquad  Z \longmapsto Z^{g} \ \coloneqq \ g\, Z \, g^{-1} -
\frac{1}{2} \, \Big(\, \frac{\diff g}{\diff s}\,\Big) \, g^{-1}
\eea
for $g \in \mathcal{G}$, where $\mathcal{G}$ is the subgroup of gauge transformations $g:(-\infty,0)\to
\mathrm{GL} (r,\C)$ that 
also preserve the equivariance conditions; this allows for application
of techniques used in the study of Nahm equations~\cite{donaldson1984,kronheimer1990hyper}.

On the one hand, one may start from a local gauge in which $Z^g$
vanishes, i.e. $Z=\tfrac{1}{2}\, g^{-1} \, \frac{\diff g}{\diff s}$ is pure gauge.
From the complex equations it then follows that the complex matrices
$W_{\a}$ are constant, i.e.~one has the local solution 
\bea
Z \= \frac{1}{2} \, g^{-1} \, \frac{\diff g}{\diff s} \qquad
\text{and} \qquad W_{\a} \= g^{-1}\, T_{\a}\, g \qquad \text{with} \quad \com{T_{\a}}{T_{\b}} \=0
\eea 
for \emph{constant} matrices $T_{\a}$ obeying the equivariance constraints \eqref{eq:equivariance_condition}. An obvious choice for these
matrices is as elements of a Cartan subalgebra of $\mathfrak{gl}(r,\C)$. To explicitly solve the 
instanton equations, one also has to include the real equation and
take into account the domain on which the gauge transformation is applicable. 
The real equation follows as equation of motion for a suitably chosen Lagrangian~\cite{Sperling:2015sra,donaldson1984}
\bea
\mathcal{L} \= \mathrm{tr} \Big(\big|Z^g+Z^{g}\,^{\+}\big|^2 + 2 \, (
-6s)^{-{14}/{9}} \, \sum_{\a=1}^3\, |W_{\a}^g|^2 \Big) \ .
\eea 
The real equation is therefore solved as a variational problem. For uniqueness of the solution and to apply this approach over the entire 
range $-\infty<s<0$, one restricts to \emph{framed} instantons and imposes boundary conditions for $s \rightarrow -\infty$, i.e. at the conical
singularity $r=0$. One therefore finds
$g_0\in\mathcal{G}$ such that
\bea
\lim_{s\rightarrow - \infty} \, W_{\a} (s) \= g_0^{-1}\, T_{\a}\, g_0
\ .
\eea
Consequently, the moduli space $\mathcal{M}$ can be described in terms of coadjoint
orbits of $\mathrm{GL}(r,\C)$ with suitable boundary conditions:
\bea
\mathcal{M} \= \mathcal{O}_{T_1} \times \mathcal{O}_{T_2} \times \mathcal{O}_{T_3} \  ,
\label{eq:modorbit}\eea
where the orbits $\mathcal{O}_{T_\alpha}$ are generally not regular with closures given by nilpotent cones.

On the other hand, the moduli space also admits a description as a K\"ahler quotient, making \eqref{eq:modorbit} into a K\"ahler manifold. Denoting the space of solutions 
$\mathcal{A}$ to the complex equations $\cf^{2,0} = 0= \cf^{0,2}$ and
the equivariance conditions as $\mathbb{A}^{1,1} $, the real
equation can be interpreted in terms of the moment map $\m: \ \mathbb{A}^{1,1}
\to \mathrm{Lie}(\mathcal{G})$ with $\m(\ca)=\Omega^{1,1} \haken \cf$,
and the moduli space is given as the K\"ahler quotient
\bea
\mathcal{M} \= {\m^{-1}(0)} \, \big/\!\!\big/ \, {\mathcal{G}} \ .
\eea

\paragraph{Sasakian quiver gauge theory on odd-dimensional spheres.}

By comparing the various examples of quivers associated to $S^7$, as well as the general realization of odd-dimensional spheres as Sasaki-Einstein spaces
$S^{2n+1} \cong \mathrm{SU}(n{+}1)/\mathrm{SU}(n)$, we suggest that
the Sasakian quiver gauge theory is universal on spheres in \emph{all}
odd dimensions. We will not attempt to give a rigorous proof of this
fact here, but  the general construction of the local section, the coset space and the representations of 
$\mathrm{SU}(n{+}1)$ stongly support this claim. Of course,
exceptional isomorphisms for low-dimensional cases, such as the 
representation $\mbf{\underline{6}}$\,, do not allow for a general treatment. 
This statement is further supported by the existence of general expressions for Hermitian Yang-Mills instantons on metric cones over \emph{generic}
Sasaki-Einstein manifolds \cite{Sperling:2015sra,Ivanova:2012vz},
which we used for the description of the moduli space $\mathcal{M}$ above.

\section{3-Sasakian quiver gauge theory on the squashed seven-sphere}
\label{sect:3Sasaki_QGT}
In this section we construct quiver gauge theories on the squashed seven-sphere, making use of its 3-Sasakian structure (see e.g. \cite{Boyer:1998sf}).
 A general treatment of 3-Sasakian seven-manifolds
and their geometry is given, for instance, in \cite{agricola20103},
while a description of the representations 
of $\mathrm{Sp}(2)$ can be again found in
\cite{fulton2013representation}. For the geometry and supergravity applications of
the squashed seven-sphere, see also \cite{DUFF19861}.

\subsection{Geometry of  ${S}^7 \cong \mathrm{Sp}(2)/\mathrm{Sp}(1)$}
\label{sect:geometry_sp2}
\paragraph{Local section.} Similarly to Section
\ref{sect:geometry_s7}, we start our description by providing local
coordinates and a basis of 1-forms, again by using certain particular 
fibrations. As the squashed seven-sphere is a fibration of $\mathrm{SU}(2)\cong \mathrm{Sp}(1)$ over a quaternionic K\"ahler 
manifold \cite{boyer19943}, we can construct 
a local section by considering the fibration 
\bea
\label{eq:fibrations_sp2_s4}
\mathrm{Sp}(2) \ \longrightarrow \ S^4 \cong \mathrm{Sp}(2)\,\big/\,
\mathrm{Sp}(1) \times \mathrm{Sp}(1) \ ,
\eea
and following the prescription in \cite{Popov:2009nx}. A local section
of this bundle can be realized by
\bea
\mathrm {Sp}(2)\ni Q \ \coloneqq \, f^{-1/2} \, \begin{pmatrix}
                                         \id_2  & -x\\
                                        x^{\+}  & \id_2
                                        \end{pmatrix} \qquad
                                        \text{with} \quad x\=x^{\m} \,
                                        \tau_{\m} \ , \quad \lb \tau_{\m}\rb\= \lb -\im \sigma_i,\id_2\rb
\eea
and $f\coloneqq 1 + x^{\+}\, x=1 + \delta_{\m \n}\, x^{\m}\, x^{\n}$;
here $\sigma_i$ for $i=1,2,3$ are the standard Pauli spin matrices. The canonical flat connection 
\bea
\label{eq:can_flat_connection}
A_0 \= Q^{-1} \, \diff Q \ \eqqcolon \ \begin{pmatrix}
                                A^{-} & -\phi \\
                                \phi^{\+} & A^{+}
                               \end{pmatrix} 
\qquad \text{with}  \quad
\phi \= \frac{1}{f} \, \diff x \ \eqqcolon \ 
\begin{pmatrix}
\chi^2 & \chi^1\\
\bar{\chi}^{\bar1}& - \bar{\chi}^{\bar2}
\end{pmatrix}
\eea
 provides complex 1-forms $\chi^1$ and $\chi^2$. An element of
 $\mathrm{Sp}(1) \cong \mathrm{SU}(2)$ can be written in local coordinates
 as
\bea
\label{eq:section_su2}
\mathrm{Sp}(1) \ni g \cdot h \ \coloneqq \ \lb 1+ z \, \bar{z}\rb^{-1/2}
\, \begin{pmatrix}
                                                           1 & - \bar{z}\\
                                                           z & 1
                                                          \end{pmatrix} \cdot
\begin{pmatrix}
\e^{-\im \varphi} & 0\\
 0             & \e^{\im \varphi} 
\end{pmatrix} \ ,
\eea 
so that a section of $\mathrm{Sp}(2) \rightarrow \mathrm{Sp}(2) / \mathrm{Sp}(1)$ can be obtained by 
\bea
\mathrm{Sp}(2) \ni \tilde{Q} \ \coloneqq \ Q\, \begin{pmatrix}
                                         g\, h& 0\\
                                         0    & \id_2
                                       \end{pmatrix} \ .
\eea
Consider first the flat connection on the twistor space $\mathrm{Sp}(2)/\mathrm{Sp}(1) \times \mathrm{U}(1)$
given by the Maurer-Cartan form of $\hat{Q}\coloneqq Q \, g$:
\bea
\label{eq:can_flat_conn_twistor}
\hat A_0 \= \hat{Q}^{-1} \, \diff \hat{Q} \= g^{-1} \, A_0 \, g +
g^{-1}\, \diff g \ \eqqcolon \
 \begin{pmatrix}
  g^{-1} \, A^{-} \, g + g^{-1}\, \diff g & - g^{-1}\, \phi \\
 \phi^{\+}\, g & A^+ 
 \end{pmatrix} \ .
\eea
Then a section of the bundle $\mathrm{Sp}(2) \rightarrow
\mathrm{Sp}(2)/\mathrm{Sp}(1)$ is given by $\tilde{Q} \coloneqq
\hat{Q} \, h$, giving
\bea
\label{eq:can_flat_conn_sp2}
\nonumber \tilde A_0 &=& h^{-1} \, \hat A_0\, h + h^{-1}\, \diff h
\\[4pt] &=& \nonumber 
\begin{pmatrix}
   h^{-1} \, \hat{A}^{-}\, h + h^{-1} \, \diff h & -h^{-1} \,
   \hat{\phi}\, \\
 \hat{\phi}^{\+} \, h & A^+
\end{pmatrix}\\[4pt]
&\eqqcolon&
\begin{pmatrix}
         \im  e^7		&	 \Theta^3		& \Theta^2		&  \Theta^1\\
         -\bar{\Theta}^{\bar3}	& 	-\im  e^7 		& \bar{\Theta}^{\bar1}	&  -\bar{\Theta}^{{\bar2}}\\
	- \bar{\Theta}^{\bar2}	& 	- {\Theta}^1    	& -\im e^8	        & - \bar{\Theta}^{\bar4}\\
        - \bar{\Theta}^{\bar1}    	& 	{\Theta}^2		& \Theta^4   		& \im e^8
        \end{pmatrix} \ ,
\eea
where $\hat{A}^-:=g^{-1} \, A^{-} \, g + g^{-1}\, \diff g$ and
$\hat{\phi}:= g^{-1}\, \phi$.
This provides a basis of left-invariant 1-forms for $\mathrm{Sp}(2)$ with  structure equations (\ref{eq:struct_eq_sp2_compl}) and 
structure constants (\ref{eq:struc_const_sp2}). We have defined the
1-forms in \eqref{eq:can_flat_conn_sp2} in such a way that
they define a 3-Sasakian structure with metric
\bea
\diff s^2 \= \sum_{\m=1}^7\, e^{\m} \otimes e^{\m} \ .
\eea
To show this, one uses
the structure equations 
\bea
 \nonumber \diff e^1 &=& e^{27}-e^{28}-e^{35}-e^{46}+e^{39}+e^{4\,10}
 \ , \quad
  \diff e^2 \= -e^{17}+e^{18} +e^{45}-e^{36}+e^{49}-e^{3\,10} \ ,\\[4pt]
 \nonumber \diff e^3 &=& e^{47}+e^{48}+e^{15}+e^{26}-e^{19}+e^{2\,10}
 \ , \quad
  \diff e^4 \= -e^{37}-e^{38}-e^{25}+e^{16}-e^{29}-e^{1\,10} \ ,\\[4pt]
  \diff e^5 &=& 2e^{67}-2e^{13}+2e^{24} \ , \quad \diff e^6 \=
  -2e^{57}-2e^{14} -2e^{23} \ , \quad \diff e^7 \= 2e^{56}+2e^{12}+2e^{34}
\eea
together with
\bea
\diff e^8 \= -2e^{12}+2e^{34}+2e^{9\,10} \ , \quad \diff e^9\=
2e^{13}+2e^{24}-2e^{8\,10} \ , \quad \diff e^{10}\= 2e^{14}-2e^{23}+2e^{89}
\eea
for the real 1-forms $e^{\m}$ defined as
\bea
e^1 - \im e^2 \ \coloneqq \ \Theta^1 \ , \quad
e^3 - \im e^4 \ \coloneqq \ \Theta^2 \ , \quad
e^5 - \im e^6 \ \coloneqq \ \Theta^3 \quad \text{and} \quad
e^9 - \im e^{10} \ \coloneqq \ \Theta^4 \ .
\eea
From these structure equations, we see that the triple $(\eta^5, \eta^6,\eta^7)\coloneqq (e^5,e^6,e^7)$ satisfies the defining
relations of a 3-Sasakian structure (see e.g.~\cite{Harland:2011zs})
\bea
\label{eq:3Sasaki_def}
\diff \eta^{\a}\= \epsilon_{\a \beta \gamma} \, \eta^{\beta} 
\wedge \eta^{\gamma} +2\, \omega^{\a} \ , \qquad
\diff \omega^{\alpha} \= 2\, \epsilon_{\a \beta \gamma} \,
\eta^{\beta} \wedge \omega^{\gamma} \qquad \text{with} \quad
\a,\beta,\gamma \=5,6,7 \ , 
\eea
where we identify the 2-forms
\bea
\label{eq:3Sasaki_def_forms}
\omega^5 \ \coloneqq \ -e^{13}+e^{24} \ , \qquad \omega^6 \ \coloneqq
\ -e^{14}-e^{23} \qquad \text{and} \qquad \omega^7 \ \coloneqq \
e^{12}+e^{34} \ .
\eea
Alternatively, one may also check the closure of some defining
forms, as we did in Section \ref{sect:geometry_s7} for the Sasaki-Einstein structure; 
see also Appendix \ref{sec:3sasaki_app}.
\paragraph{Orbifolds.} As for the coset $\mathrm{SU}(4)/\mathrm{SU}(3)$ in Section \ref{sect:geometry_s7}, one can introduce 
an action of the cyclic group $\mathbb{Z}_{k}$ on the squashed seven-sphere. For this, we embed the action in the $\mathrm{U}(1)$ factor $h$ of
\eqref{eq:section_su2} as
\bea
\label{eq:discrete_eq_sp2_1}
h_{k} \ \coloneqq \ 
\begin{pmatrix}
\e^{\im \varphi/k} 	& 0				\\
 0             			& \e^{-\im \varphi/k} 
\end{pmatrix} \ .
\eea
The action $\pi$ on 1-forms can be deduced from the action of $g$ in the final section \eqref{eq:can_flat_conn_sp2}, giving
\bea
\label{eq:discrete_eq_sp2_2}
\pi(h_k) \Theta^{\a} \= \zeta_{k}\, \Theta^{\a} \ , \ \a\=1,2 \ , \qquad \pi(h_k) \Theta^{3} \= \zeta_{k}^2\, \Theta^{3}
\qquad \text{and} \qquad \pi(h_k) e^7 \= e^7 \ .
\eea
\subsection{Instanton equations}
We shall now describe the instanton equations on both
$\mathrm{Sp}(2)/\mathrm{Sp}(1)$ and its metric cone. We will also
describe the canonical connection that will appear in the general
form of the gauge connection.
\paragraph{Canonical connection and instanton equation on Sp(2)/Sp(1).}
According to the general results of \cite{Harland:2011zs}, the torsion of the canonical connection of a 3-Sasakian manifold is given by
\bea
T^{\a} \= 3\, P_{\a \m \n} \, e^{\m \n} \ , \ \a\=5,6,7 \qquad
\text{and} \qquad  T^{a}\= \tfrac{3}{2}\, P_{a \m \n}\, e^{\m\n} \ , \ a\=1,2,3,4
\eea
with the 3-form $P \coloneqq \frac{1}{3} \, \eta^{567}+\frac{1}{3} \,
\sum_{\a=5}^7\, \eta^{\a} \wedge \omega^{\a}$. In our case we obtain
\bea
P\= \tfrac{1}{3}\, \lb
e^{567}-e^{135}+e^{245}-e^{146}-e^{236}+e^{127}+e^{347}\rb \ ,
\eea
so that the structure equations $\diff e^{\mu}= -\Gamma_{\nu}^{\m} \wedge e^{\n} +T^{\m}$ can be written as
\bea
\diff \begin{pmatrix}
       e^1\\
       e^2\\
       e^3\\
       e^4
      \end{pmatrix}
 \= \begin{pmatrix}
                 0 & e^8 & -e^9 & -e^{10}  \\
                 -e^8 & 0 & e^{10} & -e^9  \\
                e^9 & -e^{10} & 0 & -e^8  \\
                e^{10} & e^9 & e^8 & 0 
                \end{pmatrix}
\wedge \begin{pmatrix}
       e^1\\
       e^2\\
       e^3\\
       e^4
      \end{pmatrix}
+
\begin{pmatrix}
       T^1\\
       T^2\\
       T^3\\
       T^4
      \end{pmatrix} \qquad  \text{and} \qquad \diff e^{\a} \= T^{\a} \
      .
\eea
Using the adjoint representation of the generators, one sees that the canonical connection is given as
\bea
\Gamma \= I_8 \otimes e^8 + I_9 \otimes e^9 +I_{10} \otimes e^{10} \ .
\eea
In particular, as mentioned in Section \ref{sec:insthomsp}, the canonical connection on $\mathrm{Sp}(2)/\mathrm{Sp}(1)$ has a particularly simple form because
the representation of the homogeneous space coincides with the quotient of its isometry and structure groups.\footnote{The torsion again coincides with that 
obtained by setting $T(X,Y)=-[X,Y]_{\mathfrak{m}}$ for vectors $X,Y \in \mathfrak{m}$.} Its curvature reads
\bea
\label{eq:curv_can_sp2}
\cf_{\Gamma} \= \diff \Gamma +\Gamma \wedge \Gamma \= 2\, I_8 \otimes
\big(-e^{12}+e^{34} \big)  + 2\, I_9 \otimes \big(e^{13}+e^{24} \big)  +
 2\, I_{10} \otimes \big(e^{14}-e^{23}\big) 
\eea
and it solves the instanton equation \eqref{eq:eq_inst} 
for the 4-form \cite{Harland:2011zs} $Q=\tfrac{1}{6}\, \sum_{\a=5}^7\, \omega^{\a} \wedge \omega^{\a} = e^{1234}$. Written in components of the field 
strength $\cf_\Gamma = \tfrac{1}{2}\, \cf_{\m \n}\, e^{\m \n}$, this instanton equation reads
\begin{subequations}
 \beq
\label{eq:inst_sp2a}
\cf_{12}\=-\cf_{34} \ , \qquad \cf_{13}\=\cf_{24} \ , \qquad
\cf_{14}\=-\cf_{23} \ ,
\eeq
\beq
\label{eq:inst_sp2b}
\cf_{a\a} \= 0\=\cf_{\a\b} \qquad \text{for} \quad  a\=1,2,3,4 \ , \
\a,\b\=5,6,7 \ .
\eeq
\end{subequations}
\paragraph{Instantons on the metric cone.}
The metric cone over $\mathrm{Sp}(2)/\mathrm{Sp}(1)$ carries, by
definition, a hyper-K\"ahler structure, and the instanton equation
\eqref{eq:eq_inst} is written with the 4-form $Q$ defined as \cite{Harland:2011zs}
\bea
\label{eq:four_form_sp2}
Q\= \frac{1}{6}\, \Big(\, \sum_{\a=5}^7\, \omega^{\a} \wedge \omega^{\a}+ \epsilon_{\a \b \g}\, \omega^{\a} \wedge \eta^{\b \g} + 
2 \, \diff \tau \wedge \sum_{\a=5}^7\, \eta^{\a} \wedge \omega^{\a} +
6\, \diff \tau \wedge \eta^{567}\, \Big) \ .
\eea
Here $\diff \tau= \frac{\diff r}{r}$ is the 1-form associated to the cone/cylinder direction.
In components this instanton equation yields the algebraic conditions
\begin{align}
\label{eq:inst_hK1}
 \cf_{12}\=-\cf_{34} \ , \qquad \cf_{13}\=\cf_{24} \ , \qquad \cf_{14}\=-\cf_{23}
\end{align}
and the flow equations
\begin{align}
\label{eq:inst_hK2}
\nonumber &\cf_{1\tau}\=- \cf_{35}\=- \cf_{46}\= \cf_{27} \ , & &
                                                                  \cf_{3\tau}\=
                                                                  \cf_{15}\=
                                                                  \cf_{26}\=\cf_{47}
                                                                  \ ,\\[4pt]
 & \cf_{2\tau} \= \cf_{45}\=-\cf_{36}\=-\cf_{17} \ , &  & \cf_{4\tau}\= -\cf_{25}\=\cf_{16}\=-\cf_{37}
\end{align}
together with the triplet of equations
\bea
\label{eq:inst_hK3}
 \cf_{5\tau}\= \cf_{67} \ , \qquad \cf_{6\tau}\=-\cf_{57} \ , \qquad
 \cf_{7\tau}\=\cf_{56} \ . 
\eea
The first set of conditions \eqref{eq:inst_hK1} are again those of the 3-Sasakian manifold in \eqref{eq:inst_sp2a}, 
while the flow equations \eqref{eq:inst_hK2} and \eqref{eq:inst_hK3} demonstrate the $\mathrm{SU}(2)$ symmetry  of the 
structure. The canonical connection of the 3-Sasakian manifold is also an instanton on the metric cone (cf. \eqref{eq:curv_can_sp2}), 
and therefore it can be used  as starting point in the general form
for the gauge connection on the metric cone.

Since 3-Sasakian manifolds form a special class of Sasaki-Einstein
manifolds, one may also expect a corresponding embedding of the instantons, and indeed
a gauge connection satisfying the conditions \eqref{eq:inst_hK1} and \eqref{eq:inst_hK2} is also a solution to the Hermitian Yang-Mills equations. 
Thus, when studying 3-Sasakian quiver gauge theories, one  also obtains
implicitly results related to Sasakian quiver gauge theories.\footnote{An
  example is given in \cite{Geipel:2016hpk}, where the connection used as starting point for Sasakian quiver gauge theory is the canonical connection of the 
3-Sasakian geometry; it was used because it is better adapted to the
structure of the relevant homogeneous space.} In fact, when we describe the moduli space of 
instantons on the hyper-K\"ahler cone in Section \ref{sec:mod_space_HK}, we show that the description is based on intersections of 
moduli spaces of instantons on cones over Sasaki-Einstein 
manifolds.

\subsection{Quivers}
The general form for the gauge connection on
$\mathrm{Sp}(2)/\mathrm{Sp}(1)$ is given in \eqref{eq:equ_conn}, i.e. we express it as
\bea
\label{eq:equ_conn_sp2}
\ca \= A + \Gamma + \sum_{a=1}^7\, X_a \otimes e^a\= A + \sum_{j=8}^{10}\, I_j \otimes e^j + \sum_{a=1}^7\, X_a \otimes e^a 
\eea
where again $A$ denotes a connection on the vector bundle $E$ over $M^d$. On the metric cone, one can use exactly the same approach, where 
the endomorphisms then may depend on the radial coordinate, $X_a =X_a(r)$. As before, equivariance requires the vanishing of
the mixed terms (or, equivalently, the condition \eqref{eq:equivariance_condition}), which here implies that
\begin{align}
\label{eq:eq_cond_sp2}
\nonumber &\com{\hat{I}_8}{\phi_{(1)}} \= \phi_{(1)} \ ,
  & &\com{\hat{I}_8}{\phi_{(2)}} \= -\phi_{(2)} \ , 
& &\com{\hat{I}_8}{X_{\a}}\=0 \ ,\\[4pt]
&\com{I_4^+}{\phi_{(1)}}\=0 \ ,  & &\com{I_4^+}{\phi_{(2)}}\=\phi_{(1)}
                                   \ ,  & &\com{I_4^+}{X_{\a}}\=0 \ ,\\[4pt]
\nonumber &\com{I_{\bar{4}}^-}{\phi_{(1)}}\=-\phi_{(2)} \ ,
  & &\com{I_{\bar{4}}^-}{\phi_{(2)}}\=0 \ , & &\com{I_{\bar{4}}^-}{X_{\a}}\=0  
\end{align}
for $\a=5,6,7$. Here we have again defined the complex matrices
$\phi_{(1)}=\frac{1}{2}\, (X_1-\im X_2)$ and $\phi_{(2)}=\frac{1}{2}\,
(X_3-\im X_4)$. Since we use the generator dual to $e^7$ as Cartan generator, the quivers might seem to distinguish between $X_7$ and 
the role of the other two matrices $X_5,X_6$, which we sometimes combine 
to $\phi_{(3)}=\frac{1}{2}\, (X_5-\im X_6)$, but from the geometry and
the way in which they will appear in the action functional, their contribution is symmetric. 
Based on the weight diagrams 
in \mbox{Appendix \ref{sec:repres_sp2}},
 we will consider some examples of quivers in this setting.

On the orbifold, one additionally has to impose equivariance with respect to the finite group $\mathbb{Z}_{k}$ embedded in
the $\mathrm{U}(1)$ subgroup generated by $I_7$. Using the action 
\eqref{eq:discrete_eq_sp2_1} and \eqref{eq:discrete_eq_sp2_2} for the analogue of \eqref{eq:eq_cond_z}, one again obtains the
condition that the Higgs fields must act in the weight diagram in the
same way that the ladder operators do, preserving their action on the charge 
associated to $I_7$, if one wants to solve the equivariance conditions
with respect to $\mathbb{Z}_{k}$ for \emph{all} integers $k$. As mentioned before, 
for a fixed value of $k$, one might also solve the condition with more general Higgs fields because the powers of the roots of unity 
$\zeta_{k}$ enter only modulo $k$. In this sense, equivariance with respect to  $\mathbb{Z}_{k}$ embedded in $I_7$ is a weaker condition
than actually imposing $I_7$-equivariance.
\paragraph{Instanton equations.}

Evaluating the curvature of the gauge connection \eqref{eq:equ_conn_sp2} and plugging the components into the instanton equations on the metric cone yields
the flow equations of the Higgs fields \eqref{eq:inst_flow_sp2},
which coincides with the equations one could have obtained from the general formulation in \cite{Ivanova:2012vz}. They can be formulated as
\begin{subequations}
\label{eq:inst_eq_sp2_cone}
 \beq
r\,\dot{\phi}_{(1)} \= -\phi_{(1)} - \com{\phi_{(2)}^\+}{X_5}\=- \phi_{(1)}+\im \com{\phi_{(2)}^\+}{X_6}
             \= -\phi_{(1)} - \im \com{\phi_{(1)}}{X_7} \ ,
\eeq
\beq
r\,\dot{\phi}_{(2)} \=-\phi_{(2)} +\com{\phi_{(1)}^\+}{X_5}\=- \phi_{(2)} -\im \com{\phi_{(1)}^\+}{X_6}
             \= -\phi_{(2)} - \im \com{\phi_{(2)}}{X_7},
\eeq
\beq
r\,\dot{X}_{\a}\= - 2 \, X_{\a} - \tfrac{1}{2}\, \epsilon^{\a\b\g}\,
\com{X_{\b}}{X_{\g}} \qquad \mbox{for} \quad \a,\b,\g \= 5,6,7
\eeq
\end{subequations}
together with the algebraic relations
\bea
\label{eq:inst_eq_sp2_cone2}
\com{\phi_{(1)}}{\phi_{(2)}} \= 2\, \phi_{(3)} \qquad \mbox{and} \qquad
\im \com{\phi_{(1)}}{\phi_{(1)}^\+}+\im
\com{\phi_{(2)}}{\phi_{(2)}^\+} \= 2\, X_7 \ .  
\eea

\paragraph{Fundamental representation $\mbf{\underline{4}}$\,.}
Using the generators in the fundamental representation of
$\mathrm{Sp}(2)$ and the weight diagram (\ref{eq:weight_sp2_fund}), one obtains as decomposition 
\bea
\mbf{\underline{4}}\,\big|_{\mathrm{Sp}(1)} \=
\mbf{\underline{(1,0)_1}} \ \oplus \ \mbf{\underline{(-1,0)_1}} \ \oplus \ \mbf{\underline{(0,-1)_2}} 
\eea
and the quiver 
\bea
\label{eq:quiver:sp2_4_deg}
\begin{tikzpicture}[->,scale=1.1]
    \node (a) at (1.5,0) {{\small${(1,0)}$}};
    \node (b) at (-1.5,0) {{\small${(-1,0)}$}};
    \node (c) at (0,-1.5) {{\small${(0,-1)_2}$}};

     \path[->](c) edge [bend right=10] node [below] {$\scriptstyle\phi_1$} (a);
     \path[->](a) edge [bend right=10] node [above] {$\scriptstyle\phi_3$} (c);
     \path[->](c) edge [bend right=10] node [above] {$\scriptstyle\phi_2$} (b);
     \path[->](b) edge [bend right=10] node [below] {$\scriptstyle\phi_4$} (c);
     \path[<->](a) edge  node [above] {$\scriptstyle\chi^{\a}$} (b);

       \draw [->](b) to [out=65,in=115,looseness=7] (b) node[above] {};
       \draw [->](a) to [out=65,in=115,looseness=7] (a) node[above] {};
       \draw [->](c) to [out=245,in=285,looseness=8] (c) node[above] {};

      \node (aa) at (1.5,1) {$\scriptstyle\psi_{1}^{\a}$};
      \node (bb) at (-1.5,1) {$\scriptstyle\psi_{-1}^{\a}$};
      \node (aa) at (0,-2.5) {$\scriptstyle\psi_{0}^{\a}$};
     
\end{tikzpicture}
\eea
The Higgs fields are given by
\begin{align}
 \phi_{(1)}\= \begin{pmatrix}
              0 & 0 & \phi_1 & 0\\
              0 & 0 & \phi_2 &0\\
               0 & 0 & 0&  0\\
            \phi_3 & \phi_4 &0 &0
            \end{pmatrix} \ , \
 \phi_{(2)}\= \begin{pmatrix}
              0 & 0 & 0 & -\phi_1\\
              0 & 0 & 0 &-\phi_2\\
              \phi_3 & \phi_4 &0 & 0\\
             0 & 0 & 0 &  0
            \end{pmatrix} \ , \
 X_{\a}\= \begin{pmatrix}
              \psi_{1}^{\a} & \chi^{\a} & 0 & 0\\
              -\chi^{\a}\,^{\+} & \psi_{-1}^{\a} & 0 & 0 \\
               0 & 0 & \psi_0^{\a} & 0\\
              0 & 0 & 0 & \psi_0^{\a}
             \end{pmatrix}
\end{align}
with $\a=5,6,7$. The form of these Higgs fields shows a typical feature: Since not all Cartan generators enter the equivariance condition (\ref{eq:eq_cond_sp2}), the allowed 
morphisms are more general than the action of the generators (\ref{eq:def_generators_sp2}).\footnote{This typically occurs for Sasakian quiver gauge
theories because, compared to those on K\"ahler cosets, one has at least one Cartan generator less in the equivariance condition. The usual description is formulated for $H$ being a parabolic
subgroup of $G$~\cite{AlvarezConsul:2001uk}, which does not apply here.}
Imposing additionally equivariance under the $\mathbb{Z}_{k}$-action for generic $k$  
requires the orbifold quiver to be the collapsed weight diagram
\bea
\label{eq:quiver_sp2_fund_restr}
\begin{tikzpicture}[->,scale=1.1]
    \node (a) at (1.5,0) {{\small${(1,0)}$}};
    \node (b) at (-1.5,0) {{\small${(-1,0)}$}};
    \node (c) at (0,-1.5) {{\small${(0,-1)}$}};

     \path[->](a) edge node [below] {$\scriptstyle\phi_3$} (c);
     \path[->](c) edge  node [below] {$\scriptstyle\phi_2$} (b);
     \path[->](a) edge node [above] {$\scriptstyle\phi_6$} (b);
    
      \draw [->](b) to [out=65,in=115,looseness=7] (b) node[above] {};
      \draw [->](a) to [out=65,in=115,looseness=7] (a) node[above] {};
      \draw [->](c) to [out=245,in=285,looseness=8] (c) node[above] {};

      \node (aa) at (1.5,1) {$\scriptstyle\chi_{1} $};
      \node (bb) at (-1.5,1) {$\scriptstyle\chi_{-1} $};
      \node (aa) at (0,-2.4) {$\scriptstyle\chi_{0} $};
     
\end{tikzpicture}
\eea
and the Higgs fields
\bea
\nonumber &&\phi_{(1)}\=\begin{pmatrix}
            0 & 0 &0\\
            0 & 0 & \phi_2 \otimes (1,0)\\
            \phi_3 \otimes (0,1)^\top & 0
 & \mbf{0}_{2}  
           \end{pmatrix} \ , \qquad
\phi_{(2)}\=\begin{pmatrix}
            0 & 0 &  0\\
            0 & 0 & -\phi_2 \otimes (0,1)\\
            \phi_3 \otimes (1,0)^\top & 0 & \mbf{0}_{2} \\
           \end{pmatrix} \ ,\\[4pt]
&&\phi_{(3)}\=\begin{pmatrix}
            0 & 0 & 0\\
            \phi_6 & 0 & 0\\
            0 & 0 & \mbf{0}_{2}\\
            \end{pmatrix} \ , \qquad
 X_7 \= \begin{pmatrix}
\chi_{-1} & 0 & 0 \\
0 & \chi_1 & 0 \\
0 & 0 & \chi_{0} \otimes \id_2
\end{pmatrix}
\eea
take the form of the generators with homomorphisms $\phi_i$ and endomorphisms $\chi_j$ as entries.
The flow equations for the generic quiver \eqref{eq:quiver:sp2_4_deg} yield the complicated system of equations
\bea
\small
\nonumber r\,\dot{\phi}_1 &=& -\phi_1 + \im \phi_3^{\+} \, \psi_0^6 -\im \psi_{-1}^6 \, \phi_3^{\+}- \im \chi^6 \, \phi_4^{\+} 
             \= -\phi_1 - \phi_3^{\+} \, \psi_0^5 + \psi_{-1}^5 \, \phi_3^{\+}+ \chi^5 \, \phi_4^{\+}\\[4pt]
\nonumber             &=& -\phi_1  - \im \phi_1 \, \psi_0^7 + \im
\psi_{-1}^7 \, \phi_1 + \im \chi^7 \, \phi_2 \ ,\\[4pt]
\nonumber r\,\dot{\phi}_2 &=& -\phi_2 + \im \phi_4^{\+} \, \psi_0^6 + \im \chi^{6\,\+} \, \phi_3^{\+} - \im \psi_1^6 \, \phi_4^{\+}
                     \= -\phi_2 - \phi_4^{\+} \, \psi_0^5 - \chi^{5\,\+} \, \phi_3^{\+} + \psi_1^5 \, \phi_4^{\+}\\[4pt]
\nonumber                    &=& - \phi_2 - \im \phi_2 \, \psi_0^7 -
\im \chi^{7\,\+} \, \phi_1 + \im \psi_1^7 \, \phi_2 \ ,\\[4pt]
r\,\dot{\phi}_3 &=& -\phi_3 -\im \phi_1^{\+} \, \psi_{-1}^6 + \im \phi_2^{\+} \, \chi^{6\,\+} + \im \psi_0^6 \, \phi_1^{\+}
\nonumber    \=-\phi_3 +\phi_1^{\+} \, \psi_{-1}^5 - \phi_2^{\+} \, \chi^{5\,\+} - \psi_0^5 \, \phi_1^{\+}\\[4pt]
\nonumber             &=& -\phi_3 - \im \phi_3 \, \psi_{-1}^7 +\im
\phi_4 \, \chi^{7\,\+} +\psi_0^7 \, \phi_3 \ ,\\[4pt]
r\,\dot{\phi}_4 &=& -\phi_4 - \im \phi_1^{\+} \, \chi^6 - \im \phi_2^{\+} \, \psi_1^6 + \im \psi_0^6 \, \phi_2^{\+}
 \nonumber   \= -\phi_4 + \phi_1^{\+} \, \chi^5 + \phi_2^{\+} \, \psi_1^5 - \psi_0^5 \, \phi_2^{\+}\\[4pt]
            &=& - \phi_4 - \im \phi_3 \, \chi^7 - \im \phi_4 \,
            \psi_1^7 + \im \psi_0^7 \, \phi_4 \ .
\eea
\normalsize
The flow equations for the entries of the remaining three matrices are given by
\bea
\small
\nonumber r\,\dot{\psi}_{-1}^{\a} &=& - 2 \, \psi_{-1}^{\a} - \tfrac{1}{2} \,
\epsilon_{\a\b\g} \, \big( \psi_{-1}^{[\b} \, \psi_{-1}^{\g]}
                - \chi^{[\b} \, \chi^{\g]\,\+} \big) \ ,\\[4pt]
\nonumber r\,\dot{\chi}^{\a} &=& - 2 \, \chi^{\a} - \tfrac{1}{2} \,
\epsilon_{\a\b\g} \, \big( \psi_{-1}^{[\b} \, \chi^{\g]} + \chi^{[\b}
\, \psi_1^{\g]}\big) \ ,\\[4pt]
\nonumber r\,\dot{\psi}_1^{\a} &=& -2 \, \psi_{1}^{\a} - \tfrac{1}{2} \,
\epsilon_{\a\b\g} \, \big( - \chi^{[\b\+} \, \chi^{\g]} +
\psi_{1}^{[\b} \, \psi_1^{\g]}\big) \ ,\\[4pt]
r\,\dot{\psi}_0^{\a} &=& - 2\, \psi_0^{\a}  -\tfrac{1}{2}\,
\epsilon_{\a\b\g} \, \psi_0^{[\b} \, \psi_0^{\g]} \ .
\eea
\normalsize
Furthermore, one has the algebraic conditions, i.e. the quiver relations
\bea
\nonumber-\im \psi_0^7 &=& -\phi_1 \, \phi_1^{\+}-\phi_2 \, \phi_2^{\+} 
              +\phi_3 \, \phi_3^{\+} +\phi_4 \, \phi_4^{\+} \ ,\\[4pt]
\nonumber-\im \psi_{1}^7 &=& \phi_1 \, \phi_1^{\+}-\phi_3^{\+} \,
\phi_3 \ ,
    \qquad -\im \psi_{-1}^7 \=  \phi_2 \, \phi_2^{\+}-\phi_4^{\+} \,
    \phi_4 \ ,\\[4pt]
-\im \chi^7 &=& \phi_1 \, \phi_2^{\+} - \phi_3^{\+} \, \phi_4 \ ,\\[4pt]
\nonumber\psi_0 &=& \phi_3 \, \phi_1 +\phi_4 \, \phi_2 \ , 
    \qquad \psi_{-1} \= \phi_1 \, \phi_3 \ ,
    \qquad \psi_{1} \= \phi_2 \, \phi_4 \ ,\\[4pt]
\nonumber \chi &=& \phi_1 \, \phi_4 \= - (\phi_2 \, \phi_3)^{\+} \ ,
\eea
where we denote $\psi_i := \frac12\,(\psi_i^5 - \im \psi_i^6)$ and $\chi := \frac12\,(\chi^5 - \im \chi^6)$.
Restricting to the case of the simpler orbifold quiver 
\eqref{eq:quiver_sp2_fund_restr} reduces the complexity somewhat, but one still has to solve 
a highly non-trivial system of matrix equations. However, when imposing the scalar form of \cite{Harland:2011zs}, the system simplifies to 
\eqref{eq:scalar_sp2}, which has the analytic solutions \eqref{eq:scalar_sp2_sols}.

\paragraph{Representation $\mbf{\underline{5}}$\,.}

Using the five-dimensional representation (\ref{eq:rep_sp2_5}) with
the Cartan generators $\hat{I}_7=\mathrm{diag}(-1,-1,0,1,1)$,
$\hat{I}_8 = \mathrm{diag}(1,-1,0,1,-1)$ and its weight diagram \eqref{eq:weight_sp2_5}, one obtains the decomposition
\bea
\mbf{\underline{5}}\,\big|_{\mathrm{Sp(1)}} \=
\mbf{\underline{(-1,-1)_2}} \ \oplus \ \mbf{\underline{(0,0)_1}} \ \oplus \ \mbf{\underline{(1,-1)_2}}
\eea
and the quiver 
\bea
\label{eq:quiver:sp2_5_deg}
\begin{tikzpicture}[->,scale=1.1]
    \node (a) at (-1.5,0) {{\small$(-1,-1)$}};
    \node (b) at (0,1.5) {{\small$(0,0)$}};
    \node (c) at (1.5,0) {{\small$(1,-1)$}};

     \path[->](b) edge [bend right=10] node [above] {$\scriptstyle\phi_1$} (a);
     \path[->](a) edge [bend right=10] node [below] {$\scriptstyle\phi_2$} (b);
     \path[->](c) edge [bend right=10] node [above] {$\scriptstyle\phi_3$} (b);
     \path[->](b) edge [bend right=10] node [below] {$\scriptstyle\phi_4$} (c);
     \path[<->](c) edge  node [below] {$\scriptstyle\chi^{\a}$} (a);
       \draw [->](b) to [out=65,in=115,looseness=7] (b) node[above] {};
       \draw [->](a) to [out=245,in=285,looseness=8] (a) node[above] {};
       \draw [->](c) to [out=245,in=285,looseness=8] (c) node[above] {};

      \node (aa) at (-1.5,-1) {$\scriptstyle\psi_{-1}^{\a}$};
      \node (bb) at (0,2.6) {$\scriptstyle\psi_0^{\a} $};
      \node (cc) at (1.5,-1) {$\scriptstyle\psi_{1}^{\a}$};
    \end{tikzpicture}
\eea
 The Higgs fields read
\begin{subequations}
\beq
\phi_{(1)}\=\begin{pmatrix}
            0 &0 & \phi_1 & 0 & 0\\
            0 &0 & 0 & 0 & 0\\
            0 &\phi_2 & 0 & 0 & \phi_3\\
            0 &0 & \phi_4 & 0 & 0\\
            0 &0 & 0 & 0 & 0\\
           \end{pmatrix} \ , \qquad
\phi_{(2)}\=\begin{pmatrix}
            0 &0 & 0 & 0 & 0\\
            0 &0 & \phi_1 & 0 & 0\\
            -\phi_2 &0 & 0 & \phi_3 & 0\\
            0 &0 & 0 & 0 & 0\\
            0 &0 & -\phi_4 & 0 & 0\\
           \end{pmatrix} \ ,
\eeq
\beq
X_{\a}\=\begin{pmatrix}
            \psi_{-1}^{\a} & 0 & 0 & \chi^{\a} & 0\\
            0 & \psi_{-1}^{\a} & 0 & 0& -\chi^{\a}\\
            0 & 0 & \psi_{0}^{\a} & 0 & 0\\
            -\chi^{\a}\,^\dag & 0 & 0 & \psi_1^{\a} & 0\\
            0 & \chi^{\a}\,^\dag & 0 & 0 & \psi_1^{\a} 
           \end{pmatrix} \ .
\eeq
\end{subequations}
As in the previous example, the number of allowed arrows is larger
than the number of entries in the generators (\ref{eq:rep_sp2_5}), and
the quiver is reduced to the 
weight diagram by further imposing equivariance under $\mathbb{Z}_{k}
\hookrightarrow \big\langle\exp(\hat{I}_7) \big\rangle$. Then the
orbifold quiver is given by 
\bea
\label{eq:quiver:sp2_5_restr}
\begin{tikzpicture}[->,scale=1.1]
    \node (a) at (-1.5,0) {{\small$(-1,-1)$}};
    \node (b) at (0,1.5) {{\small$(0,0)$}};
    \node (c) at (1.5,0) {{\small$(1,-1)$}};

       \path[->](b) edge node [above] {$\scriptstyle\phi_1$} (a);
       \path[->](c) edge  node [above] {$\scriptstyle\phi_3$} (b);
       \path[<->](c) edge  node [below] {$\scriptstyle\chi^{5},\chi^{6}$} (a);
       \draw [->](b) to [out=65,in=115,looseness=7] (b) node[above] {};
       \draw [->](a) to [out=245,in=285,looseness=8] (a) node[above] {};
       \draw [->](c) to [out=245,in=285,looseness=8] (c) node[above] {};

      \node (aa) at (-1.5,-1) {$\scriptstyle\psi_{-1}$};
      \node (bb) at (0,2.6) {$\scriptstyle\psi_0 $};
      \node (cc) at (1.5,-1) {$\scriptstyle\psi_{1}$};
    \end{tikzpicture}
\eea
with $\psi_i :=\psi_i^7$.
The two representations $\mbf{\underline{4}}$ and $\mbf{\underline{5}}$ yield quivers of the same shape, but the decomposition into irreducible representations
of $\mathrm{Sp}(1)$ differs, so that one obtains a different set of instanton equations. The flow equations read
\bea
\nonumber r\,\dot{\phi}_1 &=& - \phi_1 - \im \chi^6 \, \phi_3^{\+} \= - \phi_1 + \chi^5 \, \phi_3^{\+} 
  \= -\phi_1 - \im \phi_1 \, \psi_0 + \im \psi_{-1} \, \phi_1 \ ,\\[4pt]
\nonumber r\,\dot{\phi}_3 &=&  - \phi_3 - \im  \phi_1^{\+} \, \chi^6 \= - \phi_3 + \phi_1^{\+} \, \chi^5 
        \= -\phi_3 - \im \phi_3 \, \psi_1 + \im \psi_0 \, \phi_3 \ ,\\[4pt]
\nonumber r\,\dot{\chi}^5 &=& - 2\, \chi^5 - \chi^6 \, \psi_1 + \psi_{-1} \,
\chi^6 \= -2\, \chi^5 - \chi^6 \, \psi_{-1} + \psi_1 \, \chi^6 \ ,\\[4pt]
r\,\dot{\chi}^6 &=&- 2\, \chi^6 + \chi^5 \, \psi_1 - \psi_{-1} \, \chi^5 \=
-2\, \chi^6 + \chi^5 \, \psi_{-1} - \psi_1 \, \chi^5 \ ,\\[4pt]
\nonumber r\,\dot{\psi}_{-1} &=& -2\, \psi_{-1} + \chi^5 \, \chi^{6\,\+} -
\chi^6 \, \chi^{5\,\+} \ ,\\[4pt]
\nonumber r\,\dot{\psi}_{0} &=& - 2\, \psi_0 \ ,\\[4pt]
\nonumber r\,\dot{\psi}_{1} &=& - 2\, \psi_1 + \chi^{5\,\+} \, \chi^6 -
\chi^{6\,\+} \, \chi^5 \ ,
\eea
while the algebraic conditions from \eqref{eq:inst_hK1} yield the quiver relations
\bea
\tfrac{1}{2}\, \big(\chi^5 - \im \chi^6\big) \=  \phi_1 \, \phi_3 \ ,\quad
-2\im \psi_{-1} \= \phi_1 \, \phi_1^{\+} \ , \quad 2\im  \psi_1 \=
\phi_3^{\+} \, \phi_3 \ , \quad  \psi_0 \= \phi_3 \, \phi_3^{\+} -
\phi_1^{\+} \, \phi_1 \ .
\eea


\paragraph{Adjoint representation $\mbf{\underline{10}}$\,.}

The ten-dimensional adjoint representation of $\mathrm{Sp}(2)$ decomposes as 
\bea
\mbf{\underline{10}}\,\big|_{\mathrm{Sp(1)}} \=
\mbf{\underline{(-2,0)_1}} \ \oplus \ \mbf{\underline{(-1,-1)_2}} \
    \oplus \ \mbf{\underline{(0,-2)_3}} \ \oplus \
       \mbf{\underline{(0,0)_1}} \ \oplus \ \mbf{\underline{(1,-1)_2}} \
           \oplus \ \mbf{\underline{(2,0)_1}} \ .
\eea
For better readability, we specialize to the case when equivariance
is imposed as well with respect to the second Cartan generator
$\hat{I}_7$. The generic case can be easily recovered by the applying
the conditions \eqref{eq:eq_cond_sp2} to the weight diagram
\eqref{eq:weight_sp2_adj}, and it will involve a large number of arrows and also several maps between the same vertices, similarly to the more general quivers in the 
previous examples (\ref{eq:quiver:sp2_4_deg}) and
(\ref{eq:quiver:sp2_5_deg}). With this restriction, we obtain the
orbifold quiver
\bea
\small
\begin{tikzpicture}[->,scale=1.3]
    \node (a) at (-2.4,0) {{\small$(-2,0)$}};
    \node (b) at (-1.2,-1.2) {{\small$(-1,-1)$}};
    \node (c) at (0,-2.4) {{\small$(0,-2)$}};
    \node (d) at (0,0) {{\small$(0,0)$}};
    \node (e) at (1.2,-1.2) {{\small$(1,-1)$}};
    \node (f) at (2.4,0) {{\small$(2,0)$}};

     \path[->](b) edge node [above] {$\scriptstyle\phi_1$} (a);
     \path[->](c) edge node [above] {$\scriptstyle\phi_2$} (b);
     \path[->](d) edge node [above] {$\scriptstyle\phi_4$} (b);
     \path[->](e) edge node [above] {$\scriptstyle\phi_5$} (d);
      \path[->](e) edge node [above] {$\scriptstyle\phi_3$} (c);
      \path[->](f) edge node [above] {$\scriptstyle\phi_6$} (e);

          \path[->](d) edge node [above] {$\scriptstyle\chi_1$} (a);
          \path[->](e) edge node [above] {$\scriptstyle\chi_2$} (b);
          \path[->](f) edge node [above] {$\scriptstyle\chi_3$} (d);

       \draw [->](a) to [out=65,in=115,looseness=7] (a) node[above] {};
       \draw [->](d) to [out=65,in=115,looseness=7] (d) node[above] {};
       \draw [->](f) to [out=65,in=115,looseness=7] (f) node[above] {};
       \draw [->](c) to [out=245,in=295,looseness=7] (c) node[above] {};
       \draw [->](b) to [out=245,in=295,looseness=7] (b) node[above] {};
       \draw [->](e) to [out=245,in=295,looseness=7] (e) node[above] {};

     \node (aa) at (-2.4,.9) {$\scriptstyle\psi_{-2}$};
     \node (dd) at (0,.9) {${\scriptstyle\psi}_{0}$};
      \node (ff) at (2.4,.9) {$\scriptstyle\psi_{2} $};
     \node (cc) at (0,-3.3) {$\scriptstyle\tilde{\psi}_{0} $};
     \node (bb) at (-1.2,-2.1) {$\scriptstyle\psi_{-1} $};
     \node (ee) at (1.2,-2.1) {$\scriptstyle\psi_{1} $};

    \end{tikzpicture}
\eea
\normalsize
where $\psi_i$ and $\tilde\psi_0$ denote the endomorphisms contained
in $X_7$, while  $\chi_i := \frac12\,(\chi_i^5 - \im \chi_i^6)$ are the entries  of $X_5$ and $X_6$.
Due to the large number of arrows, we do not write out explicitly the instanton equations for this case, but they can be obtained simply by inserting the Higgs field 
matrices into the instanton equations \eqref{eq:inst_eq_sp2_cone} and \eqref{eq:inst_eq_sp2_cone2}.


\paragraph{Representation $\mbf{\underline{14}}$\,.}
As our final example we consider the 14-dimensional representation of
$\mathrm{Sp}(2)$, which decomposes under restriction to $\mathrm{Sp}(1)$ as
\bea
&\mbf{\underline{14}}\,\big|_{\mathrm{Sp}(1)} \=
\mbf{\underline{(-2,-2)_3}} \ \oplus \ \mbf{\underline{(-1,-1)_2}}  \
    \oplus \
                      \mbf{\underline{(0,-2)_3}} \ \oplus \
                        \mbf{\underline{(0,0)_1}}  \ \oplus \
                        \mbf{\underline{(1,-2)_2}} \ \oplus \
                      \mbf{\underline{(2,-2)_3}} \ , \nonumber\\
\eea
so that the quiver (after further imposing equivariance under $\mathbb{Z}_{k}
\hookrightarrow \big\langle \mathrm{exp}(\hat{I}_7)\big\rangle$) 
follows from the weight diagram (\ref{eq:weight_14}) and is given by
\bea
\begin{tikzpicture}[->,scale=1.3]
    \node (a) at (-2.4,0) {\small$(-2,0)$};
    \node (b) at (-1.2,1.2) {\small$(-1,-1)$};
    \node (c) at (0,0) {\small$(0,-2)$};
    \node (d) at (0,2.4) {\small$(0,0)$};
    \node (e) at (1.2,1.2) {\small$(1,-1)$};
    \node (f) at (2.4,0) {\small$(2,0)$};

      \path[->](b) edge node [above] {$\scriptstyle\phi_1$} (a);
      \path[->](c) edge node [above] {$\scriptstyle\phi_4$} (b);
      \path[->](d) edge node [above] {$\scriptstyle\phi_2$} (b);
      \path[->](e) edge node [above] {$\scriptstyle\phi_3$} (d);
       \path[->](e) edge node [above] {$\scriptstyle\phi_5$} (c);
       \path[->](f) edge node [above] {$\scriptstyle\phi_6$} (e);
 
           \path[->](c) edge node [above] {$\scriptstyle\chi_1$} (a);
           \path[->](f) edge node [above] {$\scriptstyle\chi_2$} (c);
           \path[->](e) edge node [above] {$\scriptstyle\chi_3$} (b);
 
	  \draw [->](b) to [out=65,in=115,looseness=7] (b) node[above] {};
          \draw [->](d) to [out=65,in=115,looseness=7] (d) node[above] {};
          \draw [->](e) to [out=65,in=115,looseness=7] (e) node[above] {};
          \draw [->](a) to [out=245,in=295,looseness=7] (a) node[above] {};
          \draw [->](c) to [out=245,in=295,looseness=7] (c) node[above] {};
          \draw [->](f) to [out=245,in=295,looseness=7] (f) node[above] {};

        \node (aa) at (-2.4,-.9) {$\scriptstyle\psi_{-2}$};
        \node (dd) at (0,-.9) {$\scriptstyle\psi_{0}$};
        \node (cc) at (2.4,-.9) {$\scriptstyle\psi_{2} $};
        \node (ee) at (0,3.3) {$\scriptstyle\tilde\psi_{0} $};
        \node (bb) at (-1.2,2.1) {$\scriptstyle\psi_{-1} $};
        \node (dd) at (1.2,2.1) {$\scriptstyle\psi_{1} $};
   \end{tikzpicture}
\eea
This quiver has the same general structure as that of the adjoint representation, 
but the multiplicities of the vertices are different, as was also 
the case for the fundamental representation in comparison with the
representation $\mbf{\underline{5}}$\,. 
These last two examples with a large number of possibly contributing
fields clearly demonstrate the advantages of the quiver approach for quickly constructing 
an equivariant connection. 

\subsection{Yang-Mills-Higgs theories}

Plugging in the components of the field strength and using the orthonormality of the basis, one obtains
the Yang-Mills action for an equivariant gauge connection on $\mathrm{Sp}(2)/\mathrm{Sp}(1)$ as
\bea
\label{eq:action_squashed2}
  \nonumber S_{\mathrm{YM}} &=& \mathrm{vol}\big(
  \mathrm{Sp}(2)/\mathrm{Sp}(1)\big) \
  \int_{M^d} \, \diff^d y \ \sqrt{g}\ \frac{1}{2}\, \mathrm{tr} \Big(
  \, \frac{1}{2} \, F_{\m\n} \, \lb F^{\m\n} \rb^{\+}
+   \sum_{\m=1}^d\ \sum_{a=1}^7\, \left|D_{\m} X_a \right|^2\\
 \nonumber &&+ \, \left|\com{X_1}{X_2} + 2\, X_7 - 2\, I_8 \right|^2 
             + \left|\com{X_1}{X_3} - 2\, X_5 + 2\, I_9 \right|^2 +
             \left| \com{X_1}{X_4} - 2\, X_6 + 2\, I_{10} \right|^2 \\
 \nonumber && +\, \left|\com{X_2}{X_3} - 2\, X_6 - 2\, I_{10} \right|^2 +
 \left| \com{X_2}{X_4} + 2\, X_5 + 2\, I_9 \right|^2
             + \left|\com{X_3}{X_4} + 2\, X_7 + 2\, I_8 \right|^2\\
\nonumber  && + \, \left|\com{X_1}{X_5}+X_3 \right|^2               + \left|\com{X_1}{X_6}+X_4 \right|^2 + \left|\com{X_1}{X_7}-X_2 \right|^2
             + \left|\com{X_2}{X_5}-X_4 \right|^2 \\
\nonumber && + \, \left|\com{X_2}{X_6}+X_3 \right|^2+                \left|\com{X_2}{X_7}+X_1 \right|^2+ \left|\com{X_3}{X_5}-X_1 \right|^2
             + \left|\com{X_3}{X_6}-X_2 \right|^2\\
\nonumber  &&+ \, \left|\com{X_3}{X_7}-X_4 \right|^2 +  \left|\com{X_4}{X_5}+X_2 \right|^2+ \left|\com{X_4}{X_6}-X_1 \right|^2
             + \left|\com{X_4}{X_7}+X_3 \right|^2\\
 &&+ \, \left|\com{X_5}{X_6}+2\, X_7 \right|^2+
\left|\com{X_5}{X_7}-2\, X_6 \right|^2+ \left|\com{X_6}{X_7}+2\, X_5
\right|^2 \, \Big) \ ,
\eea
where the covariant derivatives are defined as before for the action of the Sasakian quiver gauge theory in \eqref{eq:action_s7}. 
The 3-Sasakian geometry of the squashed seven-sphere is evident in the form of this action. 
The instanton conditions \eqref{eq:inst_sp2b} on $\mathrm{Sp}(2)/\mathrm{Sp}(1)$ imply the vanishing of the terms in the 
last four lines in \eqref{eq:action_squashed2}, so that the Yang-Mills action 
for instantons on $\mathrm{Sp}(2)/\mathrm{Sp}(1)$ is given by
\bea
\label{eq:action_squashed2_inst}
\nonumber  S_{\mathrm{YM}}^{\mathrm{inst}} &=& \mathrm{vol}\big(
\mathrm{Sp}(2)/\mathrm{Sp}(1)\big) \ \int_{M^d} \, \diff^d y\
\sqrt{g}\ \frac{1}{2}\, \mathrm{tr} \Big(\, \frac{1}{2} \,
 F_{\m\n} \, \lb F^{\m\n} \rb^{\+}
+  \sum_{\m=1}^d \ \sum_{a=1}^7\, \left|D_{\m} X_a \right|^2\\ &&
\qquad \qquad \qquad \qquad \qquad \nonumber +\,
 2\left| \com{X_1}{X_2} + 2 \, X_7 - 2\, I_8 \right|^2 +2 \left|
   \com{X_1}{X_3} - 2 \, X_5 + 2\, I_9 \right|^2 \\ && \qquad \qquad
 \qquad \qquad 
 \qquad +\, 2\left| \com{X_1}{X_4} - 2 \, X_6 + 2\, I_{10} \right|^2
 \, \Big) \ .
\eea
As a by-product one can obtain the action for an equivariant gauge connection on the
twistor space $\mathrm{Sp}(2)/\mathrm{Sp}(1) \times \mathrm{U}(1)$, analogously to the reduction from
$S^7$ to $\mathbb{C}P^3$ from Section~\ref{sec:CP3}, and to the
quaternionic space $S^4$ underlying the local section $Q$ of the
fibration \eqref{eq:fibrations_sp2_s4}. 


\subsection{$\mathrm{Sp}(2)$-instantons on the hyper-K\"ahler cone}
\label{sec:mod_space_HK}
We shall now describe the moduli space of instantons on the hyper-K\"ahler cone 
$C\big(\mathrm{Sp}(2)/\mathrm{Sp}(1)\big) = \mathbb{H}^2$; for an overview of hyper-K\"ahler geometry, see for instance \cite{Hitchin1991}. 
The defining property is the existence of  an $\mathrm{SU}(2)$-triplet of complex structures $J_{\a}$ satisfying the quaternion relations
\bea
\label{eq:quaternionic_relations_J}
J_{\a} \,J_{\b} \= -\delta_{\a\b} \, \id + \epsilon_{\a\b}{}^{\g} \,
J_{\g} \ ,
\eea
or, equivalently, a triplet of K\"ahler forms\footnote{With our explicit choices of the defining 
equations \eqref{eq:3Sasaki_def} and \eqref{eq:3Sasaki_def_forms} of the underlying 3-Sasakian manifold, they 
can be taken as \mbox{$\Omega_1 = r^2 \, (e^{12} + e^{34}+ e^{56} + e^{\tau 7})$}, 
 \mbox{$\Omega_2 = r^2 \, (e^{31} + e^{24}+ e^{67} + e^{\tau 5})$} and
\mbox{$\Omega_3 = r^2 \, (e^{32} + e^{41}+ e^{75} + e^{\tau 6})$}.} 
$\Omega_\a (X,Y)\coloneqq g_{\rm cone} (X, J_\a Y)$.
By virtue of the quaternion relations \eqref{eq:quaternionic_relations_J}, any $J=s^{\a}\, J_{\a}$ with $s=(s^\a) \in S^2$ yields a complex structure on the 
tangent bundle of the hyper-K\"ahler manifold.

We now show that the condition of being an $\mathrm{Sp}(2)$-instanton is equivalent to imposing the condition 
of being a Hermitian Yang-Mills instanton with respect to any such complex \mbox{structure $J$}. 
On the metric cone $C(\mathrm{Sp}(2)/\mathrm{Sp}(1))$ the holonomy can be reduced from the generic holonomy group $\mathrm{SO}(8)$ of an 
oriented eight-dimensional Riemannian manifold to 
the subgroup $\mathrm{Sp}(2)$. Denoting this splitting of the
corresponding Lie algebra as
\bea
\mathfrak{so}(8) \= \mathfrak{sp}(2) \oplus \mathfrak{sp}(1) \oplus \mathfrak{k}
\eea
with $\mathfrak{k}$  the orthogonal  complement, an  $\mathrm{Sp}(2)$-instanton is a connection whose curvature $\cf=(\cf_{\m\n})$, considered
as an $\mathfrak{so}(8)$ matrix, is valued in 
the subalgebra $\mathfrak{sp}(2)$ alone. On hyper-K\"ahler manifolds there is a $\C P^1$-family 
of complex structures $J$, and $\mathrm{Sp}(2)$-instantons can be
generally described by the condition \cite{capria1988yang}
\bea
\label{eq:eq_holom_sp_inst}
\cf_J^{0,2} \= 0 \= \cf_J^{2,0} \qquad \mbox{for} \quad J\=s^{\a}\,
J_{\a}\qquad \text{with} \quad s \in S^2 \ ,
\eea
where the superscripts refer to the $(0,2)$ and $(2,0)$ parts of the curvature with respect to the complex structure $J$. Recall that \eqref{eq:eq_holom_sp_inst}
is the holomorphicity condition of Hermitian Yang-Mills instantons, i.e. it ensures that in the splitting 
\bea
\mathfrak{so}(8) \= \mathfrak{u}(4) \oplus \mathfrak{p}
\eea
the curvature is valued in the subalgebra $\mathfrak{u}(4)$ alone. In
contrast to the case of a single Hermitian Yang-Mills moduli space as in 
Section \ref{sect:cone_instantons}, by imposing the holomorphicity conditions with respect to \emph{all} complex structures,
the corresponding stability conditions are automatically fulfilled as the
following argument shows. For any $J$ the conditions in \eqref{eq:eq_holom_sp_inst}
can be formulated in terms of the projection operator $\tfrac{1}{2}\,(\id + \im J)$. In components, we then have
\bea
\tfrac{1}{4}\,\big(\delta_{\m}^{\lambda}+ \im J_{\m}^{\l} \big)\,
\big(\delta_{\n}^{\sigma}+ \im J_{\n}^{\sigma} \big)\, \cf_{\lambda
  \sigma} \= 0 \ ,
\eea
so that the $\mathrm{Sp}(2)$-instanton equations are equivalent to 
\bea
\label{eq:holom_in_components}
J_{\a\m}^{\sigma} \, \cf_{\n \sigma} \= J_{\a\n}^{\sigma} \, \cf_{\m\sigma} 
\qquad \mbox{and} \qquad
J_{\a\m}^{\lambda} \, J_{\a\n}^{\sigma} \, \cf_{\lambda \sigma} \=
\cf_{\m\n} \qquad \text{(no sum on
  $\a$)} \ .
\eea
The second set of equations in \eqref{eq:holom_in_components} in fact follows from the first set, as is demonstrated in Appendix~\ref{sect:3Sasaki_vs_SE}. Imposing these conditions for all $\a=1,2,3$ 
then also implies the conditions
\bea
\Omega_{\a}^{\m\n} \, \cf_{\m\n} \= 0 \ ,
\eea
which are the stability conditions of Hermitian Yang-Mills instantons. Therefore, the $\mathrm{Sp}(2)$-instanton equations restrict the curvature
to lie in the subalgebra $\mathfrak{su}_{\a}(4) \subset \mathfrak{su}_{\a}(4) \oplus \mathfrak{u}_{\a}(1) \oplus \mathfrak{p}_{\a}= \mathfrak{so}(8)$
for each $\a=1,2,3$. Conversely, by the same arguments,\footnote{Alternatively, note that 
the 4-form $Q$ appearing in the instanton equation \eqref{eq:eq_inst}
can be written here as 
$Q= \tfrac{1}{3}\, (Q_1+Q_2+Q_3)$,
so that a connection satisfying the Hermitian Yang-Mills 
equations with respect to all three complex structures is also a solution to the $\mathrm{Sp}(2)$-instanton equation.} 
such a triplet of Hermitian Yang-Mills equations implies that the
connection is an $\mathrm{Sp}(2)$-instanton~\cite{capria1988yang}.

Consequently, the moduli space of $\mathrm{Sp}(2)$-instantons on the
hyper-K\"ahler cone is given as the intersection of the three
Hermitian Yang-Mills
moduli spaces:
\bea
\label{eq:intersection}
\mathcal{M} \= \mathcal{M}_1 \cap \mathcal{M}_2 \cap \mathcal{M}_3 \ .
\eea
For the description of each Hermitian Yang-Mills moduli space 
\bea
\mathcal{M}_{\a} \ \coloneqq \ \big\{\ca \in \mathbb{A}^{1,1} \ \big| \ \Omega_{\a} \haken \cf\=0 \quad , \quad  
    \cf_{J_\a}^{2,0} \= 0 \=  \cf_{J_\a}^{0,2} \big\}
\eea 
one can apply the techniques of Section \ref{sect:cone_instantons}, making \eqref{eq:intersection} into a hyper-K\"ahler orbit space. Note that, of course, the equivariance conditions differ, and that also the 
quiver relations imposed by the holomorphicity conditions are slightly different, as the explicit equations \eqref{eq:inst_flow_sp2} and
\eqref{eq:inst_alg_sp2}   show. The intersection \eqref{eq:intersection}
is obviously non-empty because it contains, for instance, the trivial solutions $X_{\m}=0$ and the analytic solution of the scalar form 
from \cite{Harland:2011zs}.


\section{Translationally invariant instantons on the Calabi-Yau cone\label{sec:translationinvariant}}

The metric cone over the seven-sphere with its round metric is
$\mathbb{C}^4$, and this motivates a study of translationally invariant
instantons on an orbifold $\mathbb{C}^4/\Gamma$ by a finite group
$\Gamma\subset\mathrm{SU}(4)$, as similarly done
in~\cite{Lechtenfeld:2014fza,Lechtenfeld:2015ona}. Generally, moduli
spaces of instantons on $\C^n/\Gamma$ are related to resolutions of
orbifold singularities and aspects of the McKay correspondence, see \cite{SardoInfirri:1996gb}
and references therein. For $n=4$ and $\Gamma=\Z_k$, they determine
the vacuum moduli spaces of the Chern-Simons quiver gauge theories
discussed in Section~\ref{sec:intro}.
\subsection{Equivariant connections}
On $\C^4$ we use the coordinates $(z_1,z_2,z_3,z_4)$, equipped with the standard metric and complex structure on $\C^4$, i.e. 
$J z_{\a}= \im z_{\a}$. The differentials of the coordinates provide a translationally invariant basis of 1-forms, and we write 
a connection as
\bea
\ca \= Y_{\a} \otimes \diff z^{\a} + \bar Y_{\bar{\a}} \otimes \diff \bar z^{\bar\a}
\eea
with $\bar Y_{\bar{\a}} = - Y_{\a}^{\+}$ describing the endomorphism part
of the connection (acting on the fibres $\C^r$ of the underlying vector bundle).
Translational invariance of this connection, i.e. $\diff \ca =0$, then implies as conditions
\bea
\diff Y_{\a} \= 0 \= \diff \bar Y_{\bar{\a}} \ .
\eea
The form of the endomorphism is determined by equivariance with respect to the finite subgroup $\Gamma=\Z_k$.
When introducing the orbifold action on $S^7$ previously, its action on forms followed from the way it acts on the quantities entering the 
local section. For the round seven-sphere it was induced by the
fundamental action of $\mathrm{SU}(4)$ on $\mathbb{C}^4$ in \eqref{eq:def_z-action},
and the ensuing quotient leading to the local patch in $\C P^3$. Now, however, we start directly from the action of $I_7$ in the 
fundamental representation,
so that one can define
\bea
\pi(h_k): z_a \ \longmapsto \ \zeta_{k}^{-1} \, z_a \ , \ a\=1,2,3 \ ,
\qquad z_4  \ \longmapsto \ \zeta_{k}^{3} \, z_4 
\eea
with $\zeta_{k}$ a primitive $k$-th root of unity. In \cite[Section
6.1]{Lechtenfeld:2015ona} a detailed discussion is given of the different choices 
of $\mathbb{Z}_{k}$-actions in the case of $G$-equivariant connections and translationally invariant instantons on the cone over $S^5$. Following this, 
we do not consider the weights associated to the generator $I_7$, which has been used for our discussion of $\mathrm{SU}(4)$-equivariant instantons on orbifolds of $S^7$, 
but consider the weights pertaining to the other Cartan generators, as the examples below will clarify. 
The condition of equivariance then reads \cite{Lechtenfeld:2014fza,Lechtenfeld:2015ona}
\bea
\label{eq:eq_discrete_gamma}
\g (h_k)\, Y_{\a}\, \g (h_k)^{-1} \= \pi(h_k) Y_{\a} \ , 
\eea 
where $\g(h_k)$ denotes the action of $\Gamma$ on the fibres $\C^r$. 
With its standard metric and complex structure, the K\"ahler form on $\C^4$ is given by
\bea
\Omega \= -\frac{\im}{2} \, \sum_{\a=1}^4\, \diff z^{\a} \wedge \diff
\bar z^{\bar{\a}}
\eea 
and it allows for application of the Hermitian Yang-Mills equations. While the condition $\cf_{\a \b}=0=\cf_{\bar\a\bar\b}$ again yields 
\bea
\com{Y_{\a}}{Y_{\b}} \= 0\= \com{\bar Y_{\bar{\a}}}{\bar Y_{\bar{\b}}} \qquad
\mbox{for} \quad \a, \b\=1, 2,3, 4 \ , 
\eea
one can include a Fayet-Iliopoulos term $\Xi$ in the stability
condition $\Omega \haken \cf = \Xi $ which yields
\bea
\label{eq:inst_FI}
\sum_{\a=1}^4\, \com{Y_{\a}}{\bar Y_{\bar{\a}}} \= \Xi \ , 
\eea
where $\Xi$ lies in the center of the Lie algebra $\mathfrak{u}(r)$ of
the structure group. Motivated by the study of self-dual connections in four dimensions and their
description as hyper-K\"ahler quotients \cite{Kronheimer:1989zs}, one might also consider here $\mathrm{Sp}(2)$-instantons, whose 
equations and moduli space we shall discuss in detail in Section~\ref{sec:mod_trans_inv}. 

\subsection{Examples\label{sec:transinvexamples}}
We shall now study the resulting quiver gauge theories for some representations of $\mathrm{SU}(4)$, as used for 
the description of Section \ref{section:quiver_diagrams}. Due to the
complicated and lengthy equations for hyper-K\"ahler instantons, we restrict our attention
here to Hermitian Yang-Mills instantons with respect to one of the complex structures, as higher-dimensional analogues of the 
cases considered in \cite{Lechtenfeld:2015ona}.

As action $\gamma(h_k)$ on the fibres we assign to each subspace in the decomposition of the chosen $\mathrm{SU}(4)$-representation
certain $k$-th roots of unity with weights given by the quantum numbers with respect to the Cartan generator $I_8$. 
Note that this choice of $\gamma(h_k)$ is neither unique nor necessary, but just a \emph{possible} choice for the action on the fibres 
(as in~\cite{Lechtenfeld:2015ona}). However, in order to actually get an embedding into $\mathrm{SU}(4)$, 
one has to impose a certain condition on the dimensions $r_i$ of the occuring
vector spaces in the decomposition.\footnote{Since $\mathrm{SU}(4)$ has rank three, another obvious choice is to use the weights associated to the Cartan \mbox{generator $I_9$}. 
For the four examples given here, this yields the same quivers.} 
 
\paragraph{Fundamental representation $\mbf{\underline{4}}$\,.} For
the fundamental representation of $\mathrm{SU}(4)$, the generator of
$\mathbb{Z}_{k}$ is chosen based on the decomposition \eqref{eq:dec_4_su4} and the quantum numbers with respect to $\hat{I}_8$ as
\bea
\g (h_k) \= \begin{pmatrix}
                     \zeta_{k}^{-1} \, \id_3 \otimes \id_{r_{-1}} & 0\\
                     0          & 1  \otimes \id_{r_3}
                    \end{pmatrix}\qquad \text{with} \quad 3 r_{-1}
                    \equiv 0\ \ \mathrm{mod}\ k \ ,
\eea
 where $r_{-1}$ denotes the dimension of the vector space attached to the first vertex. 
The equivariance condition \eqref{eq:eq_discrete_gamma} then yields the 3-Kronecker quiver 
\bea
\label{fig:transl4_fund}
    \begin{tikzpicture}[->,scale=1.4]
    \node (a) at (0,0) {{\small$(-1,-1,1)$}};
    \node (b) at (2.5,0) {{\small$(3,0,0)$}};
    \node (labela) at (1.35,.4) {{$\scriptstyle\Phi_{\a}$}};

    \path[->](b) edge [bend left=12] node [above]{} (a);
    \path[->](b) edge node [above]{} (a);
    \path[->](b) edge [bend left=-12] node [above]{} (a);
    \end{tikzpicture}
\eea
and the Higgs fields are of the form
\bea
Y_{{\a}} \= \begin{pmatrix}
                  0 &\Phi_{\a}  \\
                 0& 0
                 \end{pmatrix}
\qquad \text{for} \quad \a\=1,2,3 \qquad \text{and} \qquad  Y_{{4}} \=
0 \ .
\eea
As expected, this is the higher-dimensional analogue of the quiver for the fundamental representation of $\mathrm{SU}(3)$ 
in \cite{Lechtenfeld:2015ona}. The holomorphicity condition $\com{Y_{\a}}{Y_{\b}}=0$ is trivially satisfied, while the stability condition
 \eqref{eq:inst_FI} yields
\bea
\nonumber \Phi_{1} \, \Phi_{1}^{\+}+\Phi_{2} \, \Phi_{2}^{\+}+\Phi_{3} \,
\Phi_{3}^{\+} &=& -\xi_{-1} \, \id_3 \otimes \id_{r_{-1}} \ ,\\[4pt]
\Phi_{1}^{\+} \, \Phi_{1}+\Phi_{2}^{\+} \,
\Phi_{2}+\Phi_{3}^{\+} \, \Phi_{3} &=& \xi_{3} \, 1 \otimes
\id_{r_{3}} \ ,
\eea
where $\xi_i$ are the components of the decomposition of $\Xi$ according to the action $\g(h_k)$.

\paragraph{Representation $\mbf{\underline{6}}$\,.} For the six-dimensional representation of $\mathrm{SU}(4)$ we 
consider the embedding of the generator of $\Gamma=\mathbb{Z}_{k}$
given by
\bea
\g (h_k)\= \begin{pmatrix}
                     \zeta_{k}^{-2}\, \id_3 \otimes \id_{r_{-2}} & 0\\
                     0          & \zeta_{k}^{-1}\, \id_3 \otimes \id_{r_{2}},
                    \end{pmatrix} \qquad \text{with} \quad
                    6r_{-2}+3r_{2} \equiv 0\ \ \mathrm{mod}\ k \ .
\eea
The equivariance condition \eqref{eq:eq_discrete_gamma} then again yields the 3-Kronecker quiver 
\bea
\label{fig:transl4_6}
    \begin{tikzpicture}[->,scale=1.4]
    \node (a) at (0,0) {{\small$(-2,-2,0)$}};
    \node (b) at (2.5,0) {{\small$(2,-1,1)$}};
    \node (labela) at (1.35,.4) {{$\scriptstyle\Phi_{\a}$}};

    \path[->](b) edge [bend left=12] node [above]{} (a);
    \path[->](b) edge node [above]{} (a);
    \path[->](b) edge [bend left=-12] node [above]{} (a);
    \end{tikzpicture}
\eea
with the Higgs fields
\bea
Y_{{\a}} \= \begin{pmatrix}
                  0 &\Phi_{\a}  \\
                 0& 0
                 \end{pmatrix}
\qquad \text{for} \quad \a\=1,2,3 \qquad \text{and} \qquad Y_{{4}} \=
0 \ .
\eea
The corresponding instanton equations are
\bea
\nonumber \Phi_{1} \, \Phi_{1}^{\+}+\Phi_{2} \, \Phi_{2}^{\+}+\Phi_{3} \, \Phi_{3}^{\+}
&=& -\xi_{-2} \, \id_3 \otimes \id_{r_{-2}} \ ,\\[4pt]
\Phi_{1}^{\+} \, \Phi_{1}+\Phi_{2}^{\+} \,
\Phi_{2}+\Phi_{3}^{\+} \, \Phi_{3} &=& \xi_{2} \, \id_3
\otimes \id_{r_2} \ ,
\eea
which are similar to those of the fundamental representation but, of
course, the multiplicities differ, as was already the case for the 
$\mathrm{SU}(4)$-equivariant connections constructed in Section \ref{section:quiver_diagrams}.
\paragraph{Representation $\mbf{\underline{10}}$\,.} For the
ten-dimensional representation we use the embedding 
\bea
\g(h_k) \= \begin{pmatrix}
                     \zeta_{k}^{-2} \, \id_6 \otimes \id_{r_{-2}} & 0 & 0\\
                     0          & \zeta_{k}^{-1} \, \id_3 \otimes \id_{r_{2}} & 0\\
                     0          & 0        & 1 \otimes \id_{r_{6}}
                    \end{pmatrix} \qquad \mbox{with}
\quad 12r_{-2} + 3r_2 \equiv 0\ \ \mathrm{mod} \ k 
\eea
to obtain the 3-Beilinson quiver~\cite{benson}
\bea
\label{fig:transl4_10}
    \begin{tikzpicture}[->,scale=1.4]
    \node (a) at (0,0) {{\small$(-2,-2,2)$}};
    \node (b) at (2.5,0) {{\small$(2,-1,1)$}};
    \node (c) at (5,0) {{\small$(6,0,0)$}};
    \node (labela) at (1.35,.4) {{$\scriptstyle\Phi_{\a}^1$}};
    \node (labela) at (3.8,.4) {{$\scriptstyle\Phi_{\a}^2$}};

    \path[->](b) edge [bend left=12] node [above]{} (a);
    \path[->](b) edge node [above]{} (a);
    \path[->](b) edge [bend left=-12] node [above]{} (a);

    \path[->](c) edge [bend left=12] node [above]{} (b);
    \path[->](c) edge node [above]{} (b);
    \path[->](c) edge [bend left=-12] node [above]{} (b);
    \end{tikzpicture}
\eea
and the Higgs fields
\bea
Y_{{\a}} \= \begin{pmatrix}
                  0 &\Phi_{\a}^1 &0\\
                 0& 0& \Phi_{\a}^2\\
                 0 & 0 & 0
                 \end{pmatrix}
\qquad \text{for} \quad \a\=1,2,3 \qquad \text{and} \qquad  Y_{{4}} \=
0 \ .
\eea
The Hermitian Yang-Mills equations require the holomorphicity conditions
\bea
\Phi_{\a}^1\, \Phi_{\b}^2 - \Phi_{\b}^1\, \Phi_{\a}^2 \= 0 
\eea
together with the stability conditions
\bea
\sum_{\a=1}^3\, \Phi_{\a}^1 \, \Phi_{\a}^{1\,\+} &=& -\xi_{-2} \, \id_6
 \otimes \id_{r_{-2}} \ , \nonumber \\[4pt] 
\sum_{\a=1}^3\, \big(\Phi_{\a}^2 \, \Phi_{\a}^{2\,\+} -\Phi_{\a}^{1\,\+}\,
\Phi_{\a}^1\big) &=& 
-\xi_{2} \, \id_3 \otimes \id_{r_{2}} \ , \nonumber\\[4pt]
    \sum_{\a=1}^3\, \Phi_{\a}^{2\,\+}\, \Phi_{\a}^{2} &=& \xi_{6}\, 1 \otimes
    \id_{r_6} \ .
\eea
Here the arrows related to $Y_{\a}$ for $\a=1,2,3$ occur whenever the 
difference in the powers of $\zeta_{k}$ in \eqref{eq:eq_discrete_gamma} is equal to one, so that the underlying structure of
the quivers shares the main features with those of \cite{Lechtenfeld:2015ona}. While the morphisms induced by 
$Y_4$ only appear for differences of $2$ in the case of the group
$\mathrm{SU}(3)$ studied in~\cite{Lechtenfeld:2015ona}, for the 
higher-dimensional version here it requires differences of $3$ in the powers of the root of unity, so that there is no arrow $Y_4$ for the 
ten-dimensional representation. 

\paragraph{Representation $\mbf{\underline{20}}$\,.} As our final example, consider the representation
\bea
\mbf{\underline{20}}\,\big|_{\mathrm{SU}(3)} \=
\mbf{\underline{(-3,-3,3)_{10}}} \ \oplus \
\mbf{\underline{(1,-2,2)_6}} \ \oplus \ 
                    \mbf{\underline{(5,-1,1)_3}} \ \oplus \ \mbf{\underline{(9,0,0)_1}}
\eea
which leads to the quiver 
\bea
\label{fig:transl4_20}
    \begin{tikzpicture}[->,scale=1.4]
    \node (a) at (0,0) {{\small$(-3,-3,3)$}};
    \node (b) at (2.5,0) {{\small$(1,-2,2)$}};
    \node (c) at (5,0) {{\small$(5,-1,1)$}};
    \node (d) at (7.5,0) {{\small$(9,0,0)$}}; 
    \node (labelab) at (1.35,.4) {{$\scriptstyle\Phi_{\a}^1$}};
    \node (labelbc) at (3.8,.4) {{$\scriptstyle\Phi_{\a}^2$}};
    \node (labelcd) at (6.4,.4) {{$\scriptstyle\Phi_{\a}^3$}};
     \node (labelda) at (3.8,-1.2) {{$\scriptstyle\Psi$}};
    
    \path[->](b) edge [bend left=12] node [above]{} (a);
    \path[->](b) edge node [above]{} (a);
    \path[->](b) edge [bend left=-12] node [above]{} (a);

    \path[->](c) edge [bend left=12] node [above]{} (b);
    \path[->](c) edge node [above]{} (b);
    \path[->](c) edge [bend left=-12] node [above]{} (b);

     \path[->](d) edge [bend left=12] node [above]{} (c);
    \path[->](d) edge node [above]{} (c);
    \path[->](d) edge [bend left=-12] node [above]{} (c);

     \path[<-](d) edge [bend left=25] node [above]{} (a);
    \end{tikzpicture}
\eea
The underlying structure of this quiver is again that of a chain connecting adjacent vertices with three Higgs fields, as its weight diagram 
is an extension of those for the representations $\mbf{\underline{4}}$
and $\mbf{\underline{10}}$\,. As the powers of $\zeta_{k}$ here are large enough, we 
encounter also the morphism $\Psi$ induced by $Y_4$.

Since the  representations of $\mathrm{SU}(4)$ we considered here are constructed via layers of representations of 
the subgroup $\mathrm{SU}(3)$, one may expect the general shape of the
quivers with this choice of 
group action $\gamma(h_k)$ to hold also for generic groups $\mathrm{SU}(n{+}1)$. Studying other choices for the action on the fibres
even for the concrete case $\C^4/\Gamma$ is beyond the scope of
this paper, as is the inclusion of other finite subgroups $\Gamma\subset\mathrm{SU}(4)$.


\subsection{Moduli spaces}
\label{sec:mod_trans_inv}
 
In Section~\ref{sec:transinvexamples} we considered some examples of translationally invariant connections on $\mathbb{R}^8/\mathbb{Z}_{k}$, where we specialised
to the instanton equations with respect to \emph{one} complex
structure, i.e. we considered a \emph{single} Hermitian Yang-Mills moduli space
over the flat Calabi-Yau orbifold $\mathbb{R}^8/\mathbb{Z}_{k}$. 
Analogously to the discussion of the moduli spaces of such instantons on $\mathbb{R}^6/\mathbb{Z}_{k}$ in \cite{Lechtenfeld:2015ona}
and the discussion in Section \ref{sect:cone_instantons}, the moduli space can be described as a K\"ahler quotient involving the pre-image
of the Fayet-Iliopoulos term $\Xi$ under a moment map on the space of holomorphic $\Gamma$-equivariant connections. Since we now consider translationally invariant
connections, there are only algebraic holomorphicity conditions, in contrast to the $\mathrm{SU}(4)$-instantons discussed in Section~\ref{sect:sasaki_quiver}. This moduli space then resembles a noncommutative crepant resolution, defined by the path algebra of the underlying quiver, of the nilpotent orbits that appeared in \eqref{eq:modorbit}, lending a gauge theory interpretation for some of the constructions of~\cite{Hara}.
 
However, it is also interesting to study $\mathrm{Sp}(2)$-instantons in this context. Then one can include a triplet of Fayet-Iliopoulos terms $\Xi^\a$, and the 
generalized instanton equation reads
\bea
\label{eq:inst_eq_hk_deformed}
\star\, \cf + \cf \wedge \star \, Q \= \Xi^{\a} \star \Omega_{\a} +
\Xi^{\a} \, \Omega_{\a} \wedge \star\, Q
\qquad \mbox{or} \qquad 
\star\, \tilde{\cf} + \tilde{\cf} \wedge \star \, Q \= 0
\eea
for a deformed curvature $\tilde{\cf} \coloneqq \cf - \Xi^{\a} \,
\Omega_{\a}$. This generalizes the four-dimensional anti-self-duality
equation with deformations induced by Fayet-Iliopoulos terms
$\Xi^{\a}$. In this case, for constant $\Xi^{\a}$ lying in the center of the Lie
algebra $\mathfrak{u}(r)$,
the generalized instanton equations with deformations still imply the
Yang-Mills equations, but otherwise they induce sources~\cite{Ivanova:2013mea}. 
It was shown in \cite{Kronheimer:1989zs} that the moduli space of the four-dimensional instanton equation can be expressed as a hyper-K\"ahler quotient 
$\mathcal{M}_\Xi =  \big(\m_{1}^{-1}(\Xi^{1}) \cap
\m_{2}^{-1}(\Xi^{2})  \cap \m_{3}^{-1}(\Xi^{3}) \big)\, /\!/\!/\, \mathcal{G}$, and that it is closely related to
the ALE spaces which appear as resolutions of the orbifold $\mathbb{C}^2/\Gamma$. 

For vanishing Fayet-Iliopoulos terms in our eight-dimensional setup,
the moduli space can again be described as the intersection of three
Hermitian Yang-Mills moduli spaces by the general 
arguments of Section \ref{sec:mod_space_HK}:
\bea
\mathcal{M}_0 \= \mathcal{M}_1 \cap \mathcal{M}_2 \cap \mathcal{M}_3 \
.  
\eea
Since non-vanishing $\Xi^{\a}$ describe deformations of the complex structures \cite{Ivanova:2013mea}, one cannot simply impose the holomorphicity conditions
on $\cf$. Instead, by rewriting the instanton equation \eqref{eq:inst_eq_hk_deformed} in terms of the deformed curvature $\tilde{\cf}$, 
one gets formally again the ``usual" equation whose moduli space is
given as the intersection of three Hermitian Yang-Mills moduli spaces,
\bea
\mathcal{M}_\Xi \= \tilde{\mathcal{M}}_1 \cap \tilde{\mathcal{M}}_2
\cap \tilde{\mathcal{M}}_3 \ , 
\eea
where each moduli space $\tilde{\mathcal{M}}_{\a}$ is determined by imposing the usual holomorphicity (and stability) 
condition with respect to the undeformed
complex structure $J_{\a}$ on the \emph{deformed} curvature
$\tilde{\cf} = \cf - \Xi^{\a} \, \Omega_{\a}$. 

\section{Summary and conclusions\label{sec:Summary}}

In this paper we have studied three classes of quiver gauge theories:
(1) Sasakian quiver gauge theories on the round seven-sphere $\mathrm{SU}(4)/\mathrm{SU}(3)$, 
(2) 3-Sasakian quiver gauge theories on the squashed seven-sphere $\mathrm{Sp}(2)/\mathrm{Sp}(1)$, and
(3) Translationally invariant instantons on the Calabi-Yau cone $\mathbb{R}^8/\Gamma$.
In all cases we discussed the equivariance conditions, giving explicit
examples of the resulting quivers for some low-dimensional
representations of $G$, and described the corresponding moduli spaces.

The Sasakian quiver gauge theory on $S^7$ yields a higher-dimensional
analogue of that on $S^5$~\cite{Lechtenfeld:2015ona}, 
but with one completely new class due to the exceptional representation $\mbf{\underline{\mathrm{6}}}$ of $\mathrm{SU}(4)$. 
Because of the systematic construction of all odd-dimensional round spheres $S^{2n+1} \cong \mathrm{SU}(n{+}1)/\mathrm{SU}(n)$, 
one can expect the regular quivers to be the same for all cases. This fits in the framework of generic expressions for 
quiver gauge theories \cite{Ivanova:2012vz} and the general
description of moduli spaces for Hermitian Yang-Mills instantons on
metric cones over Sasaki-Einstein manifolds \cite{Sperling:2015sra}; indeed, the moduli space of instantons on the Calabi-Yau cone 
we discussed here is contained in the general description of \cite{Sperling:2015sra}.

Making use of the 3-Sasakian structure on the coset space $\mathrm{Sp}(2)/\mathrm{Sp}(1)$, we constructed new quiver gauge theories based on
representations of $\mathrm{Sp}(2)$, again giving some explicit examples of quivers. We discussed the 
more complicated instanton equations on the metric cone and showed that the moduli space can be described as the intersection 
of three Hermitian Yang-Mills moduli spaces. In contrast to a single Hermitian Yang-Mills instanton moduli space over a Calabi-Yau cone, in the hyper-K\"ahler setup the holomorphicity
conditions automatically imply the stability conditions of the Hermitian Yang-Mills moduli spaces.

Finally, we discussed some examples of quivers for tranlationally invariant instantons on $\mathbb{R}^8/\Gamma$ with the finite group 
$\Gamma=\mathbb{Z}_{k}$ embedded into $\mathrm{SU}(4)$. While the
moduli space of Hermitian Yang-Mills instantons can be described  as a K\"ahler quotient for a possibly 
non-trivial Fayet-Iliopoulos term, as in the case of translationally invariant instantons on $\mathbb{R}^6/\mathbb{Z}_{k}$~\cite{Lechtenfeld:2015ona}, we attributed 
the moduli space of instantons with respect to the hyper-K\"ahler
structure again to the intersection of three Hermitian Yang-Mills moduli spaces. Hereby, one has 
to consider deformed curvatures if non-trivial Fayet-Iliopoulos terms are present.


\section*{Acknowledgements}

This paper is based upon work from COST Action MP1405 QSPACE, supported by COST (European Cooperation in Science and Technology). This work was supported by the Deutsche Forschungsgemeinschaft (DFG,
Germany) under the grant LE 838/13, by the Research Training Group RTG
1463 ``Analysis, Geometry and String Theory" (DFG), and by the
Consolidated Grant ST/L000334/1 from the UK Science and Technology
Facilities Council (STFC).

\newpage

\appendix

\section{Technical details for $\mbf{S^7 \cong \mathrm{SU}(4)/\mathrm{SU}(3)}$}
This appendix contains some technical details of calculations
involving the round seven-sphere, including the derivation of the
parameter dependences in the structure equations, the canonical 
connection and weight diagrams, and some
explicit realizations of $\mathrm{SU}(4)$-representations. 
\subsection{Structure equations and Sasaki-Einstein geometry}
By the choice of the 1-forms in (\ref{eq:can_flat_s7}), the generators of the Lie algebra of $\mathrm{SU}(4)$ in the fundamental representation
take the form
\small
\begin{align}
\nonumber  I_{1}^+ & \ \coloneqq \ \zeta_1 \, \begin{pmatrix}
                           0 & 1 & 0 &0\\
                           0 & 0 & 0 &0\\
                           0 & 0 & 0 &0\\
                           0 & 0 & 0 &0
                          \end{pmatrix} \ , \ 
&   I_{2}^+ & \ \coloneqq \ \zeta_2 \, \begin{pmatrix}
                           0 & 0 & 1 &0\\
                           0 & 0 & 0 &0\\
                           0 & 0 & 0 &0\\
                           0 & 0 & 0 &0
                          \end{pmatrix} \ , \ 
& I_{3}^+ & \ \coloneqq \ \zeta_3 \, \begin{pmatrix}
                           0 & 0 & 0 &1\\
                           0 & 0 & 0 &0\\
                           0 & 0 & 0 &0\\
                           0 & 0 & 0 &0
                          \end{pmatrix} \ ,\\[4pt]
I_{4}^+ & \ \coloneqq \ \l_4 \, \begin{pmatrix}
                           0 & 0 & 0 &0\\
                           0 & 0 & 1 &0\\
                           0 & 0 & 0 &0\\
                           0 & 0 & 0 &0
                          \end{pmatrix} \ , \ 
\nonumber & I_{5}^+ & \ \coloneqq  \ \l_5 \, \begin{pmatrix}
                           0 & 0 & 0 &0\\
                           0 & 0 & 0 &1\\
                           0 & 0 & 0 &0\\
                           0 & 0 & 0 &0
                          \end{pmatrix} \ , 
& I_{6}^+ & \ \coloneqq \ \l_6 \, \begin{pmatrix}
                           0 & 0 & 0 &0\\
                           0 & 0 & 0 &0\\
                           0 & 0 & 0 &1\\
                           0 & 0 & 0 &0
                          \end{pmatrix} \ , \\[4pt]
 I_{7} & \ \coloneqq \ \im \mu_7 \, \begin{pmatrix}
                           3 & 0 & 0 &0\\
                           0 & -1 & 0 &0\\
                           0 & 0 & -1 &0\\
                           0 & 0 & 0 &-1
                          \end{pmatrix} \ , \ 
& I_{8} & \ \coloneqq \ \im \mu_8 \, \begin{pmatrix}
                           0 & 0 & 0 &0\\
                           0 & 2 & 0 &0\\
                           0 & 0 & -1 &0\\
                           0 & 0 & 0 &-1
                          \end{pmatrix} \ , \ 
& I_{9} & \ \coloneqq \ \im \mu_9 \, \begin{pmatrix}
                           0 & 0 & 0 &0\\
                           0 & 0 & 0 &0\\
                           0 & 0 & -1 &0\\
                           0 & 0 & 0 &1
                          \end{pmatrix} \ ,
\label{eq:def_generators}
\end{align}
\normalsize
with real parameters $\zeta_{\a}$, $\l_{\b}$ and $\m_i$. The flatness of the connection (\ref{eq:can_flat_s7}) yields the structure equations 
\bea
\label{eq:str_s7_parameter}
\nonumber \diff \Theta^1 &=& 4\, \m_7 \, \Theta^1 \wedge \im e^7 - 2\,
\m_8 \, \Theta^1 \wedge \im e^8 + {\zeta_2 \, \l_4} \, {\zeta_1^{-1}}
\, \Theta^{2\bar{4}}
    + {\zeta_3\, \l_5}\, {\zeta_1^{-1}} \, \Theta^{3\bar{5}} \ ,\\[4pt]
\nonumber \diff \Theta^2 &=& 4\, \m_7 \, \Theta^2 \wedge \im e^7 +
\m_8 \, \Theta^2 \wedge \im e^8 + \m_9 \, \Theta^2 \wedge \im e^9 -
{\zeta_1\, \l_4}\, {\zeta_2^{-1}} \, \Theta^{14}
       + {\zeta_3\, \l_6}\, {\zeta_2^{-1}} \, \Theta^{3\bar{6}} \ ,\\[4pt]
\nonumber \diff \Theta^3 &=& 4 \, \m_7\, \Theta^3 \wedge \im e^7 +
\m_8 \, \Theta^3 \wedge \im e^8 - \m_9 \, \Theta^3 \wedge \im e^9
 - {\zeta_2 \, \l_6}\, {\zeta_3^{-1}}\, \Theta^{26} - {\zeta_1 \,
   \l_5}\, {\zeta_3^{-1}} \, \Theta^{15} \ ,\\[4pt]
\diff e^7 &=& -\tfrac{\im}{3 \m_7} \, \big(\zeta_1^2\, \Theta^{1
  \bar{1}}+ \zeta_2^2\, \Theta^{2 \bar{2}} +\zeta_3^2\, \Theta^{3
  \bar{3}} \big) \ ,
\eea
together with the equations for the complex conjugates $\bar\Theta^{\bar{\a}}$ for $\a=1,2,3$. The necessary dependences of the parameters in \eqref{eq:se_cond1}
follow from imposing closure of the fundamental form $\Omega^{1,1}$:
\bea
\nonumber 2 \im \diff \Omega^{1,1} &=& r^2 \,\left(\l_4 \, \Big(\,
  \frac{\zeta_2}{\zeta_1} - \frac{\zeta_1}{\zeta_2}\, \Big) \, \big( \Theta^{\bar{1}2\bar{4}} 
                    -\Theta^{1\bar{2}4}\big)
+ \l_5\, \Big(\, \frac{\zeta_3}{\zeta_1}- \frac{\zeta_1}{\zeta_3}\,
\Big) \, \big( \Theta^{\bar{1}3\bar{5}} - \Theta^{1\bar{3}5}\big) \right.\\
&&\qquad + \left. \l_6 \, \Big(\,
  \frac{\zeta_2}{\zeta_3}-\frac{\zeta_3}{\zeta_2} \, \Big) \, \big( \Theta^{2\bar{3}6} -\Theta^{\bar{2}3\bar{6}}\big)
     + \Big(\, 1-\frac{\zeta_1^2}{3\m_7}\, \Big) \,
     \Theta^{1\bar{1}}\wedge \big( \Theta^0+ \bar\Theta^{\bar{0}}\big) \right.\\
 \nonumber &&\qquad + \left. \Big(\, 1-\frac{\zeta_2^2}{3\m_7}\, \Big) \, \Theta^{2\bar{2}} \wedge \big( \Theta^0+ \bar\Theta^{\bar{0}}\big)
         + \Big(\, 1-\frac{\zeta_3^2}{3\m_7}\, \Big) \,
         \Theta^{3\bar{3}} \wedge \big( \Theta^{0}+
         \bar\Theta^{\bar{0}}\big) \right) \ .
\eea
The exterior derivative of the top-degree holomorphic form reads
\bea
\diff \Omega^{4,0} \= r^4 \, ( 2-6\m_7) \, \Theta^{1230\bar{0}} \ ,
\eea
leading to the condition \eqref{eq:se_cond2}. With these
parameter values one obtains the structure equations \eqref{eq:structure_s7_1}
together with
\begin{align}
\label{eq:structure_s7}
\nonumber \diff \Theta^4 & \= - \tfrac{1}{2} \im e^8 \wedge \Theta^4 -
            \tfrac{1}{2} \im e^9 \wedge \Theta^4 + \Theta^{\bar{1}2}+
            \Theta^{5\bar{6}} \ , \\[4pt]
\nonumber \diff \Theta^5 & \= -\tfrac{1}{2} \im e^8 \wedge \Theta^5 +
     \tfrac{1}{2} \im e^9 \wedge \Theta^5 + \Theta^{\bar{1}3}-
     \Theta^{46} \ ,\\[4pt]
\nonumber \diff \Theta^6 & \=  \im e^9 \wedge \Theta^6 +
            \Theta^{\bar{2}3} + \Theta^{\bar{4}5} \ , \\[4pt]
\nonumber \diff e^8 & \= 2 \im \Theta^{1\bar{1}} - \im \Theta^{2\bar{2}} -
      \im \Theta^{3\bar{3}} - 3 \im \Theta^{4\bar{4}} - 3 \im
      \Theta^{5\bar{5}} \ ,\\[4pt]
\diff e^9 & \= -\im \Theta^{2\bar{2}} + \im \Theta^{3\bar{3}} - \im
   \Theta^{4\bar{4}} + \im \Theta^{5\bar{5}} + 2 \im \Theta^{6\bar{6}}
   \ . 
\end{align}
With respect to the rescaled Cartan generators $\hat{I}_j \coloneqq - \im \m_j^{-1}\, I_j$, i.e.
\bea
\hat{I}_7 \= \mathrm{diag}\lb 3,-1,-1,-1\rb \ , \quad \hat{I}_8 \= \mathrm{diag}\lb 0,2,-1,-1\rb \quad \text{and} \quad
\hat{I}_9 \= \mathrm{diag}\lb 0,0,-1,1\rb \ ,
\eea
the non-vanishing structure constants read
\small
\begin{align}
\label{eq:structure_const_s7}
 \nonumber & C_{84}^4\= C_{85}^5\=3 \ , \quad C_{94}^4\=1 \ , \quad
             C_{95}^5\=-1 \ , \quad C_{96}^6\=-2 \ , \quad
             C_{4\bar{4}}^8\=C_{5\bar{5}}^8\=-\tfrac{1}{2} \ , \quad
 C_{4\bar{4}}^9\= -\tfrac{1}{2} \ , \\[4pt] \nonumber & C_{5\bar{5}}^9\=\tfrac{1}{2}
             \ ,\quad
   C_{6\bar{6}}^9\=1 \ , \quad C_{5\bar{6}}^4\=-1 \ ,
               \quad C_{46}^5\=1 \ , \quad C_{\bar{4}5}^6\=-1 \ ,\quad
               C_{71}^1 \= C_{72}^2 \=C_{73}^3\=4 \ ,\\[4pt]
\nonumber &  C_{81}^1\=-2 \ , \quad C_{82}^2\=C_{83}^3\=1 \ , \quad
            C_{92}^2\=1 \ , \quad C_{93}^3 \=-1 \ , \quad
C_{1\bar{1}}^7\=C_{2\bar{2}}^7 \= C_{3\bar{3}}^7\=-\tfrac{1}{3} \ ,
            \\[4pt] \nonumber &  C_{1\bar{1}}^8\= \tfrac{1}{3} \ , \quad C_{2\bar{2}}^8\= -\tfrac{1}{6} \ , \quad
            C_{3\bar{3}}^8\=-\tfrac{1}{6} \ , \quad C_{2\bar{2}}^9 \=
            -\tfrac{1}{2} \ , \quad C_{3\bar{3}}^9\=\tfrac{1}{2} \ , \quad
C_{2\bar{4}}^1 \=-1 \ , \quad C_{3\bar{5}}^1\=-1 \ , \\[4pt] & C_{14}^2\=1
            \ ,  \quad C_{3\bar{6}}^2\=-1 \ ,\quad C_{15}^3\=1 \ , \quad C_{26}^3\=1 \ ,\quad C_{\bar{1}2}^4\=-1 \ ,
   \quad C_{\bar{1}3}^5 \= -1 \ , \quad C_{\bar{2}3}^6\=-1
\end{align}
\normalsize
plus the conjugated ones. 

Explicit evaluation of the torsion of the canonical connection yields
the components
\bea
\nonumber && T^1 \= \tfrac{4}{3}\, e^{27} \ , \qquad T^2 \=
-\tfrac{4}{3}\, e^{17} \ , \qquad T^3 \= \tfrac{4}{3}\, e^{47} \ , \\[4pt]
\nonumber && T^4 \= -\tfrac{4}{3}\, e^{37} \ , \qquad
T^5 \= \tfrac{4}{3}\, e^{67} \ ,\qquad T^6 \= -\tfrac{4}{3}\, e^{57} \ , \\[4pt]
&& T^7 \= 2 \lb e^{12}+e^{34}+e^{56}\rb \ ,
\eea
which leads to the results in the main text.
\subsection{Representations of $\mathrm{SU}(4)$}

We shall now provide the weight diagrams and some of the generators which are used for the explicit examples of quivers.
To this end, recall that the ladder operators act, according to the structure constants \eqref{eq:structure_const_s7}, 
on the quantum numbers as
\bea
\nonumber && I_{\bar{1}}^-: \lb \n_7,\n_8,\n_9\rb \ \longmapsto \ \lb
\n_7 -4,\n_8+2, \n_9\rb \ , \qquad\qquad\!
I_{\bar{4}}^-: \lb \n_7,\n_8,\n_9\rb \ \longmapsto \ \lb \n_7,\n_8-3,
\n_9-1\rb \ ,\\[4pt]
\nonumber && I_{\bar{2}}^-: \lb \n_7,\n_8,\n_9\rb \ \longmapsto \ \lb
\n_7 -4,\n_8-1, \n_9-1\rb \ , \qquad
I_{\bar{5}}^-: \lb \n_7,\n_8,\n_9\rb \ \longmapsto \ \lb \n_7,\n_8-3,
\n_9+1\rb \ ,\\[4pt]
&& I_{\bar{3}}^-: \lb \n_7,\n_8,\n_9\rb \ \longmapsto \ \lb \n_7
-4,\n_8-1, \n_9+1\rb \ , \qquad
I_{\bar{6}}^-: \lb \n_7,\n_8,\n_9\rb \ \longmapsto \ \lb \n_7 ,\n_8,
\n_9+2\rb \ ,
\eea
which yields the root system\footnote{For better readability, we restrict to the generators $I_{\bar\a}^{-}$ and do not depict the adjoint generators.}
\bea
\footnotesize
    \begin{tikzpicture}[->,scale=1.2]
    \node (a) at (0,0) {};
    \node (b) at (-1,1) {${(-4,-1,1)}$};
    \node (c) at (-1,-1) {${(-4,-1,-1)}$};
    \node (d) at (2,0) {${(-4,2,0)}$};
    \node (e) at (0,2) {${(0,0,2)}$};
    \node (f) at (-3,-1) {${(0,-3,-1)}$};
    \node (g) at (-3,1) {${(0,-3,1)}$};
   
     \path[->](a) edge node [above]{$I_{\bar{1}}^-$} (d);
     \path[->](a) edge node [above]{$I_{\bar{3}}^-$} (b);
     \path[->](a) edge node [below right]{$I_{\bar{2}}^-$} (c);
     \path[blue, ->](a) edge node [right]{$I_{\bar{6}}^-$} (e);
     \path[blue, ->](a) edge node [below]{$I_{\bar{4}}^-$} (f);
     \path[blue, ->](a) edge node [above]{$I_{\bar{5}}^-$} (g);
\end{tikzpicture}
\eea
\normalsize
Here we depict the ladder operators of the subalgebra $\mathfrak{h}=\mathfrak{su}(3)$, along which one has to collapse the weight diagram, by blue arrows.
By using this root system, one can easily construct the representations we list in the following. For details on the representation theory of
$\mathrm{SU}(4)$ (or $\mathrm{SL}(4,\mathbb{C})$), see \cite{fulton2013representation}.
\paragraph{Fundamental representation $\mbf{\underline{4}}$\,.}
The generators are those of the chosen defining representation (\ref{eq:def_generators}), and the weight diagram is the tetrahedron
\bea \label{fig:weight_4}
\footnotesize
\begin{tikzpicture}[->,scale=1.1]
    \node (a) at (2,0) {${(-1,2,0)}$};
    \node (b) at (-1,1) {${(-1,-1,1)}$};
    \node (c) at (-1,-1) {${(-1,-1,-1)}$};
    \node (d) at (0,0) {${(3,0,0)}$};
   
     \path[->](d) edge node [above]{} (a);
     \path[->](d) edge node [below right]{} (c);
     \path[->](d) edge node [above right]{} (b);
     \path[blue, ->](a) edge node {} (b);
     \path[blue, ->](a) edge node {} (c);
     \path[blue, ->](c) edge node {} (b);
        
\end{tikzpicture}
\eea
\normalsize
\paragraph{Representation $\mbf{\underline{6}}$\,.}
The generators of the six-dimensional irreducible representation can
be chosen as
\bea
I_{\bar{\a}}^- \= \begin{pmatrix}
                             \mbf{0}_3 & \mbf{0}_3\\
                             \tilde{I}_{\bar{\a}} & \mbf{0}_3
                             \end{pmatrix} \= -\big( I_{{\a}}^+
                                                    \big)^{\+}  
\qquad\textrm{for}\quad \alpha=1,2,3
\eea
with
\bea 
\tilde{I}_{\bar{1}} \= \begin{pmatrix}
                0 & 0 & 0\\
                0 & 1 & 0\\
                0 & 0 & -1
               \end{pmatrix} \ , \qquad
\tilde{I}_{\bar{2}} \= \begin{pmatrix}
                0 & 1 & 0\\
                0 & 0 & 0\\
                1 & 0 & 0
               \end{pmatrix} \ , \qquad 
\tilde{I}_{\bar{3}} \= \begin{pmatrix}
                0 & 0 & -1\\
                -1 & 0 & 0\\
                0 & 0 & 0
               \end{pmatrix} \ ,
\eea
and
\bea
I_{\bar{\b}}^- \= \begin{pmatrix}
                    \tilde{I}_{\bar{\b}}^1 & \mbf{0}_3\\
                   \mbf{0}_3  &  \tilde{I}_{\bar{\b}}^2
                   \end{pmatrix} \=-\big( I_{{\b}}^+ \big)^{\+}  
\qquad\textrm{for}\quad \beta=4,5,6
\eea
with
\begin{subequations}
\bea
\tilde{I}_{\bar{4}}^1 \= \begin{pmatrix}
                         0 & 0 &0\\
                         0 & 0 &0\\
                        -1 & 0 &0      
                       \end{pmatrix} \ , \qquad
\tilde{I}_{\bar{4}}^2 \= \begin{pmatrix}
                         0 & -1 &0\\
                         0 & 0 &0\\
                         0 & 0 &0      
                       \end{pmatrix} \ , \qquad
\tilde{I}_{\bar{5}}^1 \= \begin{pmatrix}
                         0 & 0 &0\\
                         -1 & 0 &0\\
                         0 & 0 &0      
                       \end{pmatrix} \ ,
\eea
\bea
\tilde{I}_{\bar{5}}^2 \= \begin{pmatrix}
                         0 & 0 &-1\\
                         0 & 0 &0\\
                         0 & 0 &0      
                       \end{pmatrix} \ , \qquad
\tilde{I}_{\bar{6}}^1 \= \begin{pmatrix}
                         0 & 0 &0\\
                         0 & 0 &-1\\
                         0 & 0 &0      
                       \end{pmatrix} \ , \qquad
\tilde{I}_{\bar{6}}^2 \= \begin{pmatrix}
                         0 & 0 &0\\
                         0 & 0 &1\\
                         0 & 0 &0      
                       \end{pmatrix} \ ,
\eea
\bea
&& \hat{I_7} \= \mathrm{diag}\lb 2 \id_3, -2 \id_3 \rb \ , \quad
\hat{I}_8 \= \mathrm{diag} \lb 2,-1,-1,-2,1,1\rb \ , \quad
\hat{I}_9\ = \mathrm{diag}\lb 0,1,-1,0,1,-1\rb \ . \nonumber\\ &&
\eea
\end{subequations}
The weight diagram is the octahedron
\bea
\footnotesize
\label{fig:weight_6}
    \begin{tikzpicture}[->,scale=1.3]
    \node (a) at (0,0) {{${(2,2,0)}$}};
    \node (b) at (-1,1) {${(-2,1,1)}$};
    \node (c) at (-1,-1) {${(-2,1,-1)}$};
    \node (d) at (-3,-1) {${(2,-1,-1)}$};
    \node (e) at (-3,1) {${(2,-1,1)}$};
    \node (f) at (-4,0) {${(-2,-2,0)}$};

     \path[->](e) edge node [above]{} (b);
     \path[->](d) edge node [below]{} (c);
     \path[->](a) edge node [below right] {} (c);
     \path[->](e) edge node [above left] {} (f);
     \path[->](a) edge node [above right] {} (b);
     \path[->](d) edge node [below left] {} (f);

     \path[blue, ->](a) edge node [above]{} (d);
     \path[blue, ->](a) edge node [below]{} (e);
     \path[blue, ->](d) edge node [left]{} (e);

     \path[blue, ->](b) edge node [below]{} (f);
     \path[blue, ->](c) edge node [above]{} (f);
     \path[blue, ->](c) edge node [right]{} (b);
\end{tikzpicture}
\eea
\normalsize
Collapsing this diagram along the action of the ladder operators of $\mathrm{SU}(3)$ (blue arrows) then yields the quiver (\ref{fig:quiver_6}).      
\paragraph{Representation $\mbf{\underline{10}}$\,.}
The weight diagram of the ten-dimensional representation reads
\bea
\footnotesize
\begin{tikzpicture}[->,scale=1.1]
    \node (a) at (4,0) {${(-2,4,0)}$};
    \node (b) at (2,0) {${(2,2,0)}$};
    \node (c) at (1,1) {${(-2,1,1)}$};
    \node (d) at (1,-1) {${(-2,1,-1)}$};
    \node (e) at (0,0) {${(6,0,0)}$};
    \node (f) at (-1,1) {${(2,-1,1)}$};
    \node (g) at (-1,-1) {${(2,-1,-1)}$};
    \node (h) at (-2,2) {${(-2,-2,2)}$};
    \node (i) at (-2,0) {${(-2,-2,0)}$};
    \node (j) at (-2,-2) {${(-2,-2,-2)}$};
   
     \path[blue, ->](a) edge node {} (c);
     \path[blue, ->](c) edge node {} (h);
     \path[blue, ->](b) edge node {} (f);
     \path[blue, ->](d) edge node {} (i);

     \path[blue, ->](d) edge node {} (c);
     \path[blue, ->](g) edge node {} (f);
     \path[blue, ->](j) edge node {} (i);
     \path[blue, ->](i) edge node {} (h);

     
     \path[blue, ->](a) edge node {} (d);
     \path[blue, ->](d) edge node {} (j);
     \path[blue, ->](b) edge node {} (g);
     \path[blue, ->](c) edge node {} (i);

     \path[->](b) edge node [above]{} (a);
     \path[->](b) edge node [below right]{} (c);
     \path[->](b) edge node [above right]{} (d);

    \path[->](e) edge node [above]{} (b);
     \path[->](e) edge node [below right]{} (f);
     \path[->](e) edge node [above right]{} (g);

     \path[->](f) edge node [above]{} (c);
     \path[->](f) edge node [below right]{} (h);
     \path[->](f) edge node [above right]{} (i);

     \path[->](g) edge node [above]{} (j);
     \path[->](g) edge node [below right]{} (d);
     \path[->](g) edge node [above right]{} (i);
        
\end{tikzpicture}
\eea
\normalsize
It can be obtained from the representation $\mbf{\underline{6}}$ by adding four \emph{fundamental tetrahedra} to the octahedron of 
$\mbf{\underline{6}}$\,. Each layer with a fixed quantum number $\n_7 \in \{6,2,-2\}$ corresponds to one $\mathrm{SU}(3)$-representation, so that collapsing along 
the $\mathrm{SU}(3)$ generators leads to a quiver with 
three vertices. The generators in this representation can be chosen so that their only non-vanishing components are
\begin{align}
\label{eq:rep_10}
\nonumber &\lb I_{\bar{1}}^- \rb_{21}\=\sqrt{2} \ ,&  &\lb
                                                       I_{\bar{1}}^-
                                                       \rb_{52}\=-\sqrt{2}
                                                       \ , 
	    &   &\lb I_{\bar{1}}^- \rb_{63}\=-1,& &\lb I_{\bar{1}}^-
                                                   \rb_{74}\=-1 \ ,\\[4pt]
\nonumber &\lb I_{\bar{2}}^- \rb_{31}\=\sqrt{2} \ ,&  &\lb
                                                       I_{\bar{2}}^-
                                                       \rb_{62}\=-1 \
                                                       ,& &\lb
                                                            I_{\bar{2}}^-
                                                            \rb_{93}\=-\sqrt{2}
                                                            \ ,& 
             &\lb I_{\bar{2}}^- \rb_{84}\=-1 \ ,\\[4pt]
\nonumber &\lb I_{\bar{3}}^- \rb_{41}\=\sqrt{2} \ ,& &\lb
                                                       I_{\bar{3}}^-
                                                       \rb_{72}\=-1 \
                                                       ,& &\lb
                                                            I_{\bar{3}}^-
                                                            \rb_{83}\=-1
                                                            \ ,&
             &\lb I_{\bar{3}}^- \rb_{10\,4}\=-\sqrt{2} \ ,\\[4pt]
 &\lb I_{\bar{4}}^- \rb_{32}\=-1 \ ,& &\lb I_{\bar{4}}^- \rb_{87}\=-1
                                        \ ,& &\lb I_{\bar{4}}^-
                                               \rb_{65}\=-\sqrt{2} \ ,&
             &\lb I_{\bar{4}}^- \rb_{96}\=-\sqrt2 \ ,\\[4pt]
\nonumber &\lb I_{\bar{5}}^- \rb_{42}\=-1 \ ,& &\lb I_{\bar{5}}^-
                                                 \rb_{75}\=-\sqrt{2} \
                                                 ,& &\lb I_{\bar{5}}^-
                                                      \rb_{86}\=-1 \ ,& 
             &\lb I_{\bar{5}}^- \rb_{10\,7}\=-\sqrt{2} \ ,\\[4pt]
\nonumber &\lb I_{\bar{6}}^- \rb_{43}\=-1 \ ,& &\lb I_{\bar{6}}^-
                                                 \rb_{76}\=-1 \ ,&
                &\lb I_{\bar{6}}^- \rb_{10\,8}\=-\sqrt{2} \ ,&
             &\lb I_{\bar{6}}^- \rb_{89}\=-\sqrt{2}
\end{align}
and
\bea
\nonumber I_7 &=& \mathrm{diag} \lb 6,2,2,2,-2,-2,-2,-2,-2,-2\rb \ , \qquad
I_8 \= \mathrm{diag} \lb 0,2,-1,-1,4,1,1,-2,-2,-2\rb \ ,\\[4pt]
I_9 &=& \mathrm{diag} \lb 0,0,-1,1,0,-1,1,0,-2,2\rb \ .
\eea
\paragraph{Adjoint representation $\mbf{\underline{15}}$\,.}
The adjoint representation is described by the weight diagram
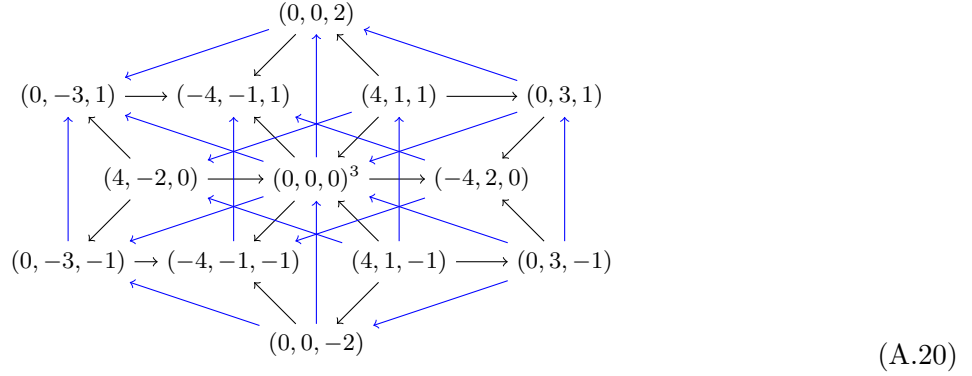
\bea
\footnotesize
\label{fig:weight_adj}
    \begin{tikzpicture}[->,scale=1.1]
    \node (a) at (0,2) {${(0,0,2)}$};
    \node (b) at (-3,1) {${(0,-3,1)}$};
    \node (c) at (-1,1) {${(-4,-1,1)}$};
    \node (d) at (1,1) {${(4,1,1)}$};
     \node (e) at (3,1) {${(0,3,1)}$};
    \node (f) at (-2,0) {${(4,-2,0)}$};
    \node (g) at (0,0) {${(0,0,0)^3}$};
    \node (h) at (2,0) {${(-4,2,0)}$};
     \node (i) at (-3,-1) {${(0,-3,-1)}$};
    \node (j) at (-1,-1) {${(-4,-1,-1)}$};
    \node (k) at (1,-1) {${(4,1,-1)}$};
    \node (l) at (3,-1) {${(0,3,-1)}$};
    \node (m) at (0,-2) {${(0,0,-2)}$};

\path[->,black](b) edge node {} (c);
\path[->,black](d) edge node {} (e);
\path[->,black](f) edge node {} (g);
\path[->,black](g) edge node {} (h);
\path[->,black](i) edge node {} (j);
\path[->,black](k) edge node {} (l);
     
\path[->](a) edge node {} (c);
\path[->](f) edge node {} (i);
\path[->](d) edge node {} (g);
\path[->](g) edge node {} (j);
\path[->](e) edge node {} (h);
\path[->](k) edge node {} (m);
 
\path[->](m) edge node {} (j);
\path[->](f) edge node {} (b);
\path[->](k) edge node {} (g);
\path[->](g) edge node {} (c);
\path[->](l) edge node {} (h);
\path[->](d) edge node {} (a);

\path[blue, ->](e) edge node {} (a);
\path[blue, ->](a) edge node {} (b);
\path[blue, ->](l) edge node {} (e);
\path[blue, ->](l) edge node {} (g);
\path[blue, ->](l) edge node {} (m);
\path[blue, ->](g) edge node {} (a);
\path[blue, ->](m) edge node {} (g);
\path[blue, ->](g) edge node {} (b);
\path[blue, ->](g) edge node {} (i);
\path[blue, ->](m) edge node {} (i);
\path[blue, ->](i) edge node {} (b);
\path[blue, ->](e) edge node {} (g);

\path[blue, ->](h) edge node {} (c);
\path[blue, ->](h) edge node {} (j);
\path[blue, ->](j) edge node {} (c);

\path[blue, ->](d) edge node {} (f);
\path[blue, ->](k) edge node {} (f);
\path[blue, ->](k) edge node {} (d);

\end{tikzpicture}
\eea
\normalsize
where the center node at $(0,0,0)$ has multiplicity $3$. Collapsing this diagram yields four vertices representing one trivial, one fundamental, one anti-fundamental, and one adjoint 
representation of $\mathrm{SU}(3)$. The generators are given by the structure constants (\ref{eq:structure_const_s7}) as
$\big( I_{\hat{\l}}\big)_{\hat{\mu}}^{\hat{\nu}}  = 
C_{\hat{\l} \hat{\m}}^{\hat{\nu}}$.



\section{Technical details for $\mbf{S^7 \cong \mathrm{Sp}(2)/\mathrm{Sp}(1)}$} 

We shall now provide some details on the calculations needed for the squashed seven-sphere $\mathrm{Sp}(2)/\mathrm{Sp}(1)$, in particular 
the defining properties of a 3-Sasakian manifold, and also representations of $\mathrm{Sp}(2)$ where we again refer to~\cite{fulton2013representation}.
\subsection{Structure equations and 3-Sasakian geometry}

\label{sec:3sasaki_app}
The choice of 1-forms in \eqref{eq:can_flat_conn_sp2} yields the fundamental representation of the generators as
\small
\bea
\label{eq:def_generators_sp2}
I_{\bar{1}}^- &=& \begin{pmatrix}
             0	&0 & 0 & 0\\
             0	&0 & 1 & 0\\
             0	&0 & 0 & 0\\
             -1	&0 & 0 & 0\\
            \end{pmatrix} \ , \qquad
I_{\bar{2}}^- \= \begin{pmatrix}
             0	&0 & 0 & 0\\
             0	&0 & 0 & -1\\
             -1	&0 & 0 & 0\\
             0	&0 & 0 & 0\\
            \end{pmatrix}, \qquad
I_{\bar{3}}^- \= \begin{pmatrix}
             0	&0 & 0 & 0\\
             -1	&0 & 0 & 0\\
             0	&0 & 0 & 0\\
             0	&0 & 0 & 0\\
            \end{pmatrix} \ , \\[4pt]
\nonumber 
I_{\bar{4}}^- &=& \begin{pmatrix}
            0	&  0  & 0 & 0\\ 
            0   &  0  & 0 & 0\\
            0	&  0  & 0 & -1\\
            0	&  0  & 0 & 0\\
            \end{pmatrix} \ , \quad
\hat{I}_7 \ \coloneqq \ - \im I_7 \= \mathrm{diag}\lb 1,-1,0,0\rb \ , \quad
\hat{I}_8 \ \coloneqq \ - \im I_8 \= \mathrm{diag} \lb 0,0,-1,1\rb \ ,
\eea
\normalsize
and the (complex) structure equations read
\begin{align}
\label{eq:struct_eq_sp2_compl}
\nonumber  &\diff \Theta^1 \= - \im e^7 \wedge \Theta^1 + \im e^8
             \wedge \Theta^1 - \Theta^{\bar{2}3} + \Theta^{2\bar{4}} \
             ,
 & &\diff \Theta^2 \= -  \im e^7 \wedge \Theta^2 -\im e^8 \wedge
     \Theta^2 + \Theta^{\bar{1}3}-\Theta^{14} \ ,\\[4pt]
          & \diff \Theta^3 \= -2\im e^7 \wedge \Theta^3- 2 \,
            \Theta^{12} \ ,
& &\diff \Theta^4 \= -2 \im e^8 \wedge \Theta^4 + 2 \,
    \Theta^{\bar{1}2} \ ,\\[4pt]
\nonumber  &\diff e^7 \= - \im
             \big(\Theta^{1\bar{1}}+\Theta^{2\bar{2}}+\Theta^{3\bar{3}}
             \big) \
             ,
&  &\diff e^8 \= \im
     \big(\Theta^{1\bar{1}}-\Theta^{2\bar{2}}-\Theta^{4\bar{4}} \big)
     \ .
\end{align}
This yields the non-vanishing structure constants
\begin{align}
\label{eq:struc_const_sp2}
\nonumber &f_{71}^1\= - f_{7\bar{1}}^{\bar{1}}\=1 \ , & &f_{72}^2\= -
                                                          f_{7\bar{2}}^{\bar{2}}\=1
                                                          \ , 
           &  &f_{73}^3\= - f_{7\bar{3}}^{\bar{3}}\=2 \ , &
  &f_{81}^1\= - f_{8\bar{1}}^{\bar{1}}\=-1 \ ,\\[4pt]
\nonumber &f_{82}^2 \= - f_{8\bar{2}}^{\bar{2}}\=1 \ , & &f_{84}^4 \=
                                                           -
                                                           f_{8\bar{4}}^{\bar{4}}\=2
                                                           \ ,&
              &f_{2\bar{3}}^{\bar{1}}\= f_{\bar{2}3}^{{1}}\=1 \ ,
           & &f_{2\bar{4}}^{1}\=f_{\bar{2}{4}}^{\bar{1}}\=-1 \ , \\[4pt]
\nonumber &f_{12}^3\=f_{\bar{1}\bar{2}}^{\bar{3}}\=2 \ ,&
                                                        &f_{14}^2\=f_{\bar{1}\bar{4}}^{\bar{2}}\=1
                                                          \ ,&
              &f_{\bar{1}3}^2\=f_{1\bar{3}}^{\bar{2}}\=-1 \ , 
          &   &f_{\bar{1}2}^4\=f_{1\bar{2}}^{\bar{4}}\=-2 \ ,\\[4pt]
          &f_{1\bar{1}}^7\=f_{2\bar{2}}^7\=f_{3\bar{3}}^7\=-1 \ ,&
                                                        &f_{1\bar{1}}^8\=1
                                                          \ ,&
              &f_{2\bar{2}}^8\=-1 \ , & &f_{4\bar{4}}^8\=-1 \ . 
\end{align}
The generators $\big\{I_4^+, I_{\bar{4}}^-, \hat{I}_8 \big\}$ span the subalgebra $\mathfrak{sp}(1) \cong \mathfrak{su}(2)$ which is factored when
forming the homogeneous space.

The fact that the orthonormal basis $\{e^1, \ldots, e^7\}$ describes a 3-Sasakian structure can also be shown by considering the metric cone. By definition,
a manifold $M$ is 3-Sasakian if its metric cone $C(M)$ is hyper-K\"ahler. For this, one can again introduce a fourth holomorphic 1-form,
\bea
\Theta^0 \ \coloneqq \ \frac{\diff r}{r} - \im e^7 \ ,
\eea 
as in Section \ref{sect:geometry_s7}, and establish the
Sasaki-Einstein property because the forms
\bea
\Omega^{1,1} \ \coloneqq \ -\tfrac{\im}{2} \, r^2\, \big( \Theta^{1\bar{1}}+\Theta^{2\bar{2}}+\Theta^{3\bar{3}}+\Theta^{0\bar{0}}\big)
\qquad \text{and} \qquad \Omega^{4,0} \ \coloneqq \ r^4 \, \Theta^{1230}
\eea
are closed. For the holonomy to be further reduced from
$\mathrm{SU}(4)$ to $\mathrm{Sp}(2)$, one additonally requires closure of the form
\cite{Haupt:2011mg}
\bea
\Omega^{2,0} \ \coloneqq \ r^2 \, \big( \Theta^{12}+\Theta^{30}\big) \
,
\eea
which follows from the structure equations \eqref{eq:struct_eq_sp2_compl}.
\subsection{Instanton equations on hyper-K\"ahler and Calabi-Yau cones}
\label{sect:3Sasaki_vs_SE}
We will now provide some technical details of the relation between instanton equations on the metric cones over 3-Sasakian manifolds and 
those over Sasaki-Einstein manifolds. 
In components the quaternion relations \eqref{eq:quaternionic_relations_J} read 
\bea
\label{eq:eq_quater_components}
J_{\a\m}^{\s}\,  J_{\b\s}^{\n} \= - \delta_{\a\b} \, \delta_{\m}^{\n}
+ \epsilon_{\a\b}{}^{\g} \, J_{\g\m}^{\n} \ ,
\eea
and the associated K\"ahler forms
\bea
\label{eq:eq_kaehler_components1}
\Omega_{\a\m\n} \= g_{\m\s}\,  J_{\a\n}^{\s}\=-g_{\n\s}\,
J_{\a\m}^{\s} \= - g_{\s\n}\,  J_{\a\m}^{\s} \ , \qquad
J_{\a\m}^{\n} \= g^{\n\s} \, \Omega_{\a\s\m}
\eea
satisfy
\bea\nonumber
J_{\a\m}^{\s}\, \Omega_{\b\s\n}&=& \delta_{\a\b}\, g_{\m\n} +
\epsilon_{\a\b}{}^{\g} \, \Omega_{\g\m\n} \ ,\\[4pt]
\label{eq:eq_kaehler_components2}
J_{\a\s}^{\m} \, \Omega_{\b}^{\s\n} &=& \delta_{\a\b}\, g^{\m\n} -
\epsilon_{\a\b}{}^{\g} \, \Omega_{\g}^{\m\n} \qquad \text{with} \quad \Omega_{\a}^{\m\n} 
\ \coloneqq \ g^{\m\rho} \, J_{\a\rho}^{\n} \ . 
\eea
Using \eqref{eq:eq_quater_components}, one obtains from the first
equation of \eqref{eq:holom_in_components} the relation
\bea
J_{\a\m}^{\l}\, J_{\b\n}^{\s}\, \cf_{\l\s} \= \delta_{\a\b}\,
\cf_{\m\n} + \epsilon_{\a\b}{}^{\g}\,  J_{\g\m}^{\s} \, \cf_{\n\s} \ , 
\eea
which includes the second equation of \eqref{eq:holom_in_components}
upon setting $\a=\b$. To show that the stability condition of
Hermitian Yang-Mills instantons automatically follows from 
the holomorphicity conditions, one contracts
\eqref{eq:holom_in_components}  with $\Omega^{\m\n}_{\b}$ for $\b \neq
\a$ to get
\bea
J_{\a\m}^{\l}\,  J_{\a\n}^{\s} \, \Omega^{\m\n}_{\b} \, \cf_{\lambda\sigma} \=
\Omega^{\m\n}_{\b} \, \cf_{\m\n} \qquad \text{(no sum on
  $\a$)} \ .
\eea
With the help of the properties \eqref{eq:eq_kaehler_components1} and \eqref{eq:eq_kaehler_components2}, one can show that the left-hand side gives the 
negative of the right-hand side, so that the stability condition
\bea
0 \= \Omega^{\m\n}_{\b} \, \cf_{\m\n}
\eea
follows.\footnote{As an example, consider the Hermitian Yang-Mills equations associated to $\Omega_7$ where the holomorphicity conditions yield the relations
\bea
\nonumber && \cf_{13} = \cf_{24} \ , \quad \cf_{15} = \cf_{26} \ ,
\quad \cf_{35} = \cf_{46} \ , \quad \cf_{23} = - \cf_{14} \ ,
\quad \cf_{25} = - \cf_{16} \ , \quad \cf_{45} = - \cf_{36} \ ,\\
\nonumber && \cf_{1\tau} = \cf_{27} \ , \quad \cf_{2\tau} = -
\cf_{17} \ , \quad \cf_{3\tau} = \cf_{47} \ , \quad
\cf_{4\tau} = -\cf_{37} \ , \quad \cf_{5\tau} = \cf_{67} \ , \quad
\cf_{6\tau} = -\cf_{57} \ .
\eea
They imply the condition
 \bea \nonumber
 0 = -\cf_{13} +\cf_{24} + \cf_{67} - \cf_{5\tau} = \Omega_5 \haken
 \cf \ ,
 \eea
which is the stability condition associated to $\Omega_5$.}

For our case of the cone $C(\mathrm{Sp}(2)/\mathrm{Sp}(1))$, the $\mathrm{Sp}(2)$-instanton equation explicitly gives
the flow equations
\begin{align}
\label{eq:inst_flow_sp2}
 \nonumber &\frac{\diff X_1}{\diff \tau} \= -X_1 + \com{X_3}{X_5} \=
             -X_1 +\com{X_4}{X_6} \= -X_1 - \com{X_2}{X_7} \ ,\\[4pt]
  \nonumber &\frac{\diff X_2}{\diff \tau} \= -X_2 - \com{X_4}{X_5} \=
              -X_2 + \com{X_3}{X_6} \= -X_2 + \com{X_1}{X_7} \ ,\\[4pt]
           &\frac{\diff X_3}{\diff \tau} \= -X_3 - \com{X_1}{X_5} \=
             -X_3 - \com{X_2}{X_6} \= -X_3 - \com{X_4}{X_7} \ ,\\[4pt]
 \nonumber &\frac{\diff X_4}{\diff \tau} \= -X_4 + \com{X_2}{X_5} \=
             -X_4 - \com{X_1}{X_6} \= -X_4 + \com{X_3}{X_7} \ ,\\[4pt]
 \nonumber & \frac{\diff X_5}{\diff \tau} \= - 2 X_5 - \com{X_6}{X_7}
             \ , \qquad \frac{\diff X_6}{\diff \tau} \= - 2 X_6 +
             \com{X_5}{X_7} \ , \qquad 
       \frac{\diff X_7}{\diff \tau} \= - 2 X_7 - \com{X_5}{X_6} \ ,
\end{align}
while the algebraic conditions read
\small
\bea
\label{eq:inst_alg_sp2}
&& 4X_5 \= \com{X_1}{X_3}-\com{X_2}{X_4} \ , \qquad 4X_6 \=
\com{X_1}{X_4} + \com{X_2}{X_3} \ , \qquad 
4 X_7 \= - \com{X_1}{X_2} - \com{X_3}{X_4} \ . \nonumber \\ &&
\eea
\normalsize
One recognizes immediately the form of the intersection of three
Hermitian Yang-Mills instanton equations as discussed in
Section~\ref{sect:cone_instantons}.\footnote{Note that in these
  Hermitian Yang-Mills equations one has slightly different algebraic quiver relations and a different scaling factor in the flow equations
for $X_5$, $X_6$ and $X_7$.}

\paragraph{Scalar solution.} Setting $X_a =\l(r)\, I_a$ for $a=1,  2,3, 4$ and $X_{\a} = \psi(r)\, I_{\a}$ for $\a=5,6,7$ with functions $\l$ and $\psi$, the equivariance conditions are
automatically satisfied and the instanton equations reduce to the system
\bea
\label{eq:scalar_sp2}
\nonumber \dot{\l} &=& \l \, (\psi-1) \ ,\\[4pt]
\dot{\psi} &=& 2 \, \psi \, (\psi-1) \ ,\\[4pt]
\nonumber \l^2 &=& \psi \ .
\eea
This is exactly the scalar form for instanton equations on cones over 3-Sasakian manifolds discussed in~\cite{Harland:2011zs}, where they
give the analytic solutions
\bea
\label{eq:scalar_sp2_sols}
\psi(\tau) \= \big( 1+ \e^{2(\tau-\tau_0)}\big)^{-1} \qquad \text{and}
\qquad \l(\tau) \= \pm \, \psi(\tau)^{1/2} \ .
\eea

\subsection{Representations of $\mathrm{Sp}(2)$}
\label{sec:repres_sp2}
In the following we collect the weight diagrams and some explicit choices for the generators used in the main text.
Due to the structure constants \eqref{eq:struc_const_sp2}, the root system of the Lie algebra of $\mathrm{Sp}(2)$
 is spanned by
\bea
\footnotesize
    \begin{tikzpicture}[->,scale=1.35]
    \node (a) at (0,0) {};
    \node (b) at (-1,1)  {{\small${(-1,1)}$}};
    \node (c) at (-2,0)  {{\small${(-2,0)}$}};
    \node (d) at (-1,-1) {{\small${(-1,-1)}$}};
    \node (e) at (0,-2)  {{\small${(0,-2)}$}};

     \path[->](a) edge node [above right]{$I_{\bar{1}}^-$} (b);
     \path[->](a) edge node [below]{$I_{\bar{3}}^-$} (c);
     \path[->](a) edge node [below right]{$I_{\bar{2}}^-$} (d);

     \path[blue, ->](a) edge node [right]{$I_{\bar{4}}^-$} (e);
     
\end{tikzpicture}
\eea
\normalsize
together with the conjugate operators, where the blue arrow represents the ladder operator $I_{\bar{4}}^-$ of the 
subalgebra $\mathfrak{sp}(1)\cong \mathfrak{su}(2)$.

\paragraph{Fundamental representation $\mbf{\underline{4}}$\,.}
The generators are those in \eqref{eq:def_generators_sp2}, and the weight diagram of the fundamental representation $\mbf{\underline{4}}$ is given by
\bea
\label{eq:weight_sp2_fund}
    \begin{tikzpicture}[->,scale=1.1]
    \node (a) at (1,0)  {{\small${(1,0)}$}};
    \node (b) at (0,1)  {{\small${(0,1)}$}};
    \node (c) at (-1,0) {{\small${(-1,0)}$}};
    \node (d) at (0,-1) {{\small${(0,-1)}$}};

     \path[<-](b) edge node [below left]{} (a);
     \path[<-](c) edge node [above]{} (b);
     \path[<-](c) edge node [above left]{} (a);
     \path[<-](c) edge node [above left]{} (d);
     \path[<-](d) edge node [above left]{} (a);

     \path[blue, <-](d) edge node [left]{} (b);
     
\end{tikzpicture}
\eea
\paragraph{Representation $\mbf{\underline{5}}$\,.}
The five-dimensional representation is characterized by the weight diagram
\bea
\label{eq:weight_sp2_5}
    \begin{tikzpicture}[->,scale=.96]
    \node (a) at (0,0)  {{\small${(0,0)}$}};
    \node (b) at (1,1)  {{\small${(1,1)}$}};
    \node (c) at (-1,1) {{\small${(-1,1)}$}};
    \node (d) at (-1,-1){{\small${(-1,-1)}$}};
    \node (e) at (1,-1) {{\small${(1,-1)}$}};

     \path[<-](c) edge node {} (b);
     \path[<-](d) edge node {} (e);
     \path[<-](a) edge node {} (b);
     \path[<-](c) edge node {} (a);
     \path[<-](d) edge node {} (a);
     \path[<-](a) edge node {} (e);

     \path[blue, <-](e) edge node [left]{} (b);
     \path[blue, <-](d) edge node [left]{} (c);
     
\end{tikzpicture}
\eea
and the generators can be chosen as 
\small
\bea
\label{eq:rep_sp2_5}
\nonumber I_{\bar{1}}^- &=& \begin{pmatrix}
           0 & 0 &\sqrt{2} & 0 &0 \\
           0 & 0 &0 & 0 &0 \\
           0& 0& 0& 0&-\sqrt{2}\\
           0 & 0 &0 & 0 &0  \\
           0 & 0 &0 & 0 &0
          \end{pmatrix} \ , \quad
I_{\bar{2}}^- \= \begin{pmatrix}
           0 & 0 &0 & 0 &0 \\
           0 & 0 &\sqrt{2} & 0 &0 \\
           0 & 0& 0& -\sqrt{2}&0\\
           0 & 0 &0 & 0 &0  \\
           0 & 0 &0 & 0 &0
          \end{pmatrix} \ , \quad
I_{\bar{3}}^- \=  \begin{pmatrix}
           0 & 0 &0 & -1 &0 \\
           0 & 0 &0 & 0 &1 \\
           0 & 0 & 0& 0&0\\
           0 & 0 &0 & 0 &0  \\
           0 & 0 & 0 & 0 &0
          \end{pmatrix} \ ,\\[4pt]
I_{\bar{4}}^- &=&  \begin{pmatrix}
           0 & 0 &0 & 0 &0 \\
           -1 & 0 &0 & 0 &0 \\
           0 & 0& 0& 0&0\\
           0 & 0 &0 & 0 & 0  \\
           0 & 0 &0 & 1 &0
           \end{pmatrix} \ , \quad
I_7 \= \mathrm{diag} \lb -1,-1,0,1,1\rb \ , \quad I_8 \= \mathrm{diag}
\lb 1,-1,0,1,-1\rb \ .
\eea
\normalsize
\paragraph{Adjoint representation $\mbf{\underline{10}}$\,.}
The generators of the adjoint representation are determined by the structure constants (\ref{eq:struc_const_sp2}), and one obtains the weight diagram
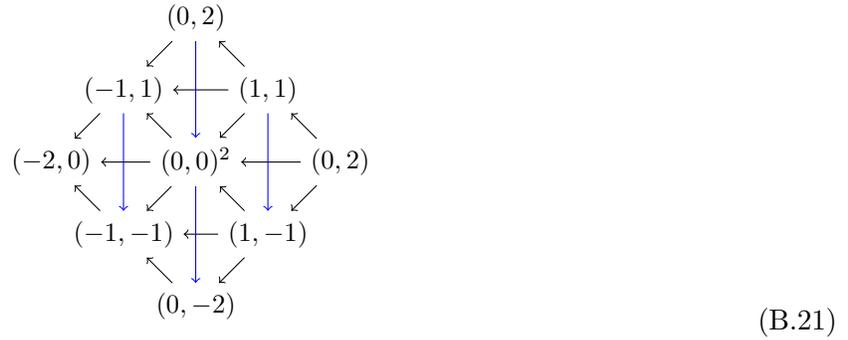
\bea
\label{eq:weight_sp2_adj}
 \begin{tikzpicture}[->,scale=.96]
    \node (a) at (0,0) {{\small${(0,0)^2}$}};
    \node (b) at (1,1) {{\small${(1,1)}$}};
    \node (c) at (0,2) {{\small${(0,2)}$}};
    \node (d) at (-1,1) {{\small${(-1,1)}$}};
    \node (e) at (-2,0) {{\small${(-2,0)}$}};
    \node (f) at (-1,-1) {{\small${(-1,-1)}$}};
    \node (g) at (0,-2) {{\small${(0,-2)}$}};
    \node (h) at (1,-1) {{\small${(1,-1)}$}};
    \node (i) at (2,0) {{\small${(0,2)}$}};

     \path[<-](a) edge node {} (b);
     \path[<-](f) edge node {} (a);
     \path[<-](g) edge node {} (h);
     \path[<-](h) edge node {} (i);
     \path[<-](e) edge node {} (d);
     \path[<-](d) edge node {} (c);

     \path[<-](c) edge node {} (b);
     \path[<-](b) edge node {} (i);
     \path[<-](d) edge node {} (a);
     \path[<-](a) edge node {} (h);
     \path[<-](e) edge node {} (f);
     \path[<-](f) edge node {} (g);

     \path[<-](d) edge node {} (b);
     \path[<-](e) edge node {} (a);
     \path[<-](a) edge node {} (i);
     \path[<-](f) edge node {} (h);

     \path[blue, <-](a) edge node [left]{} (c);
     \path[blue, <-](g) edge node [left]{} (a);
     \path[blue, <-](f) edge node [left]{} (d);
     \path[blue, <-](h) edge node [left]{} (b);
     
\end{tikzpicture}
\eea
where the center node at $(0,0)$ has multiplicity $2$.
\paragraph{Representation $\mbf{\underline{14}}$\,.}
We do not determine the generators explicitly here, but restrict our
attention to the weight diagram of the 14-dimensional representation
which is given by
\bea
\label{eq:weight_14}
 \begin{tikzpicture}[->,scale=.96]
    \node (a) at (2,2)  {{\small${(2,2)}$}};
    \node (b) at (0,2)  {{\small${(0,2)}$}};
    \node (c) at (-2,2) {{\small${(-2,2)}$}};
    \node (d) at (1,1)  {{\small${(1,1)}$}};
    \node (e) at (-1,1) {{\small${(-1,1)}$}};
    \node (f) at (2,0)  {{\small${(2,0)}$}};
    \node (g) at (0,0)  {{\small${(0,0)^2}$}};
    \node (h) at (-2,0) {{\small${(-2,0)}$}};
    \node (i) at (1,-1) {{\small${(1,-1)}$}};    
    \node (j) at (-1,-1){{\small${(-1,-1)}$}};
    \node (k) at (2,-2) {{\small${(2,-2)}$}};  
    \node (l) at (0,-2) {{\small${(0,-2)}$}};  
    \node (m) at (-2,-2){{\small${(-2,-2)}$}};

     \path[->](a) edge node {} (b);
     \path[->](b) edge node {} (c);
     \path[->](d) edge node {} (e);
     \path[->](f) edge node {} (g);
     \path[->](g) edge node {} (h);
     \path[->](i) edge node {} (j);
     \path[->](k) edge node {} (l);
     \path[->](l) edge node {} (m);

     \path[->](d) edge node {} (b);
     \path[->](e) edge node {} (c);
     \path[->](j) edge node {} (h);
     \path[->](g) edge node {} (e);
     \path[->](f) edge node {} (d);
     \path[->](i) edge node {} (g);
     \path[->](k) edge node {} (i);
     \path[->](l) edge node {} (j);

      \path[->](a) edge node {} (d);
     \path[->](b) edge node {} (e);
     \path[->](e) edge node {} (h);
     \path[->](g) edge node {} (j);
     \path[->](j) edge node {} (m);
     \path[->](i) edge node {} (l);
     \path[->](f) edge node {} (i);
     \path[->](d) edge node {} (g);

     \path[blue, ->](a) edge node [left]{} (f);
     \path[blue, ->](f) edge node [left]{} (k);
     \path[blue, ->](d) edge node [left]{} (i);
     \path[blue, ->](b) edge node [left]{} (g);
     \path[blue, ->](g) edge node [left]{} (l);
     \path[blue, ->](e) edge node [left]{} (j);
     \path[blue, ->](c) edge node [left]{} (h);
     \path[blue, ->](h) edge node [left]{} (m);

\end{tikzpicture}
\eea

\newpage

\bibliographystyle{utphys}
{\small 
\providecommand{\href}[2]{#2}\begingroup\raggedright\endgroup
}

\end{document}